\documentclass[useAMS,usenatbib]{mnras}
\usepackage{deluxetable}
\usepackage{amssymb}

\title[Searching for fading CSS sources]{Young but fading radio sources: searching for remnants among compact steep-spectrum radio sources}
\author[M. Orienti et al.]
  {M. Orienti$^{1}$\thanks{E-mail: orienti@ira.inaf.it},
M. Murgia$^{2}$,
D. Dallacasa$^{1,3}$
G. Migliori$^{1}$,
F. D'Ammando$^{1}$\\
$^{1}$INAF - Istituto di Radioastronomia, Via P. Gobetti 101, I-40129 Bologna, Italy\\
$^{2}$INAF - Osservatorio Astronomico di Cagliari, 
Via della Scienza 5, I09047 Selargius, Italy\\
$^{3}$Dipartimento di Fisica e Astronomia, Universit\`a di Bologna,
Via Gobetti 93/2, I-40129 Bologna, Italy\\
}
\date{Received \today; accepted ?}

\pagerange{\pageref{firstpage}--\pageref{lastpage}} \pubyear{2002}

\def\LaTeX{L\kern-.36em\raise.3ex\hbox{a}\kern-.15em
    T\kern-.1667em\lower.7ex\hbox{E}\kern-.125emX}

\begin{document}

\label{firstpage}

\maketitle

\begin{abstract}
 The incidence of young but fading radio sources provides important
 information on the life cycle of radio emission in radio-loud
 active galactic nuclei. Despite its
 importance for constraining the models of radio source evolution,
 there are no systematic studies of remnants in complete samples of young radio
 sources. We report results on the study of 18 compact
 steep-spectrum (CSS) radio sources, selected from the statistically
 complete B3-VLA CSS sample, characterized by a steep optically-thin spectrum
 ($\alpha \geq 1.0$) and no
 core detection in earlier studies.
 Our deep multi-frequency Very Large Array (VLA), 
  pc-scale Very Long Baseline Array (VLBA), and eMERLIN observations
  allowed us to locate the core component in 10 objects.
  In 3 CSS sources there is no clear evidence of present-time
    active regions, suggesting
  they are likely in a remnant phase.
  Among sources with core detection, we find 3 objects that have no
  clear active
  regions (hotspots) at the edges of the radio structure, suggesting that the
  radio emission may have just restarted.
  Our results support a power-law distribution of the
  source ages, although the poor statistics prevents us from setting
  solid constraints on the percentage of remnants and restarted
  sources in 
  sub-populations of radio sources.

 \end{abstract}


\section{Introduction}

Understanding how radio emission originates and evolves in
extragalactic radio sources is one of the
greatest challenges faced by modern astrophysics.
The improvement in sensitivity recently achieved at radio wavelengths,
together with the availability of multi-band information from infrared
to gamma-ray observations is allowing a major step forward in our
understanding of
the physics of radio sources and the global properties of their host
galaxies.\\
It appears that radio-loud active galactic nuclei (AGN) represent
  only a small fraction (about 10 per cent) of the total population of
  their host galaxies, suggesting that radio emission may be a
  transient phase in the lives of these systems.
The typical duration of the active phase is 10$^{7-8}$ yr, much shorter than
the age of the host galaxy
\citep{parma07,orru10,harwood17}. Detection of dying radio emission
implies that at some point the injection of relativistic plasma switches
off and the radio emission rapidly fades due to energy losses
\citep[][]{komissarov94,murgia99,murgia11,hardcastle18}. For this reason, search for remnants
is not an easy task in flux-density-limited catalogs
\citep[e.g.][]{slee01,parma07}, and many relics have been discovered
serendipitously \citep[e.g.][]{brienza16,randria20}.\\
After the radio emission has switched off, it is possible that a second
epoch of radio activity starts. If the time elapsed is
  comparable, or longer than the radiative time of the outer lobes,
  only the new emission could be visible 
precluding us to identify the radio source as a restarted object.
On the other hand, if the time elapsed between
  two consecutive episodes of radio emission is about 10 per cent of the
  total source age,
  we observe double-double
  radio sources \citep[e.g.,][]{lara99,schoenmakers00,konar06}.
Radio sources
showing more than two activity periods are very rare
\citep{brocksopp07,hota11,singh16}. However, remnants and restarted
objects are key to understand the life cycle of the radio emission and
how active and quiescent periods are interleaved \citep{morganti17}.\\
Systematic studies of remnants and restarted sources have been
recently boosted mainly at low frequencies,
thanks to the availability of high-sensitivity
deep-field observations
\citep[e.g.][]{saripalli12,hardcastle16,brienza17,godfrey17,mahatma18,hardcastle19,
mahatma19,jurlin20,morganti21,jurlin21}. The
relatively high fraction of remnants and restarted objects found in
these works requires a dominant population of short-lived radio
sources \citep{shabala20}.\\ 
There are indications that a significant fraction of compact steep spectrum
(CSS) and peaked spectrum (PS) sources, 
which are radio sources in an early
evolutionary stage \citep[see e.g.][]{odea21} will
not be able to evolve into classical extended radio galaxies. First,
there is a count excess of CSS/PS sources in flux-limited
samples, which cannot be explained even when
luminosity evolution is taken into account
\citep[e.g.][]{odea21}. Secondly, 
the age distribution estimated in a sample of CSS/PS sources peaks
below 500 yr \citep{gugliucci05}. Finally, the fast evolution of the
radio spectrum that we
observe in some PS candidates suggests that not all the objects would
become classical Fanaroff-Riley radio sources \citep{mo20}. 
To explain these short-lived
objects, \citet{czerny09} suggested
a radiation pressure instability scenario which may cause an
intermittent activity of the central
engine: outbursts of radio emission,
with a duration of 10$^{3-4}$ years,
repeat regularly every 10$^{5-6}$
years. The detection of quasars that have
transitioned from radio-quiet to radio-loud state on decadal time,
may provide another indication that intermittent episodes of
short-lived radio emission may be common in AGN \citep{nyland21}. 
Extended arcsecond scale radio emission (i.e. tens to hundred kpc and
ages of 10$^{7-8}$ yr)
likely associated with young radio sources has been found only in a handful
of PS objects \citep{cstan05,tinti05}.  
However, the discovery of relics from previous
radio activities at parsec-scale
distance from the young radio sources J1511+0518 \citep{mo08}, and
OQ\,208 \citep{luo07}, suggests that the intermittency of short-lived active
phases may have duty cycles of the order of 10$^{3-4}$ yr.
If the distribution of the source age follows a power-law as
suggested by \citet{shabala20}, one would expect a larger incidence of
remnants among compact symmetric objects (CSOs), i.e. sources with
linear size (LS) $\lesssim$ 1 kpc, rather than
middle-sized symmetric objects (MSOs), i.e. objects with 1 kpc
$\lesssim$LS$\lesssim$20 kpc\footnote{For details on the populations of
CSOs and MSOs see the review by \citet{odea21}}. \\
So far, not many studies searching for young but dying objects have
been carried out \citep[e.g.,][]{kunert07}, and only a handful of young but
fading sources are known \citep[e.g.,][]{mo10,callingham15,osullivan21}. \\
In this paper we aim at inferring, for the first time, the fraction of
remnants in a complete sample of CSS
radio sources. This will allow us to investigate the
incidence of short-lived objects and to shed a light on the life cycle
of the radio emission in its early evolutionary stages after the outburst.
Multi-frequency very large array (VLA), very long baseline array
(VLBA), and eMERLIN observations of the target sources were performed
in order to constrain the synchrotron radio spectrum and unveil the
presence of active cores that may have been missed in previous
observations with lower sensitivity. \\
The paper is organized as follows: in Section 2 we present the source
sample, in Section 3 we describe the radio
observations. Results are presented in Section 4 and discussed in
Section 5. In Section 6 we draw our summary.\\
Throughout this paper, we assume the following cosmology: $H_{0} =
70\; {\rm km/s\, Mpc^{-1}}$, 
$\Omega_{\rm M} = 0.27$ and $\Omega_{\rm \Lambda} = 0.73$,
in a flat Universe. The spectral index $\alpha$
is defined as 
$S {\rm (\nu)} \propto \nu^{- \alpha}$. \\

\section{The sample}

To look for young but fading sources we constructed a complete sample
of candidate remnants from the B3-VLA CSS sample \citep{fanti01}. 
The B3-VLA CSS sample consists of 87 CSS/PS sources with
flux density $>$ 0.8 Jy at 0.4 MHz, a steep radio spectrum
($\alpha_{\rm 4.8}^{\rm 8.4} \gtrsim 0.7$),
and an angular size $\lesssim$2 arcsec (i.e. linear size
$\lesssim$ 20 kpc). 
Multi-frequency VLA and VLBA
observations were performed to study the radio structure:
25 out of the 87 sources of the sample with largest angular scale
(LAS) $<$ 0.25 arcsec were
target of VLBA observations \citep{dd02a,mo04}, whereas the
remaining 62 sources were target of VLA, EVN and
MERLIN observations \citep{fanti01,dd02b,rossetti06}.\\
Among the sources of the B3-VLA CSS sample we selected all the objects
with a spectral index $\alpha_{4.8}^{8.4} >$ 1.0 
and without any core detection in
earlier observations \citep{fanti01,dd02a,dd02b,mo04,rossetti06}.
We end up with a sample of 18 CSS sources: 6 with LAS$<$0.25 arcsec, and
12 with LAS$>$0.25 arcsec.
The former were target of VLBA observations
while
the latter of VLA observations.
For 6 of them we also got high-resolution eMERLIN
observations.
We complement our observations with
archival high angular resolution VLA observations
already published in
\citet{fanti01} and \citet{rossetti06}.\\ 
The source sample and the source properties are 
reported in Table \ref{sample}\footnote{We still consider MSOs the
three sources with LS between 20 and 30 kpc.}.\\

\begin{table*}
\caption{The source sample. Column 1: source name; column 2: optical
  identification (G = galaxy; E: empty field); 
column 3: redshift (p = photometric redshift); 
columns 4 and 5: flux density at
4.8 and 8.4 GHz, respectively, from \citet{fanti01}; column 6: spectral
index between 4.8 and 8.4 GHz; column 7: angular size from
\citet{fanti01}; column 8: projected linear size. When the redshift is unknown
we assume $z$=1.05 \citep[see,][]{fanti01};
column 9: reference for the redshift: 1:
\citet{thompson00}; 2: \citet{fanti01}; 3: \citet{jarvis01}; 4:
\citet{cruz06}; 5: \citet{pahre95}; 6: \citet{fanti11}; 7:
\citet{allington88}; 8: \citet{thompson94}; 9: \citet{rawlings01}; 10:
\citet{wegner03}; 11: Sloan Digital Sky Survey Data Release 13
\citep{albareti17}.} 
\begin{center}
\begin{tabular}{cccccccccc}
\hline
Source Name & ID & z & S$_{\rm 4.8 GHz}$ & $S_{\rm 8.4 GHz}$&
$\alpha_{4.8}^{8.4}$&LAS&LS&Ref.\\
    &  &  & mJy & mJy &   & arcsec & kpc & \\
\hline
B3-0003+387 & G & 1.47 &  52.8 &  22.4 & 1.4 & 1.3 & 11.2 & 1 \\
B3-0034+444 & G & 2.79 & 187.7 &  96.5 & 1.0 & 3.2 & 25.6 & 2 \\ 
B3-0128+394 & G & 0.929& 128.9 &  42.8 & 1.1 & 2.7 & 21.5 & 3 \\
B3-0140+387 & G & 2.9p &  71.0 &  27.5 & 1.5 & 0.7 &  5.5 & 2 \\
B3-0748+413B\tablenotemark{a}& E &  -   &  42.5 &  23.4 & 1.2 & 0.4 &  3.2 & - \\
B3-0754+396\tablenotemark{a} & G & 2.119& 143.1 &  74.7 & 1.1 & 2.2 & 19.6 & 2 \\
B3-0810+460B& G & 0.620& 466.4 & 153.9 & 1.1 & 0.63&  4.3 & 4 \\ 
B3-0856+406 & G & 2.285&  37.8 &  14.0 & 1.5 & 0.8 &  6.7 & 5 \\
B3-1016+443 & G & 0.7p &  80.1 &  39.3 & 1.2 & 0.11&  0.7 & 6 \\
B3-1039+424\tablenotemark{a} & E &  -   &  66.2 &  32.3 & 1.1 & 1.5 & 12.2 & - \\
B3-1049+384 & G & 1.018& 200.6 & 104.9 & 1.1 & 0.1 &  0.8 & 7 \\
B3-1133+432 & E &  -   & 499.1 & 286.3 & 1.0 & 0.07&  0.5 & - \\
B3-1143+456 & G & 0.762& 137.1 &  64.7 & 1.4 & 0.8 &  6.0 & 8 \\ 
B3-1212+380 & G & 0.947&  63.8 &  33.3 & 1.2 & 0.3 &  2.4 & 9 \\
B3-1216+402 & G & 0.756& 161.6 &  61.5 & 1.0 & 3.8 & 28.3 & 8 \\
B3-1225+442 & G & 0.348&  96.3 &  45.7 & 1.2 & 0.2 &  1.0 &10 \\
B3-1340+439 & E &   -  & 130.4 &  67.0 & 1.2 & 0.07&  0.5 & - \\
B3-1449+421 & G & 0.86p& 147.4 &  66.5 & 1.6 & 0.08&  0.6 &11 \\
\hline
\end{tabular}
\end{center}
\tablenotetext{a}{Source re-classified as a large symmetric object.}
\label{sample}
\end{table*}

\section{Radio observations}

\subsection{VLA observations}
\label{vla-section}

Multi-frequency VLA observations of the 12 CSS sources 
with LAS$>$0.25 arcsec (LS$> $ 1 kpc) 
were carried out during four runs between 2015 and 2017 (Table
\ref{radio-log}).  
Observations were performed in L (1$-$2 GHz) and S
(2$-$4 GHz) bands with the array in A-configuration, while
observations in U band (13$-$15 GHz)
were performed with the array in B-configuration. The average angular
resolution is about 1.2 arcsec, 0.6 arcsec, and 0.5
arcsec in L, S, and U bands, respectively.
L-band observations have a bandwidth of 1 GHz, whereas 
observations in S and U bands have a bandwidth of 2 GHz.
In each frequency band the bandwidth was
divided into 16 spectral windows. Depending on the local sidereal time
range of
the observing runs, either 3C\,48 or 3C\,286 was used as
primary, band pass, and electric vector position angle (EVPA) calibrator.
Each source was
observed for about 8 min per frequency. Secondary calibrators were
chosen to minimize the antenna slewing. In two observing runs (15A-140
in L and S bands, and 16B-107 in U
band) we observed as secondary calibrators J0713+4349 and J0111+3906,
respectively, which we also used as leakage (D-term) calibrators.\\
Calibration was performed using the \texttt{CASA} software
\citep{mcmullin07}
following
the standard procedure for the VLA. Parts of L and S bands were highly
affected by RFI. In particular in L band we had to fully
flag some spectral windows. 
Errors on the amplitude calibration are conservatively 3 per cent in
L and U bands, and 5 per cent in S band. 
After the a-priori calibration,
imaging was done with the \texttt{CASA} task \texttt{TCLEAN}. When 
other objects were present in the field of view, mainly in L-band, we
made use of the task \texttt{CLEAN} which allows imaging with several
phase centers. \\
Few iterations of imaging and self-calibration were performed before
creating the final images, which were produced with Briggs weighting
and using Multi-frequency synthesis algorithm with 2 Taylor
terms. Flux
densities were measured using the \texttt{CASA} task \texttt{IMFIT}
which performs a Gaussian fit on the image plane. If the source (source
components) could not be fitted by a Gaussian, we extracted the flux
density on a selected polygonal area using the \texttt{CASA} task
\texttt{VIEWER}. 
In Table \ref{vla-flux} we report the flux densities at 1.4, 3 and 14
GHz. To maximize the spectral coverage, we produced an image 
for each spectral window (see Section \ref{synage}). Errors on the
flux densities are estimated by $\sigma = \sqrt{\sigma_{\rm
    cal}^{2} + \sigma_{\rm fit}^{2} + \sigma_{\rm rms}^{2}}$, 
where $\sigma_{\rm cal}$ is the uncertainty on the
amplitude calibration, $\sigma_{\rm fit}$ is the error on the Gaussian
fit, and $\sigma_{\rm rms}$ is the
1-$\sigma$ noise level measured on the image plane, which is
$\sigma_{\rm rms} = {\rm rms} \times \sqrt{\theta_{\rm source}/
  \theta_{\rm beam}}$, where $\theta_{\rm source}$ is the source area,
and $\theta_{\rm beam}$ is the area of the Gaussian restoring beam.\\
Final images of the CSS sources are presented in Fig. A1, while
images of the sources falling within the VLA field
of view showing a resolved structure are presented in
Fig. B1. Properties of field sources are reported in Table B1. \\

\begin{table}
  \caption{Log of radio observations. Column 1: date of observation;
    column 2: project code; column 3: array; column 4: observing band;
  column 5: polarization calibration.}
  \begin{center}
\begin{tabular}{lcccc}
\hline
Date & Project & Array & Band & Polarization \\
\hline
&&&&\\
2015-02-15 & 15A-140 & VLA-B & U  & Y   \\
2015-08-14 & 15A-140 & VLA-A & L,S & N \\
2017-01-06 & 16B-107 & VLA-A & L,S & Y \\
2017-10-19 & 17B-153 & VLA-B & U   & N \\
2010-12-31 & BO038A  & VLBA  & C,X & N \\
2011-01-16 & BO038C  & VLBA  & C,X & N \\
2011-01-21 & BO038B  & VLBA  & C,X & N \\
2020-08-03/09&CY10208&eMERLIN& L,C& N \\
&&&&\\
\hline
\end{tabular}
\end{center}
\label{radio-log}
\end{table}

\begin{table*}
\caption{Multi-frequency VLA flux density of CSS sources with LAS$>0.25$
  arcsec. 
Column 1:
  source name; column 2: source component; columns 3-5: flux density (in
  mJy) at
  1.4, 3, and 14 GHz, respectively;
  column 6: spectral index between 3 and 14 GHz.}
  \begin{center}
\begin{tabular}{cccccc}
\hline
Source & Comp. & S$_{1.4}$ & S$_{3}$ & S$_{14}$ & $\alpha_{\rm 3 GHz}^{\rm 14 GHz}$ \\
(1) & (2) & (3) & (4) & (5) & (6) \\
\hline
B3-0003+387 & N & 262.8$\pm$ 7.9 & 114.5$\pm$3.4 & 11.3$\pm$0.4 &
1.5$\pm$0.1 \\ 
            & S & 227.0$\pm$ 6.8 & 108.8$\pm$3.3 & 12.4$\pm$0.4 &
1.4$\pm$0.1 \\  
            &Tot& 487.4$\pm$14.6 & 214.6$\pm$6.4 & 23.4$\pm$0.7 &
1.4$\pm$0.1 \\
B3-0034+444 & N & 581.8$\pm$17.5 & 314.3$\pm$9.4 & 56.1$\pm$1.7& 1.1$\pm$0.1\\
            & S &  99.1$\pm$6.0  &  26.4$\pm$0.8 & 2.0$\pm$0.1 &
1.7$\pm$0.1 \\
            & A &     -          & 243.6$\pm$7.5 & 51.8$\pm$1.5& 1.1$\pm$0.2\\
            & B &     -          &  45.1$\pm$2.7 &  4.8$\pm$0.2&
1.6$\pm$0.2 \\
            & C &     -          &   9.3$\pm$1.7 &  1.1$\pm$0.2& 1.4$\pm$0.2\\
            & D &     -          &  17.8$\pm$2.0 &  1.0$\pm$0.2& 1.9$\pm$0.2\\
            &Tot&  707.8$\pm$21.2& 347.6$\pm$10.4& 58.1$\pm$1.7&
1.2$\pm$0.1\\
B3-0128+394 & N & 173.2$\pm$ 5.4 &  86.4$\pm$4.0 & 13.4$\pm$0.5&
1.2$\pm$0.1 \\
            & S & 163.0$\pm$5.1  &  75.5$\pm$4.0 & 11.2$\pm$0.6& 1.2$\pm$0.1\\
            &Tot& 334.3$\pm$10.0 & 162.1$\pm$4.9 & 25.0$\pm$0.8&
1.2$\pm$0.1\\
B3-0140+387 & N &     -          &  90.6$\pm$2.7 &  6.3$\pm$0.2& 1.7$\pm$0.1\\
            & S &     -          &  77.2$\pm$2.3 &  7.9$\pm$0.3& 1.5$\pm$0.1\\
            &Tot&  450.2$\pm$13.6& 167.1$\pm$5.0 & 14.0$\pm$0.4&
1.6$\pm$0.1\\
B3-0748+413B& N &  200.3$\pm$6.1 &  83.5$\pm$4.4 & 13.6$\pm$0.5& 1.2$\pm$0.1\\
            & C &    2.9$\pm$0.2 &  1.5$\pm$0.3 &  0.7$\pm$0.1& 0.6$\pm$0.1\\
            & S &   20.5$\pm$1.2 &   9.1$\pm$0.7 &  2.6$\pm$0.2&
0.9$\pm$0.2\\ 
            &Tot&  231.7$\pm$7.0 &  94.1$\pm$4.9 & 17.6$\pm$0.5&
1.1$\pm$0.1\\
B3-0754+396 & N &  525.7$\pm$15.8& 247.8$\pm$13.1& 45.7$\pm$1.4& 1.1$\pm$0.1\\
            & S &    7.8$\pm$ 0.3&   3.4$\pm$0.3 &  0.5$\pm$0.1& 1.2$\pm$0.1\\
            &Tot&  533.5$\pm$16.0& 251.2$\pm$13.3& 46.2$\pm$1.5& 1.1$\pm$0.1\\
B3-0810+460B& N &          -     & 216.1$\pm$10.8& 35.0$\pm$1.1&
1.2$\pm$0.1\\       
            & S &           -    & 276.6$\pm$13.8& 54.9$\pm$1.6& 1.1$\pm$0.1\\
            &Tot& 1146.5$\pm$34.3& 492.7$\pm$24.6& 90.9$\pm$2.8&
1.1$\pm$0.1\\
B3-0856+406 & E &      -         &  54.3$\pm$3.1 &  4.2$\pm$0.2& 1.7$\pm$0.1\\
            & W &      -         &  26.3$\pm$1.5 &  3.8$\pm$0.2& 1.3$\pm$0.1\\
            &Tot&  225.5$\pm$ 6.8&  80.6$\pm$4.5 &  7.7$\pm$0.3& 1.6$\pm$0.1\\
B3-1039+424 & N &    7.5$\pm$0.5 &   2.7$\pm$0.3 &  0.4$\pm$0.1& 1.3$\pm$0.1\\
            & C &    0.2$\pm$0.1 &  0.15$\pm$0.03&  0.05$\pm$0.02 & 0.4$\pm$0.7\\
            & S &  273.6$\pm$8.2 & 118.2$\pm$6.5 & 17.8$\pm$0.6& 1.3$\pm$0.1\\
            &Tot&  281.1$\pm$8.4 & 120.9$\pm$6.6 & 18.0$\pm$0.6&
1.3$\pm$0.1\\
B3-1143+456 & N &        -       &  45.8$\pm$2.6 &  4.2$\pm$0.2& 1.6$\pm$0.1\\
            & S &        -       & 230.8$\pm$12.0& 27.8$\pm$0.8& 1.4$\pm$0.1\\
            &Tot&  693.6$\pm$23.8& 276.6$\pm$13.8& 31.8$\pm$1.1&
1.5$\pm$0.1\\
B3-1212+380 &Tot& 304.6$\pm$ 9.1 & 121.6$\pm$6.1 & 17.5$\pm$0.6&
1.3$\pm$0.1\\
B3-1216+402 & N &  155.3$\pm$4.7  &  71.5$\pm$4.0 & 14.4$\pm$0.4& 1.1$\pm$0.1\\ 
           & S &  211.4$\pm$6.3  & 107.2$\pm$5.4 & 22.3$\pm$0.7& 1.0$\pm$0.1\\
           &Tot&  364.3$\pm$10.9 & 178.7$\pm$9.0 & 37.2$\pm$1.1& 1.0$\pm$0.1\\
\hline
\end{tabular}
\end{center}
\label{vla-flux}
\end{table*}

\subsection{VLBA observations and data reduction}

VLBA observations in C and X bands of the CSS sources with LAS$<$0.25 arcsec 
were carried out between 2010 December and 2011 January in dual
polarization (Table \ref{radio-log}) at 32-MHz bandwidth, with the
exception of B3-1133+432 (both bands) and B3-1049+384 (C band only)
whose data were already published in \citet{mo04}.\\
Each source was observed for about 2 hr spread into twelve 10-min scans.
No fringes were detected for B3-1225+442 in C band.
Calibration and data reduction were performed following the standard
procedures described in the Astronomical Image Processing System
(\texttt{AIPS}) cookbook. J0927+3902 was used to generate the band pass
correction. Amplitudes were calibrated using antenna system
temperatures and antenna gains. 
Uncertainties on the amplitude calibration are 
approximately 10 per cent at both frequencies.\\
Images were produced in \texttt{AIPS} with the task \texttt{IMAGR}.
Few iterations of imaging and phase-only self-calibration were
performed before producing final images. In addition to
full-resolution images, we produced a set of images in C and X bands
with the same {\it uv}-range, beam size, and pixel size, with the aim of
computing the spectral index.
Flux densities were measured
using \texttt{TVSTAT} in \texttt{AIPS} which allows the selection of a
polygonal region on the image plane. VLBA images are shown in
Fig. A2, while flux densities and spectral indexes are
reported in Table \ref{vlba-flux}. Errors on the flux density are
computed using the formula reported in Section
\ref{vla-section}, while errors on the spectral index are computed
assuming the error propagation theory. \\

\begin{table*}
\caption{VLBA flux density of the CSS sources with LAS$<$0.25 arcsec. Column
  1: source name; column 2: source component; columns 3 and 4: flux
  density at 4.8 and 8.3 GHz, respectively; column 5: spectral
  index between 4.8 and 8.3 GHz (see Section 3.2).
  For the source B3-1049+384 the flux
  density at 4.8 GHz is measured on the image from \citet{mo04}.
  For B3-1225+442 the
  spectral index is computed between 1.7 GHz \citep{dd02a} and
  8.3 GHz.}
\begin{center}
\begin{tabular}{cccccc}
\hline
Name & Comp. & S$_{4.8}$ & S$_{8.3}$ &
$\alpha_{4.8}^{8.3}$\\
     &       & mJy & mJy & \\
\hline
B3-1016+443 & N & 45.4$\pm$4.6 & 18.7$\pm$1.9 & 1.6$\pm$0.3 \\ 
            & S &  6.6$\pm$1.0 &     -        &   -     \\
            &Tot& 52.0$\pm$5.2  & 18.7$\pm$1.9 &   -     \\
B3-1049+384 & W1& 74.6$\pm$7.5 & 33.9$\pm$3.4 & 1.3$\pm$0.3 \\ 
            & W2& 83.8$\pm$8.4 & 51.5$\pm$5.2 & 0.9$\pm$0.3 \\ 
            & C& 15.1$\pm$1.5 & 19.0$\pm$1.9 &-0.4$\pm$0.3 \\ 
            & E&  4.2$\pm$0.5 &  2.2$\pm$0.3 & 1.2$\pm$0.3 \\
            &Tot&180.7$\pm$18.1&106.6$\pm$10.7& 1.0$\pm$0.3 \\
B3-1225+442 & W1A &     -       & 7.1$\pm$0.7 & 0.1$\pm$0.1 \\ 
            & W1B &     -       & 2.4$\pm$0.2 & 0.6$\pm$0.1 \\ 
            & E &      -       &  8.6$\pm$1.0 & - \\ 
            &Tot&      -       & 18.1$\pm$1.9 &   -         \\
B3-1340+439 & N1& 84.7$\pm$8.5 & 29.7$\pm$3.0 & 1.2$\pm$0.3 \\ 
            & N2& 20.6$\pm$2.1 & 13.6$\pm$1.4 & 0.9$\pm$0.3 \\ 
            & C &  8.0$\pm$0.9 &  6.4$\pm$0.7 & 0.4$\pm$0.3 \\ 
            & S &  8.6$\pm$0.9 &  2.6$\pm$0.4 & - \\ 
            &Tot&124.8$\pm$12.5& 52.3$\pm$5.3 & - \\ 
B3-1449+421 & N & 66.5$\pm$6.6 & 21.5$\pm$2.2 & 1.7$\pm$0.3 \\ 
            & C &  8.1$\pm$1.0 &  5.0$\pm$0.5 & 0.9$\pm$0.3 \\
            & S3& 14.3$\pm$1.5 &  5.2$\pm$0.6 & 1.7$\pm$0.3 \\ 
            & S2& 21.0$\pm$2.1 & 11.5$\pm$1.2 & 0.9$\pm$0.3 \\
            & S1& 22.2$\pm$2.2 &  5.7$\pm$0.7 & - \\ 
            &Tot&132.4$\pm$13.3& 48.8$\pm$5.0 & - \\ 
\hline
\end{tabular}
\end{center}
\label{vlba-flux}
\end{table*}

\subsection{eMERLIN observations}
\label{merlin-section}

eMERLIN observations were carried out between 3 and 9 August 2020 in
L and C bands. 
The target sources were observed for about 4 and 3.5 hr
in L and C bands, respectively. Initial flagging and data calibration 
were done using the eMERLIN \texttt{CASA} pipeline. After the a-priori
calibration some data were still highly affected by RFI, especially in
L-band, and additional
flagging was done before starting the imaging process. Final images
were produced after a few self-calibration iterations and are reported
in Fig. A3. In addition to the {\it full-resolution}
images, we also produced a set of {\it low-resolution} images with
{\it uv}-range (50--1250 k$\lambda$), 
pixel size and restoring beam common to L and C
bands. These images were used to estimate the spectral index between L
and C bands. Owing to the
severe flagging, the final rms of the L-band images is of a few mJy
beam$^{-1}$, much worse than the planned rms.\\ 
Flux
densities are reported in Table \ref{merlin-flux} and were measured
with the same \texttt{CASA} tasks described for the analysis of VLA
data (see Section \ref{vla-section}). Errors on the flux density are
computed using the formula reported in Section \ref{vla-section}.\\

\begin{table*}
\caption{eMERLIN flux density of CSS sources. Column 1: source name;
  column 2: source component; columns 3 and 4: flux density at 1.5 and
  4.8 GHz, respectively, measured on the {\it full-resolution} images; 
  column 5: spectral index between 1.5 and 4.8 GHz, computed
  using images constructed with the same {\it uv}-range, pixel size and
  restoring beam (see Section \ref{merlin-section}).} 
\begin{center}
\begin{tabular}{ccccc}
&&&&\\
\hline
Name & Comp. & S$_{1.5}$ & S$_{4.8}$& $\alpha_{1.5}^{4.8}$\\
     &       &    mJy    &    mJy   & \\
\hline
B3-0003$+$387  & N   & 221.9$\pm$11.2  & 29.8$\pm$1.5 & 1.7$\pm$0.1 \\
               & S   & 171.0$\pm$8.8   & 32.9$\pm$1.7 & 1.4$\pm$0.1  \\
               & Tot & 392.9$\pm$19.6  & 62.7$\pm$3.1 & 1.6$\pm$0.1\\
B3-0140+387    & N   & 211.4$\pm$10.6  & 11.0$\pm$0.6 & 1.8$\pm$0.1 \\
               & C   &  52.8$\pm$2.9   &  3.7$\pm$0.5 & 1.7$\pm$0.1 \\
               & S1  &  88.4$\pm$4.4   &  4.7$\pm$0.5 & 1.9$\pm$0.1 \\
               & S2  &  15.2$\pm$1.3   &      -       &     -       \\
               & Tot & 371.0$\pm$18.6  & 19.4$\pm$1.0 & 1.9$\pm$0.1 \\
B3-0810+460B   & N   & 439.4$\pm$22.0  & 79.7$\pm$4.0 & 1.2$\pm$0.1 \\
               & S1  & 258.9$\pm$13.1  & 59.1$\pm$3.0 & 1.1$\pm$0.1 \\
               & S2  & 262.5$\pm$13.5  & 11.7$\pm$5.2 & 1.4$\pm$0.2 \\  
               & Tot & 943.0$\pm$47.2  & 150.5$\pm$7.6& 1.3$\pm$0.2 \\
B3-0856+406    & E   & 154.1$\pm$7.8   &  13.5$\pm$0.7& 1.8$\pm$0.1 \\
               & C   &    -            &   1.0$\pm$0.3&    -        \\
               & W   &  46.2$\pm$2.4   &   5.1$\pm$0.3& 1.4$\pm$0.1 \\
               & Tot & 202.8$\pm$10.4  &  19.6$\pm$1.0& 1.7$\pm$0.1 \\
B3-1143+456    & N   & 115.2$\pm$6.8   &   7.0$\pm$1.1& 2.4$\pm$0.1 \\
               & S   & 526.8$\pm$26.6  &  64.8$\pm$3.4& 1.8$\pm$0.1 \\
               & Tot & 628.3$\pm$32.0  &  71.8$\pm$3.6& 1.9$\pm$0.1 \\
B3-1212+380    & N   & 138.0$\pm$7.0   &  18.9$\pm$1.0& 1.5$\pm$0.1 \\
               & S   & 137.3$\pm$7.0   &  14.0$\pm$0.7& 1.5$\pm$0.1 \\ 
               & S1  &    -            &   5.1$\pm$0.3&   -  \\
               & S2  &    -            &   8.9$\pm$0.6&   -  \\
               & Tot & 275.3$\pm$13.8  &  32.9$\pm$1.7& 1.5$\pm$0.1 \\
\hline
\end{tabular}
\end{center}
\label{merlin-flux}
\end{table*}

\section{Results}

\subsection{Spectral fit} 
\label{synage}

For the sources observed with the wide bandwidth available
at the VLA we measured the flux
density for each spectral window, in order to exploit the spectral
coverage as much as possible.
Flux densities at 4.8 and 8.4 GHz are from historical VLA
observations with narrow bandwidth \citep{fanti01}. Flux densities in U
band are from our new low-resolution wide band observations which
guarantee a better characterization of the high frequency part of the
spectrum, preventing the artificial steepening that may be caused by
the lack of short spacings of earlier high resolution observations
\citep{rossetti06}, unable to pick up extended low-surface brightness
emission. \\ 
For the compact sources targeted by VLBA observations we constructed the
integrated radio spectrum making use of VLA flux density from the
literature where the sources are unresolved. 
This prevents us to miss flux density that may be
resolved out and/or poorly sampled in our VLBA observations. \\
We complemented the information in the GHz regime by including flux
density at 74 MHz from the VLA low-frequency sky survey
\citep[VLSS;][]{cohen07} and VLSS redux \citep{lane14}, 150 MHz from
the TIFR GMRT sky survey \citep[TGSS;][]{intema17}, 
and at 408 MHz \citep{ficarra85} in order to characterize
the radio spectrum at low frequencies.\\

To determine the shape of the synchrotron emission we fit the radio
spectrum of the sources assuming two different models. The first model
implies a continuous injection of fresh relativistic particles
(CI model), while in the second one the
injection of new particles has switched off and the source is already
in a relic phase (CI OFF model). In the latter model a second break
appears in the spectrum at high frequencies and beyond that the
spectrum cuts off exponentially. In both models radiative losses
dominate over the other energy losses \citep[see][]{murgia03}
and are characterized by a)
the injection spectral index ($\alpha_{\rm inj}$), b) the break
frequency ($\nu_{\rm b}$), c) and the flux normalization. In addition
to these parameters, the CI OFF model is also characterized by the
relic to total source age ratio ($t_{\rm OFF}/{t_{\rm s}}$, i.e. the
relic period and the source age, respectively). For $t_{\rm OFF}
\rightarrow 0$, CI OFF model approaches CI model\footnote{
In addition to the CI/CI OFF models, we fitted the spectrum
  considering a passive evolution of a synchrotron spectrum without
  any injection of fresh particles and either with a continuous
  re-isotropization of the pitch angle \citep[JP,][]{jaffe73} or without
  \citep[KP,][]{kardashev62}. Both models are usually comparable and
  their $\chi_{\rm red}^{2}$ is in good agreement with that of CI OFF
  model, since the CI OFF model
tends to the JP for $t/t_{\rm OFF}
\rightarrow 1$.}.\\

When the turnover frequency falls within the frequency range covered
by the observations, we fit the optically-thick part of the spectrum
with a pure synchrotron self-absorption (SSA) model ($\alpha_{\rm thick} =
  2.5$). A detailed description
of the spectral models can be found in \citet{murgia11}.\\

\begin{figure*}
  \begin{center}
 \includegraphics{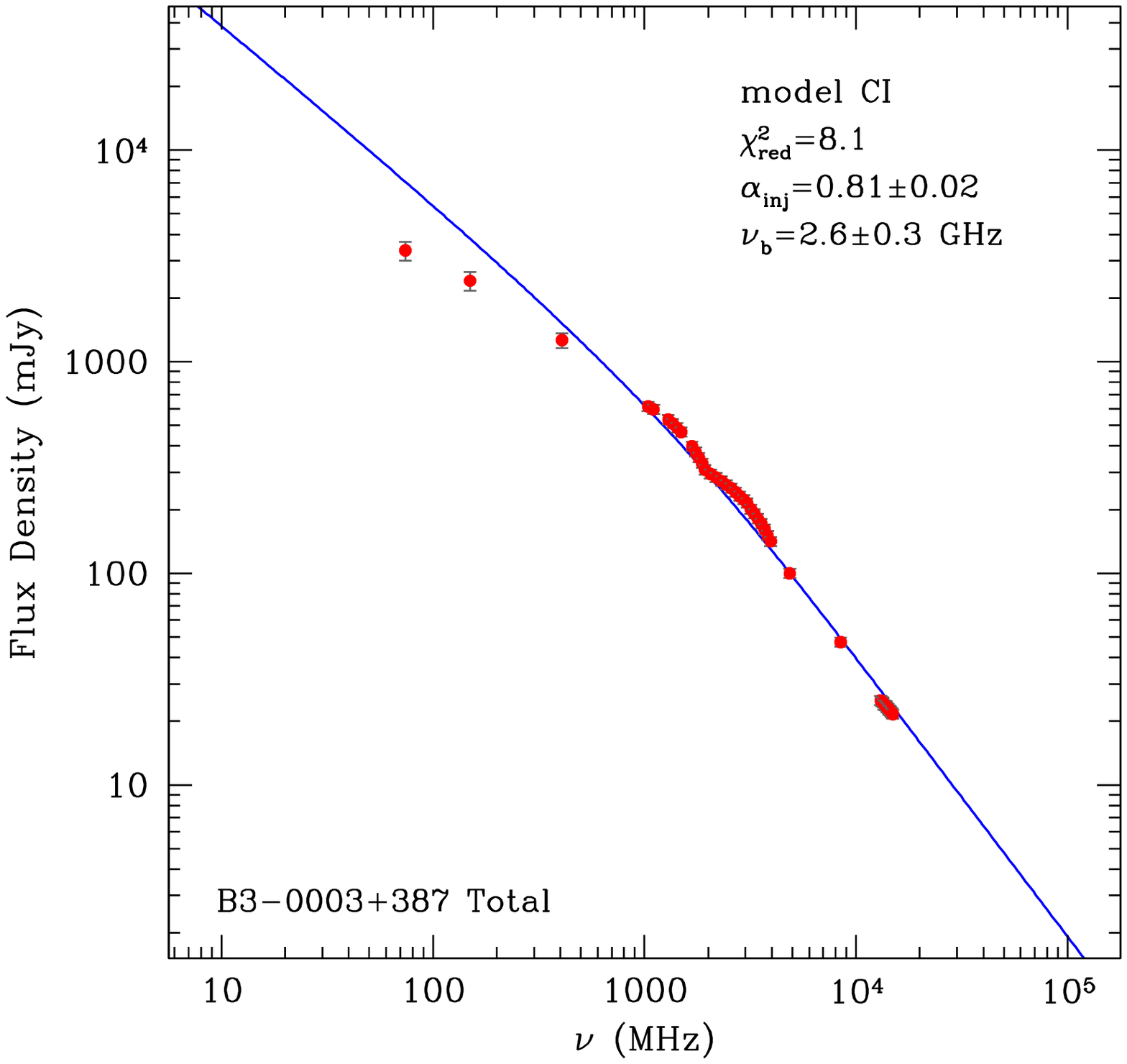}
\includegraphics{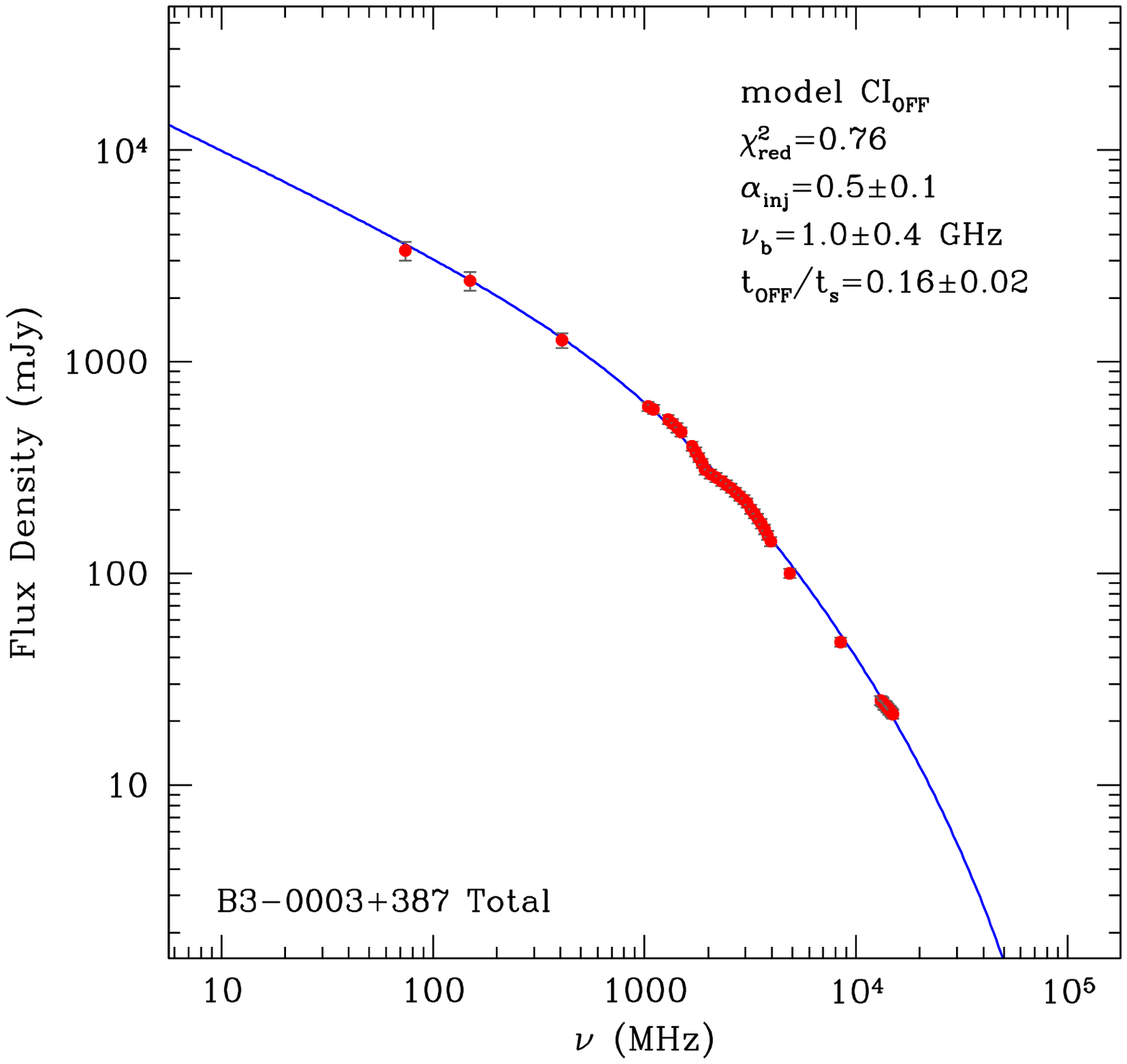}
\vspace{8cm}
\caption{Integrated spectrum of the source B3-0003+387
  along with the best-fit CI model
  ({\it left}) and best-fit CI OFF model ({\it right}).}
\label{synage-example}
\end{center}
\end{figure*}

To evaluate the best model that fits the spectrum we compare the value
of the injection spectral index and the reduced chi-square. An example
of the fits for B3-0003+387 is shown in Fig. \ref{synage-example}. Fits of
the overall spectra for the whole sample are presented in Fig. C1,
and the parameters are reported
in Table C1.\\ 

\subsection{Polarization}
\label{polarization}

For the sources observed with the VLA we could study the polarization
either in L-S bands or in U band. In all the sources we detect
polarized emission with the exception of B3-1143+456 and
B3-1212+380 which are unpolarized (polarization percentage $m < 0.1$
per cent) in
both L and S bands, and B3-0003+387, unpolarized in U band ($m < 0.5$
per cent).
To estimate the rotation measure (RM) we complemented our data with the
information in C and X bands from \citet{fanti01}. In three sources
(B3-0003+387, B3-0034+444, and B3-1216+402) we find $|$RM$|$ $>$ 100
rad m$^{- 2}$, whereas in B3-0748+413B, B3-0754+396, and B3-1039+424,
$|$RM$|$ $<$12 rad m$^{- 2}$,
consistent with the Galactic value \citep{taylor09}.
The latter sources are also those showing the highest 
polarization percentage $m \gtrsim$ 10 per cent.
Our deep VLA observations point out that
these objects are components of larger
sources which extend far beyond the host galaxy, and therefore are not
genuine CSS sources (see Section \ref{notes}). \\
For some objects our RM differs from those reported in
\citet{rossetti08} and this is likely due to the lower resolution of
the WSRT data (8 arcsec) which could not resolve the polarized emission
from different components, implying some level of beam depolarization. \\ 

\begin{table*}
\caption{Polarization properties. Column 1: source name; column 2:
  source component; columns 3 and 4: polarization percentage and polarization
angle in L band, respectively; columns 5 and 6: polarization
percentage and polarization angle in S band, respectively; columns 7
and 8: polarization percentage and polarization angle in U band,
respectively; column 9: rotation measure; column 10: intrinsic
polarization angle.}
\begin{center}
\begin{tabular}{cccccccccc}
\hline
Source name & Comp. & $m_{\rm L}$ & $\chi_{\rm L}$ & $m_{\rm S}$ & $\chi_{\rm S}$ & $m_{\rm U}$ & $\chi_{\rm U}$ & RM & $\chi_{0}$\\
  &  & \% & deg & \% & deg & \% & deg & rad/m$^{2}$ & deg\\
\hline
B3-0003+387 & N & -  & - & - & - & $<$0.5 & - & - & -\\
            & S & -  & - & - & - &    6.5 & 23$\pm$4 & -135 & 26 \\
B3-0034+444 & A & -  & - & - & - &    4.3 & 14$\pm$7 & 109 & 11 \\
            & B & -  & - & - & - &    4.0 & -47$\pm$2 & - & - \\
B3-0128+394 & N & -  & - & - & - &    2.9 & -62$\pm$5 & - & -\\
            & S & -  & - & - & - &    4.7 & -60$\pm$5 & - & - \\
B3-0140+387 & N & -  & - & - & - &    2.5 &  66$\pm$2 & - & - \\
            & S & -  & - & - & - &    5.0 & -50$\pm$2 & - & - \\
B3-0748+413B& N &0.1 & -44$\pm$4 & 2.4 & -38$\pm$8 & - & - & -4 &
-34\\
            & S &11.7 & 42$\pm$4 & 16.2 & 13$\pm$3 & - & - & - & - \\
B3-0754+396 & N & 8.0 & 85$\pm$8 & 9.8 & 84$\pm$8 & - & - & 2 & 81\\
            & S & $<$0.1 & - & 3.6 & 28$\pm$8 & - & - & - & -\\
B3-0810+460B& W & $<$0.1 & - & 0.1 & 0$\pm$2& - & - & - & -\\
B3-0856+406 &Tot& $<$0.1 & - & $<$0.1 & - & - & - & - & - \\
B3-1039+424 & N & 5.6 & 36$\pm$4 & 14.1 & 17$\pm$4 & - & - & - & - \\
            & S & 7.0 & -49$\pm$2& 13.8 & -67$\pm$2 & - & - & 12 &
-76\\
B3-1143+456 &Tot& $<$0.1 & - & $<$0.1 & - & - & - & - & - \\
B3-1212+380& Tot& $<$0.1 & - & $<$0.1 & - & - & - & - & - \\
B3-1216+402 & N & 0.6 & 73$\pm$5 & 8.5 & 77$\pm$4 & - & -  & -5 & 84\\
            & S & 0.2 & 49$\pm$6 & 1.8 & -42$\pm$8& - & - & -151 &
41\\
\hline
\end{tabular}
\end{center}
\label{tab-pol}
\end{table*}

\subsection{Notes on individual sources}
\label{notes}

Here we provide a brief description of the sources, and their
  classification: active, remnant, restarted, large symmetric object (LSO).
  The classification is based on the
  detection, or not, of active regions, like core and hotspots, and results
  from the synchrotron fit to the integrated spectrum.
  A question mark indicates that the
  information is not adequate for a secure classification of the radio
  emission. \\
  In general, the spectral indices were computed using the flux
  densities measured on images 
that were produced using a {\it uv}-range, image sampling, and restoring beam
common to both frequencies. For VLA data, we report the
spectral index between 3 and 14 GHz, since the structure of many
sources is unresolved or marginally resolved at 1.4 GHz. For some
sources with eMERLIN observations, in addition to the spectral index
between 1.7 and 4.8 GHz from the eMERLIN data sets, we also computed
the spectral index using the 4.8-GHz eMERLIN data and the
high-resolution 15-GHz VLA data from \citet{rossetti06}. The use of
these data sets obtained with different arrays may affect
the determination of the spectrum, mainly for extended
structures. Therefore, we
provide the spectral index using the 4.8-GHz eMERLIN data and the
high-resolution 15-GHz VLA data only 
for compact regions.\\
VLA images of the sources are shown in Fig. A1: contours
show total intensity emission, while polarized emission (and
E-vectors) are shown in colors (for data sets with available
polarization information). Total intensity images from VLBA
observations are shown in Fig. A2, while eMERLIN images
are presented in Fig. A3.\\

\subsubsection{B3-0003+387 - remnant}
The source has a double structure with no evidence of the source
core. Both components, unresolved with the VLA, have a steep
  spectral index ($\alpha > 1.0$, Tables \ref{vla-flux} and
  \ref{merlin-flux}) and are elongated in 
  the C-band eMERLIN image, with no evidence of any compact region. No
  significant total-intensity flux-density asymmetry between the two components is
  observed. The polarized emission is highly asymmetric, with
  the southern component showing a fractional polarization of about
  6.5 per cent and an
  RM of -135 rad/m$^2$, while the northern component is unpolarized.
The radio spectrum is well reproduced by CI OFF model
(Fig. \ref{synage-example}), 
supporting the relic nature of this object. This source is one of
  the two objects best fitted by a JP model in \citet{rossetti06}.
The time spent by the
source in the relic phase is about 16 per cent of the total radio source age.\\

\subsubsection{B3-0034+444 - active?}
The source has an elongated morphology that reminds of a jetted
structure. There is no clear evidence for
a flat-spectrum core component. \citet{rossetti06} suggested component
B as the candidate core based on their high resolution
15-GHz data. However, from our VLA data this component shows a rather
steep spectrum ($\alpha_{3}^{14} \sim 1.6$), while component A,
edge-brightened in the 15-GHz images presented in \citet{rossetti06}, 
has a spectral index of 1.1. Components C and D are resolved out
  in the high-resolution 15-GHz images \citep{rossetti06},
    suggesting the lack of compact regions.
At 14 GHz only components A and B are polarized.
Both CI and CI OFF models provide a
similar fit to the spectrum. Since $t_{\rm OFF}$ is very short,
the CI OFF model is rather equivalent to the CI model (see Section 4.1). 
The result from the spectral analysis together with the
  information on the source structure pointed out by high-resolution
  images make us conservatively consider B3-0034+444 as a candidate
  active source.\\

\subsubsection{B3-0128+394 - active?}
Source with a double structure characterized by a mildly steep
  spectrum ($\alpha_{3}^{14} \sim 1.2$) and no significant
  flux-density asymmetry. There is no clear evidence for a
flat-spectrum core component. In the high resolution images
  presented in \citet{rossetti06} the extended structure of both the
  northern and southern components is resolved
  out and only the hotspots
  are detected and resolved in substructures.
  In particular, the compact component in the northern
  lobe has a rather flat spectral index $\alpha_{8.5}^{15} \sim 0.5$
  when super-resolved
  images at 8.5
  GHz are considered \citep{rossetti06}.
    At 14 GHz the peak of polarized emission in the
  northern component does not
  coincide with the peak of total-intensity emission, but it is
  centered on the component labelled N2 in \citet{rossetti06}.
Both CI and CI OFF models provide a
similar fit to the spectrum, with the latter slightly preferred over
the former. The result from the spectral analysis together with the
  information on the source structure, i.e. presence of hotspots but
  lack of clear core detection
  make us conservatively consider B3-0128+394 a candidate active source.\\

\subsubsection{B3-0140+387 - active}
Source showing a double structure in VLA images. When observed with
the higher resolution of eMERLIN it shows a triple structure. If we
compare the eMERLIN data at 4.8 GHz and the VLA data at 15 GHz from
\citet{rossetti06} we find that the central component, labelled C in
Fig. A3, has a relatively flat spectrum
($\alpha_{4.8}^{15} \sim 0.5$), indicating that it may host the source
core. Similarly, the outer compact components labelled N and S1
  in eMERLIN image (Fig. A3) have a spectral index
  $\alpha_{4.8}^{15} \sim$  0.6 and 0.7, respectively, i.e. much flatter
  than those derived by lower resolution data (Tables \ref{vla-flux}
  and \ref{merlin-flux}). However, when observed with the high
 resolution provided by the European Very Long Baseline Interferometry
(VLBI) network (EVN) at 18 cm only two compact regions are detected at the
 edges of the source, while no compact component is clearly detected at the
 centre \citep{dd02b}, suggesting that the core may be absorbed at
 such frequency.
The steep spectrum derived from low-resolution data may
be due to the contamination of steep-spectrum emission from lobes and
jets, which is
resolved out at high resolution. Component N is the brightest
  in total intensity, with a flux density ratio between N and S1 of
  about 2.3 and 2.5 in 4.8-GHz eMERLIN data and in 15-GHz high-resolution
  VLA data \citep{rossetti06}, respectively,
  while component S has the highest level of
  polarization (Table \ref{tab-pol}). Both SSA+CI and SSA+CI OFF models
provide a similar fit to the spectrum. However, the detection of
  hotspots and the tentative identification of the core component
  suggest that the source is active.\\

\subsubsection{B3-0748+413B - LSO}  
The source that was classified as a CSS in \citet{fanti01} turns out
to be the northern lobe/hotspot, labelled N in Fig. A1, 
of a larger radio source with LAS $\sim$ 13.5 arcsec (LS$\sim$110 kpc
assuming $z$=1.05). The radio source
has an asymmetric structure with the core at about 4 arcsec from
component N and about 9 arcsec from component S. The RM of component N
is consistent with the Galactic value (Table \ref{tab-pol}) suggesting
that the radio emission is not embedded within the interstellar medium
(ISM) of the host
galaxy. On the basis of these new pieces of evidence we classify
B3-0748+413B as an LSO and remove it from 
the sample of CSS sources.\\

\subsubsection{B3-0754+396 - LSO}
The source that was classified as a CSS in \citet{fanti01} is likely 
the northern lobe/hotspot, labelled N in Fig. A1,
of a larger radio source with LAS $\sim$ 11 arcsec (LS$\sim$93 kpc at
$z$=2.119). Both components
are elongated at low frequencies. No evidence of the source core is
found in our VLA images. The high fractional polarization of
component N and the RM consistent with the Galactic value (Table
\ref{tab-pol}) support the idea that the emission is not embedded
within the ISM of the host galaxy. Moreover, EVN observations of
  component N could detect only a set of knots, none reminiscent of
  the core component \citep{dd02b}. On the basis of these new pieces of
evidence we classify
B3-0754+396 as an LSO, and remove it from
the sample of CSS sources.\\

\subsubsection{B3-0810+460B - active}
The source has a double structure in our VLA images, while it is
resolved into three main components in eMERLIN images. 
Component S1
might host the source core and its steep spectrum ($\alpha \sim 1.1$)
may be due to the combination of core+jet emission which
cannot be spatially resolved by our observations. If we compare the eMERLIN
flux density at 4.8 GHz with high-resolution VLA data at 15 GHz from
\citet{rossetti06} we derive a relatively flat spectrum
($\alpha_{4.8}^{15} \sim 0.5$) supporting its interpretation as
the source core. EVN observations at 18 cm could detect
  only components N and S1, while component S2 is
  completely resolved out, indicating its extended structure
  \citep{dd02b}. No significant flux density asymmetry is observed
  between the northern and southern (S1+S2) components (Tables
  \ref{vla-flux} and \ref{merlin-flux}, and \citet{rossetti06}). The
  source is slightly polarized in S band.
The CI model provides a good fit to the radio
spectrum, supporting the active nature of the source.\\

\subsubsection{B3-0856+406 - restarted?}
The source has a double structure in our VLA images, while it is
resolved into three main components in eMERLIN images. The component
labelled C in Fig. A3 is likely the source
core. 
  Components E and W have a relatively steep
spectrum both in our eMERLIN observations and in VLA data. Component E
is the brightest one with a flux density ratio ranging between 3.3, in
eMERLIN data at 1.4 GHz, and 1.1 in our VLA data at 14 GHz.
In high-resolution VLA data at 15 GHz, only component E was detected,
indicating that component W has a steep spectrum and
is likely extended \citep{rossetti06}. Although component E is
  compact in high-resolution 15-GHz images, its spectrum is steep
  ($\alpha > 1.3$) both in our VLA and eMERLIN data. 
Neither component C was picked
up in that dataset. The source is unpolarized in our images in L and
S bands.
Both CI and CI OFF models
provide a similar fit to the spectrum (though with a rather large
value of the reduced chi-square), with the latter slightly
preferred over the former. This source is the only one together with
  the relic B3-0003+387 (out of the 42 analysed) whose
  spectrum was significantly better fitted with a JP model rather
  than a CI one in \citet{rossetti06}. These pieces of evidence suggest that the
radio emission of this source has restarted not long ago. However,
owing to the presence of a compact region in component E, though
with a steep spectrum, we conservatively classify this
object as a candidate restarted.\\

\subsubsection{B3-1016+443 - remnant}
The source has a double structure in our VLBA image at 4.8 GHz
(Fig. A2) and at
1.7 GHz \citep{dd02a}.
At 8.4 GHz only the northern component is
detected. The spectrum of both components
is steep and there is no evidence of the source
core. Both SSA+CI and SSA+CI OFF models
provide a similar fit to the spectrum, with the former slightly
preferred over the latter. However, the few data points available
leave large uncertainties on the fit results and more data points
  are needed for a proper model of the synchrotron spectrum. Our
  classification of B3-1016+443 as a remnant is based on the
  multi-frequency morphology and spectral index information.

\subsubsection{B3-1039+424 - LSO}
The source that was classified as a CSS in \citet{fanti01} turns out
to be the southern lobe/hotspot, labelled S in Fig. A1,
of a larger radio source with LAS $\sim$ 19 arcsec (LS$\sim$154 kpc
assuming $z$=1.05). The faint core is
roughly midway between the northern and southern components.
The RM of component S
is consistent with the Galactic value (Table \ref{tab-pol}) suggesting
that the radio emission is not embedded within the ISM of the host
galaxy. On the basis of these new pieces of evidence we classify
B3-1039+424 as an LSO, and remove it from
the sample of CSS sources.\\

\subsubsection{B3-1049+384 - active}

The source is characterized by 4 components with an unusual
morphology. Components W1 and W2 seem a scaled-down version of a
double radio galaxy, and their elongation is misaligned with respect
to components E and C. Component C has a flat spectrum and is likely
the source core. A high flux density asymmetry is observed
  between the eastern and western (W1+W2) components (Table \ref{vlba-flux}).
The SSA+CI model provides a good fit to the radio
spectrum, supporting the active nature of the source.\\

\subsubsection{B3-1133+432 - active?}

This source is a CSO with a double structure and no evidence of the
source core. VLBA images were presented in \citet{dd02a} and
\citet{mo04}. The northern component is the brightest, with a flux
  density that is roughly double that of the southern component.
  The spectral index is $\alpha_{1.7}^{4.8} \sim$ 0.9 and 1.0 for component N and
S, respectively, and slightly steepens between 4.8 and 8.4 GHz
\citep[$\alpha_{4.8}^{8.4} \sim 1.2$,][]{mo04}.
The SSA+CI model provides a better fit
than the SSA+CI OFF model. However, the few data points available
leave large uncertainties on the fit results. Although the core
  is not detected, we conservatively consider B3+1133+432 as a
  candidate active source.\\ 

\subsubsection{B3-1143+456 - active?}
This source has a double structure in VLA and eMERLIN
images. Both components have a steep spectrum and there is no evidence
of the source core. However, if we compare the eMERLIN
flux density at 4.8 GHz with high-resolution VLA data at 15 GHz from
\citet{rossetti06} we derive $\alpha_{4.8}^{15} \sim$ 0.5 and 0.7 for
component N and S, respectively, suggesting that they may be active
hotspots. However, we notice that these are not compact regions, and
the spectral index may be artificially flat, due to missing flux
density from extended structures unrecovered by eMERLIN
observations. The flux density ratio between component N and S is
about 5-7 depending on the frequency,
with the southern being the brightest one. The source is
unpolarized in L and S bands. 
Both SSA+CI and SSA+CI OFF models
give a similar fit to the spectrum, with the former providing
  a slightly lower value of the reduced chi-square value.\\

\subsubsection{B3-1212+380 - active}
This source has a double structure in eMERLIN
images and in the high-resolution 15-GHz VLA image \citep{rossetti06},
while in our VLA observations the source is unresolved in L and S
bands, and slightly resolved in U band.
In eMERLIN images the southern component is slightly resolved
into two components elongated in the NW-SE direction. Component S1  
might host the source core, and the
steep spectrum of the whole southern component  
may arise from the blending of the core emission with that of the
steep-spectrum jet and mini-lobe.
This interpretation is
supported by the estimate of the spectral index of S1 between
eMERLIN data at 4.8 GHz and high-resolution VLA data at 15 GHz
\citep{rossetti06}, which turns out to be $\alpha_{4.8}^{15} \sim
0$. The low data quality of EVN observations could not help identify the
  core component \citep{dd02b}. In eMERLIN images where the source structure
  is resolved, there is no significant flux density 
  asymmetry between the northern and southern components (Table
  \ref{merlin-flux}).  
  The source is unpolarized in our images in L and S bands.
The CI model provides a good fit to the radio 
spectrum, supporting the active nature of the source.\\ 

\subsubsection{B3-1216+402 - remnant?}
This source has a double structure in our VLA images. 
There is no evidence
of the source core. The spectral index $\alpha$ $\sim$ 1.0,
i.e. not ultra-steep as it is found in other sources, may indicate the
co-existence of aged emission from lobes with the contribution of
active regions like hotspots, which are mixed up with the
resolution of our observations. However, in high resolution 15-GHz
  VLA observations only the southern component hosts a faint
  steep-spectrum compact
  region, while the structure of the northern component is fragmented,
  with no clear indication of a hotspot. Neither the core is detected
  in this data set \citep{rossetti06}. There is no significant flux
  density asymmetry between the northern and southern components at
  low frequencies, while it becomes roughly double at 14 GHz, with the
  southern being the brightest (Table \ref{vla-flux}).
Polarization increases from L to S
band, and is detected in both components, with component N
significantly polarized in S band (8.5 per cent). Both CI and CI OFF models
provide a similar fit to the spectrum. The lack of a clear core
  detection and the possible presence of a candidate hotspot, (though
  with a relatively steep spectrum) in
  the southern component make us conservatively classify this object as a candidate remnant.\\

\subsubsection{B3-1225+442 - restarted}
Although at 1.7 GHz the source shows a triple morphology \citep{dd02a},
at 8.4 GHz it has a double structure, being the westernmost
component, labelled W2 in
\citet{dd02a} undetected at high frequency. Only the central
  component, labelled W1 in \citet{dd02a}, and the eastern one are
  detected in our observations.
The component labelled E in Fig. A2
is severely resolved out, whereas the other component is resolved
  into two sub-components labelled W1A and W1B.  
The spectral index of W1A is
flat ($\alpha \sim 0.1$) indicating that the source core
may be hosted in this component. The lack of compact components at the
edge of the source and the possible presence of the core may indicate
that the radio emission has restarted not long ago. The SSA+CI OFF model
provides a slightly better fit than the SSA+CI model, supporting the recent
restart of the radio emission.\\

\subsubsection{B3-1340+439 - restarted?}
The source is resolved into several sub-components in our VLBA
images. The component labelled C in Fig. A2
has a rather flat spectrum and is
likely the source core. Component S is severely resolved out at 8.3
GHz, while component N1, though with a rather compact structure at 8.3
GHz, has a steep spectrum ($\alpha_{4.8}^{8.3} \sim 1.9$). The
  northern component (N1+N2+N3) is brighter than the southern one,
  with a flux density ratio $>$10.
The SSA+CI OFF model
provides a slightly better fit than the SSA+CI one. As in the case of
B3-0856+406 and B3-1225+442,
these pieces of evidence may indicate that the radio
emission in this source
has restarted not long ago. However, the morphology of the
  northern components is consistent with a jet structure, making us
  conservatively consider this source as a candidate restarted.\\

\subsubsection{B3-1449+421 - active}
The source is resolved into several sub-components in our VLBA
images. The component labelled C in Fig. A2
is likely the source core, and the
rather steep spectrum may be due to the combination of core+jet
emission that cannot be resolved by our observations. The
  morphology of the southern components is consistent with a jet
  structure. There is no
significant flux density ratio ($\sim$ 1) between the northern and southern
components.
The SSA+CI model
provides a good fit to the radio  
spectrum, supporting the active nature of the source.\\

\section{Discussion}
\label{discussion}

A possible explanation for the high number of intrinsically compact
extragalatic radio sources is the intermittency of short-lived
episodes of jet activity. Several models predict that the dominant
population of extragalactic radio sources consists in short-lived objects
\citep[e.g.][]{reynolds97,hardcastle18,shabala20} which may not evolve into classical
Fanaroff-Riley type I/II radio galaxies. The straightforward outcome
from this scenario is the existence of a population of young but dying
radio sources. To unveil the incidence of remnants in the population
of young radio sources we study the statistically complete B3-VLA CSS
sample and we searched for active regions in the 18 sources with a
steep ($\alpha_{\rm 4.8}^{\rm 8.4} > 1.0$)
integrated radio spectrum and without any evidence of the core from
earlier observations. These could be considered the signatures of an
aged population of relativistic electrons, without a significant
contribution from freshly produced/accelerated ones. \\
The wide frequency coverage and the deep
sensitivity of our new observations allowed us to classify
the sources in either active or remnants. We find that 3 CSS sources
selected by \citet{fanti01} are components of large radio
sources, and we remove them from the CSS sample. The final
sample thus consists of 84 objects divided into 25 sources with
LAS$<$0.25 arcsec, that we consider as CSOs (though the core is not
detected in all of them), while 59 sources have LAS$>$0.25 arcsec, that
we classify as MSOs.\\

\begin{table}
\caption{Summary of core detection and candidate remnants/restarted
  sources. Column 1: class of objects;
  column 2: number of sources in the B3-VLA CSS sample; column 3:
  number of sources of the sample studied in this work; column 4:
  number of core detected in the data presented here (total number of core
  detected for the whole B3-VLA CSS sample
  including information from the literature); column 5:
  number of (candidate) remnants/restarted sources.}
\begin{center}
\begin{tabular}{ccccc}
\hline
Class & B3-VLA & this & core & remnants/ \\
      &   CSS          &  sample     &       & restarted     \\
\hline
All CSS   &    84      &    15       & 8 (44) & 3/3 \\ 
CSOs  &    25      &     6       & 4 (12)  & 1/2 \\ 
MSOs  &    59      &    9       & 4 (30)  & 2/1 \\ 
\hline
LSOs  &     3      &     3       & 2 (2)   &  - \\
\hline 
\end{tabular}
\end{center}
\label{summary}
\end{table}

\subsection{Core detection}
\label{core-discussion}

We detect the core component in 10 sources of our sample
(inclusive of LSOs), while
in 6 CSS sources we 
could set an upper limit on its flux density.
The radio morphology of
B3-0034+444 precludes us from determining the location of the core or
setting a tight upper limit.

\begin{figure}
\begin{center}
\includegraphics{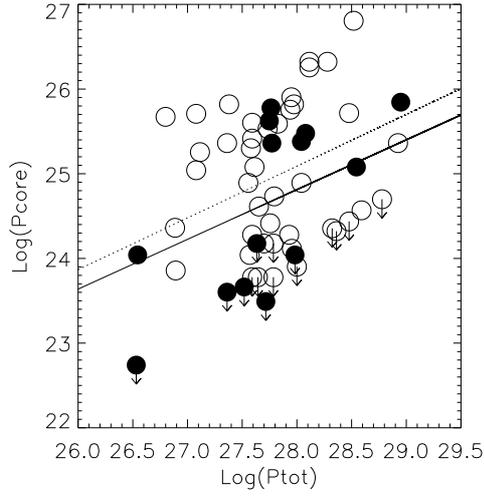}
\vspace{7cm}
\caption{Core luminosity at 5 GHz vs total luminosity at 408 MHz
  of the sources from the B3-VLA CSS sample (see Section
  \ref{core-discussion}).
  Empty circles are CSS
  sources from the B3-VLA CSS sample \citep{fanti01},
  while filled
  circles are the sources studied in this paper
  (with the exception of B3-0034+444 and LSOs). The solid line represents
  the relation obtained for 3CR radio sources \citep{gg88}
corrected for the cosmology adopted in this paper, while the dotted
line represents the best 
fit for CSS sources from the B3-VLA CSS sample (see Section
\ref{core-discussion}.) Downward arrows indicate upper limits.}
\label{lum-core-tot}
\end{center}
\end{figure}

In Fig. \ref{lum-core-tot} we plot the ratio
between the core luminosity P$_{\rm c}$ at 5 GHz vs the total source luminosity P$_{\rm
  tot}$ at 408 MHz for the sources of the B3-VLA CSS sample, with
the exception of the three LSOs (all
luminosities are k-corrected). 
When no information at 5 GHz is available for
the core component, we extrapolate the flux density assuming a flat
spectral index $\alpha = 0$. If the core is undetected at all
frequencies we assume as upper limit the three sigma noise level of
either the
high-resolution VLA images at 15 GHz reported in \citet{rossetti06}
for MSOs, or the VLBA images at 8.4 GHz for CSOs\footnote{We do not
consider all the sources from the B3-VLA CSS sample without the core detection, but only those
showing a morphology, mainly a double structure, for which it is
reasonable to set an upper limit to the flux density related to the
rms of the images. For those sources with a complex morphology, the core may be
hidden by other components, like jets and low-surface brightness
emission from lobes that bridges the outermost hotspot regions
(e.g. B3-0902+416).}. 
When the redshift is unknown, we assume $z = 1.05$ \citep[see][]{fanti01}.

In Fig. \ref{lum-core-tot} we plot the
best fit (corrected for the cosmology adopted in this paper) 
that is obtained for the 3CR radio galaxies
\citep{gg88}, i.e. 
for
objects with a luminosity between 
10$^{24 - 28}$ W Hz$^{-1}$.
CSS sources from the B3-VLA are
in general above 
the best fit. 
If we perform a regression fit that minimizes the chi-square error
statistics, we obtain for the CSS sources:\\

Log P$_{\rm c}$ = (8.0$\pm$5.3) - (0.6$\pm$0.2)$\times$logP$_{\rm tot}$ \,.\\

\noindent This empirical relation has a slope (though with
large uncertainty) that is similar to the one
found by \citet{gg88}, but it is slightly shifted towards higher
luminosities. This may be due to the slightly higher luminosities of
the B3-VLA CSS sample, i.e. about 10$^{26}-10^{29}$ W Hz$^{-1}$ \citep{fanti01}.\\
For the CSS sources with no detection of the core component we could set a tight
upper limit on the luminosity, with the exception of B3-0034+444,
which is in general well below the average trends, suggesting that
some might be remnants. However, we cannot exclude the
possibility that some level of core activity (below the sensitivity
threshold) is present. 
A very weak core and the steep
spectrum of the source may indicate that either some level of
activity of the AGN is always present, even if it is unable to launch
powerful jets, or that the radio core has just switched on
again \citep{jurlin21}. Apart from a few objects where the optical
  counterpart is unknown, all the sources of our sample are optically
  classified as
  galaxies rather than quasars (i.e. jets oriented at a large angle with respect
  to our line of sight). Therefore, we cannot exclude the possibility
  that the core flux density may
  be Doppler diminished and thus
  non-detected by our observations due to sensitivity limitation
  \citep[see e.g.,][]{saikia94}.  

\subsection{Remnants and restarted sources}

\begin{figure}
\begin{center}
\includegraphics{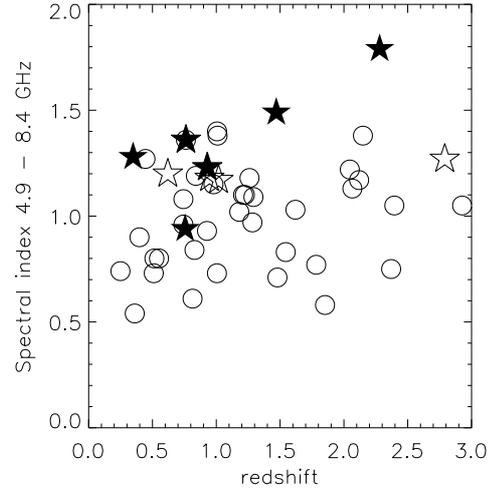}
\vspace{7cm}
\caption{Observed-frame spectral index between 4.9 and 8.4 GHz vs redshift of the
  sources from our sample either (candidate-)remnant/restarted (filled
  stars) or (candidate-)active (open stars) together with the
  remaining sources of the B3-VLA CSS. Only sources with spectroscopic
  redshift have been considered.}
\label{alpha-z}
\end{center}
\end{figure}

The analysis of the integrated radio spectrum points out that 5
sources
are best fitted by a continuous particle injection model,
while other 5 sources
require the inclusion of a relic phase. For the
remaining 5 CSS sources both models provide a quite equivalent
good fit to their spectra
(CI OFF model approaches CI model). \\
For LSOs are contaminant of our sample, they are dropped
from our analysis since they do not belong to the sample of subgalatic
size
radio sources.\\
To unveil whether a source is active or not, we need to
  correctly determine both the integrated synchrotron spectrum, and
  the nature of each sub-structure: extended vs compact regions, and
  spectral index information. For the sources of our sample
  we find that 9 are likely active: 5 with a secure classification, and 4
candidate active objects. The latter are radio sources with no core
detection, but with the presence of a hotspot (with $\alpha < 0.7$ at
least on one side). The non detection of a hotspot on both sides of the
radio emission may be an indication of an inhomogeneous
environment. 11 out of
the 15 MSOs/CSOs sources
of our sample show significant flux density asymmetry (i.e. flux
density ratio $\gg$1) between the two
sides of the radio emission, whereas polarization asymmetry is
observed in 4 of the sources with polarization information.
Asymmetries in flux density, polarization, and arm-length
ratio between the two sides of the radio emission caused by an
inhomogeneous medium have been found more often in young radio sources
than in classical double radio galaxies \citep[see
  e.g.,][]{saikia03,saikia03b,mo08,dd13,mo14,mo16,odea21}.\\ 
Among sources
with core detection, B3-1225+442 (CSO) is likely a restarted object. Its extended components
that are well imaged at low frequency \citep{dd02a} are either
undetected or highly resolved-out at the highest frequency,
and show a steep spectrum,
indicating that no compact active
regions, like hotspots, are present at the edges of the sources. On the
contrary the central component has a spectral index that is consistent
with current activity. The other two sources, B3-0856+406 (MSO) and
B3-1340+439 (CSO) are candidate restarted. Their integrated radio spectra
are better modelled when a relic period is included, suggesting that
the activity has not been continuous during the total source
age. However, the presence of an edge-brightened component in
B3-0856+406 and a jet-like structure in B3-1340+439 make us
conservatively consider
these objects candidate restarted.\\
We classify the remaining sources B3-0003+387 (MSO) and
  B3-1016+443 (CSO) as  remnants, while we conservatively consider
B3-1216+482 (MSO) a candidate
remnant.\\
The improvement in sensitivity and angular resolution that
will be achieved with 
the next generation of radio facilities, like the Square Kilometre Array
(SKA) and the next-generation VLA (ngVLA), will allow
us to study the radio spectra of each component of the radio sources
(i.e. similar to the spectral studies currently done for radio galaxies with
sizes of hundred of kpc and beyond), rather than the
integrated one, providing a robust classification of the activity
of young radio sources. \\
Detecting young radio
sources already in a remnant/restarted phase strongly supports the power-law
distribution of the source ages
\citep[e.g.][]{hardcastle18,shabala20,morganti21} 
implying that not all the
radio sources will evolve into large and old radio galaxies. 
The detection of only three possible
cases of restarted sources suggests that the remnant phase in young
objects is shorter
than the time elapsed between the end of the first epoch and the start
of the second epoch of activity (i.e. 10$^{3-5}$ yr).
In CSS sources the decay of the
radio emission may be faster than in larger sources due to the
magnetic field \citep[about a few mG,][]{fanti95,dd02a,mo04,mo08} 
that is usually orders of magnitude higher than in
extended radio galaxies \citep[about a few
  $\mu$G,][]{croston05,brienza16,ineson17}. On the other hand,
inverse-Compton losses should not play a major role since the
equivalent magnetic field of the cosmic microwave background 
for the source at the highest
redshift is about 50 $\mu$G, i.e. much lower than the magnetic field
of the radio source. Similarly, inverse Compton losses due to the upscattering of various
surrounding photon fields (disc, torus, stars) by the lobes' electrons
should be negligible \citep{stawarz08}.\\
For the
  sources of the B3-VLA CSS sample span a large redshift range, 
  the steepness of the
  spectrum may be due to the z-$\alpha$ correlation
  \citep[e.g.,][]{ker12,morabito18}, 
  rather than the ageing. In Fig. \ref{alpha-z} we plot the $\alpha$-z
  correlation for the sources of our sample divided into active (open
  stars) and (candidate) remnant/restarted (filled stars), together
  with the radio sources from the whole B3-VLA CSS (open
  circles). Only radio sources with a spectroscopic
  redshift have been considered since we notice some discrepancy
  between photometric and spectroscopic redshifts that may affect the
  results of the correlation. The majority of the sources from our
  sample are indistinguishable from those of the B3-VLA CSS,
  suggesting that the redshift should not act as a main character in the
  steepness of the spectrum. B3-0856+406 seems an outlier, with its
  steep spectrum likely due to a combination of both high redshift and
ageing of the radio emission.\\
\indent We remark that among the sources of the B3-VLA CSS sample not observed
here, there are other objects lacking a secure core detection and
showing radio structures dominated by extended low-surface brightness
features.
Although
their integrated spectrum is not ultra-steep, we
cannot exclude that there may be some sources that have either
switched off recently, or their radio emission has just restarted. For
this reason the percentages from this work should be considered lower limits.
Multi-frequency high angular
resolution observations of these remaining sources are thus
crucial for unveiling
the incidence of
restarted/remnants objects among the population of CSS sources. \\

\section{Summary}

This paper presents the first study of the incidence of remnant/restarted radio
sources in a complete sample of CSS objects, making use of 
multi-frequency VLA, pc-scale VLBA, and eMERLIN observations of 18
candidate remnants from the B3-VLA CSS complete sample.
The conclusions that we can draw from this investigation are:

\begin{itemize}

\item we detect the core region in 10 radio sources;
  
\item we find that 3 sources (2 MSOs and 1 CSO) out of the 18 studied
  in this paper are 
  likely remnants: two with a rather secure classification, while the
  remaining one is conservatively classified as a candidate remnant;

\item among the sources with core detection, two CSOs and 1
  MSO are likely restarted radio sources (one with a rather
    robust classification, while two conservatively considered
    candidates). These sources, 
  together with B3-0003+387 and B3-0128+394, are the only five objects whose
  synchrotron spectra are better reproduced by a CI+OFF
  model than a CI one;  

\item we discovered that 3 objects are not CSS sources but 
components of larger radio sources, and we remove them from the B3-VLA
CSS sample;

\item significant flux density asymmetry is observed in 11 out of
15 MSOs/CSOs of our sample, in agreement with the high fraction of
asymmetric objects
found in other samples of compact and young radio sources.

\end{itemize}

In order to extend the percentage of remnants and restarted sources
to the whole B3-VLA CSS sample we need further high-resolution
observations of those sources still lacking multi-frequency data.\\
Determining the incidence of young but fading radio sources that can
be observed by current and future facilities is paramount. 
A complete characterization of the life-cycle of the radio emission
from relativistic jets will be fundamental for the exploitation of the
data from the forthcoming facilities, like SKA and ngVLA. Thanks to the
significant improvement in sensitivity and resolution, SKA and ngVLA
will allow systematic studies of populations of fainter
and farther radio sources, providing a fundamental step forward in our
understanding of their cosmological evolution. \\

\section*{Acknowledgments}
We thank the referee, D. Saikia, for reading the manuscript carefully
and making valuable suggestions.
The VLA and VLBA are operated by the US 
National Radio Astronomy Observatory which is a facility of the National
Science Foundation operated under cooperative agreement by Associated
Universities, Inc. e-MERLIN is a National Facility operated by the
University of Manchester at Jodrell Bank Observatory on behalf of
STFC, part of UK Research and Innovation. 
This work has made use of the NASA/IPAC
Extragalactic Database NED which is operated by the JPL, Californian
Institute of Technology, under contract with the National Aeronautics
and Space Administration. AIPS is produced and maintained by the 
  National Radio Astronomy Observatory, a facility of the National
  Science Foundation
operated under cooperative agreement by Associated Universities, Inc.
Funding for the Sloan Digital Sky Survey IV has been provided by the
Alfred P. Sloan Foundation, the U.S. Department of Energy Office of
Science, and the Participating Institutions. SDSS acknowledges support
and resources from the Center for High-Performance Computing at the
University of Utah. The SDSS web site is www.sdss.org.
SDSS is managed by the Astrophysical Research Consortium for the
Participating Institutions of the SDSS Collaboration including the
Brazilian Participation Group, the Carnegie Institution for Science,
Carnegie Mellon University, Center for Astrophysics $|$ Harvard \&
Smithsonian (CfA), the Chilean Participation Group, the French
Participation Group, Instituto de Astrofísica de Canarias, The Johns
Hopkins University, Kavli Institute for the Physics and Mathematics of
the Universe (IPMU) / University of Tokyo, the Korean Participation
Group, Lawrence Berkeley National Laboratory, Leibniz Institut für
Astrophysik Potsdam (AIP), Max-Planck-Institut für Astronomie (MPIA
Heidelberg), Max-Planck-Institut f\"ur Astrophysik (MPA Garching),
Max-Planck-Institut für Extraterrestrische Physik (MPE), National
Astronomical Observatories of China, New Mexico State University, New
York University, University of Notre Dame, Observatório Nacional /
MCTI, The Ohio State University, Pennsylvania State University,
Shanghai Astronomical Observatory, United Kingdom Participation Group,
Universidad Nacional Autónoma de México, University of Arizona,
University of Colorado Boulder, University of Oxford, University of
Portsmouth, University of Utah, University of Virginia, University of
Washington, University of Wisconsin, Vanderbilt University, and Yale
University.

\section*{Data Availability}
The data uderlying this article are available in the NRAO Data Archive 
(https://science.nrao.edu/observing/data-archive) with the
  project codes BO038, 15A-140, 16B-107, and 17B-153. Calibrated
  data are available upon reasonable request.

\appendix
\section{Images of the target sources}
\label{appendix-figure}

In Figs. \ref{vla-figure}, \ref{vlba-figure}, and \ref{merlin-figure}
we show images of the sources observed with
the VLA, VLBA, and eMERLIN, respectively. On each
image, we provide the source name, the observing band, the peak
brightness (peak) and the first contour (f.c.), which is three times
the off-source noise level on the image plane. Contours increase by a
factor of 2. The beam is plotted in the bottom left-hand corner of
each image. For the VLA data sets with polarization information, the
color-scale represents the polarization intensity, while vectors
represent the EVPA.

\begin{figure*}
\begin{center}
\includegraphics{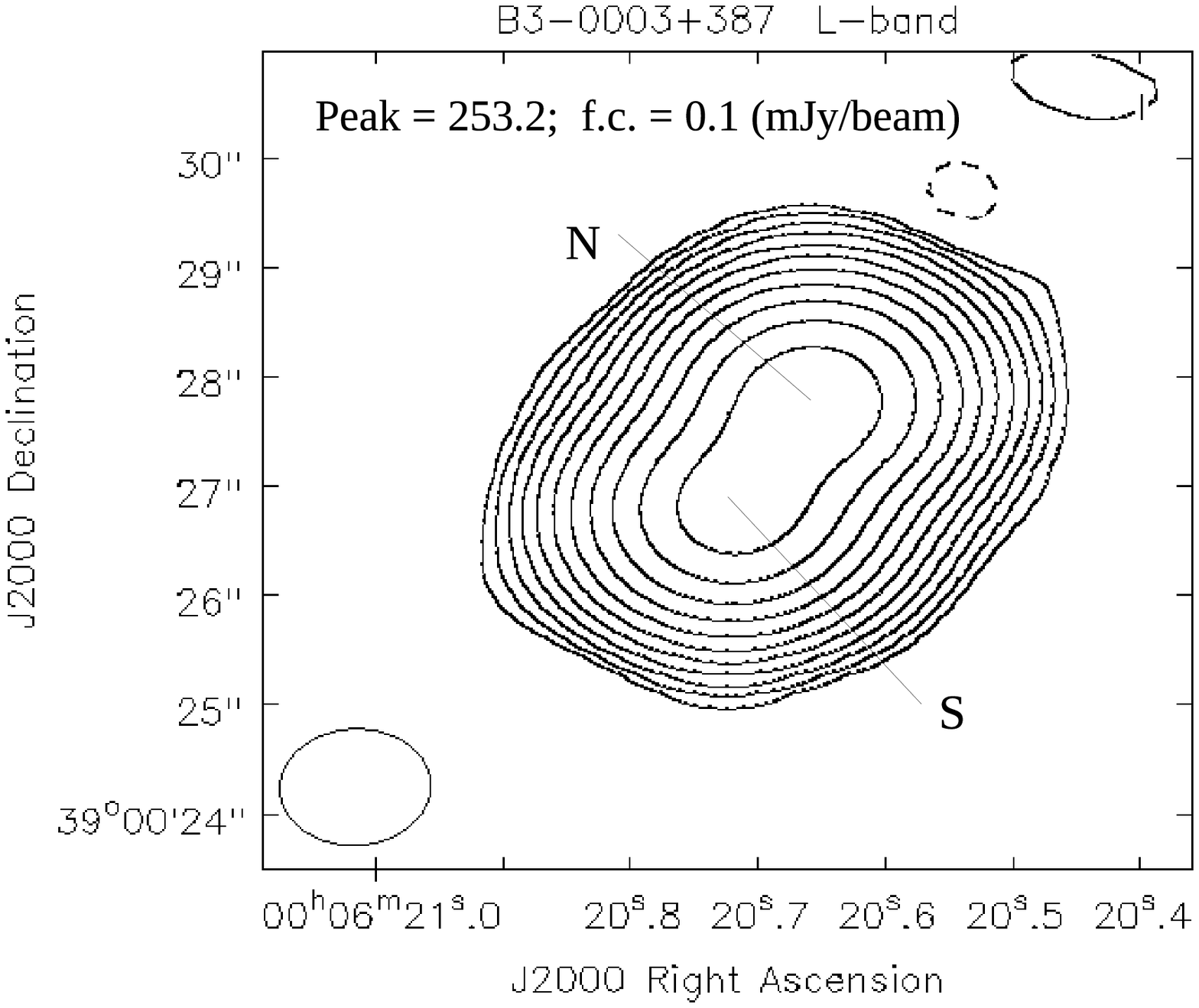}
\includegraphics{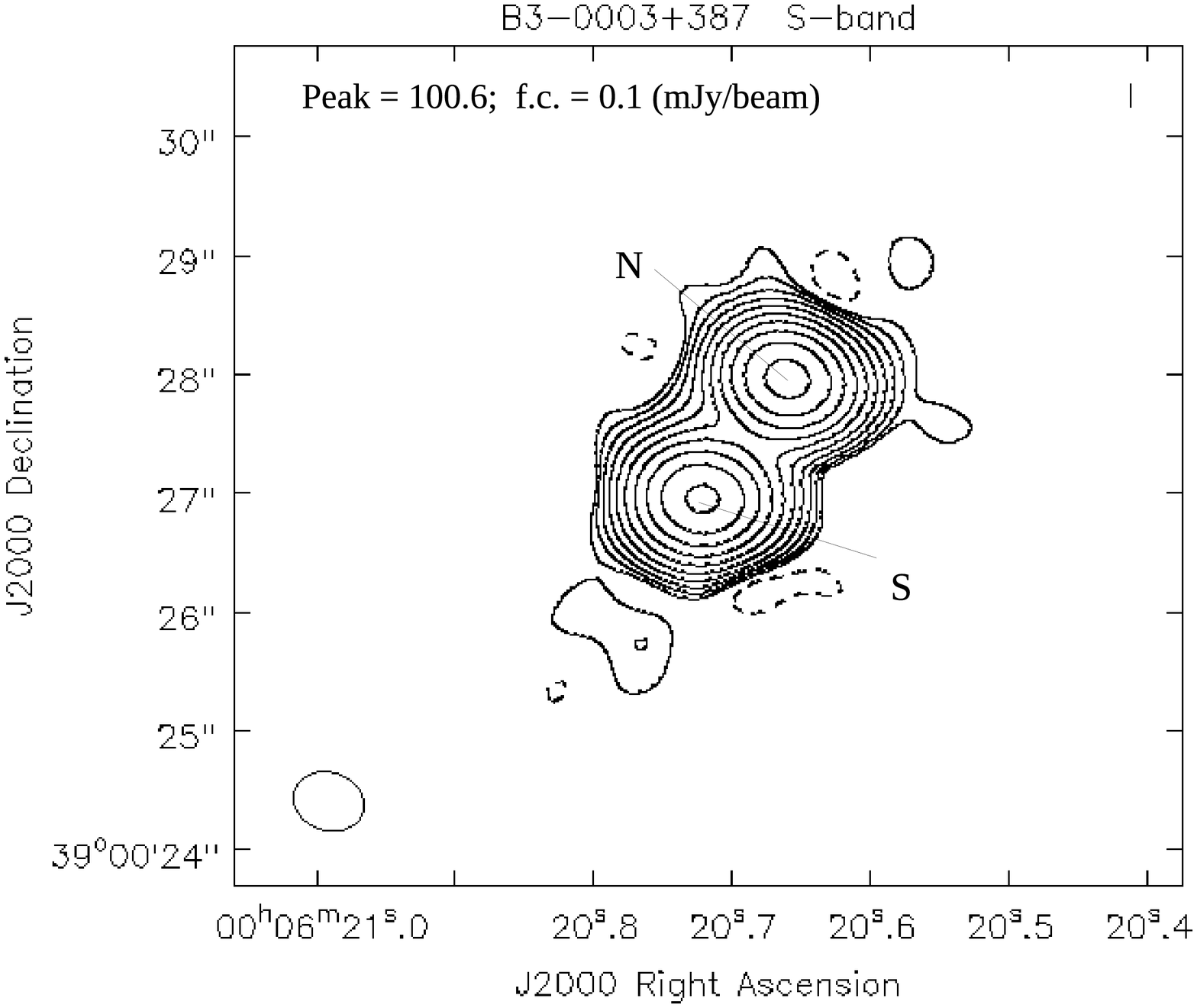}
\includegraphics{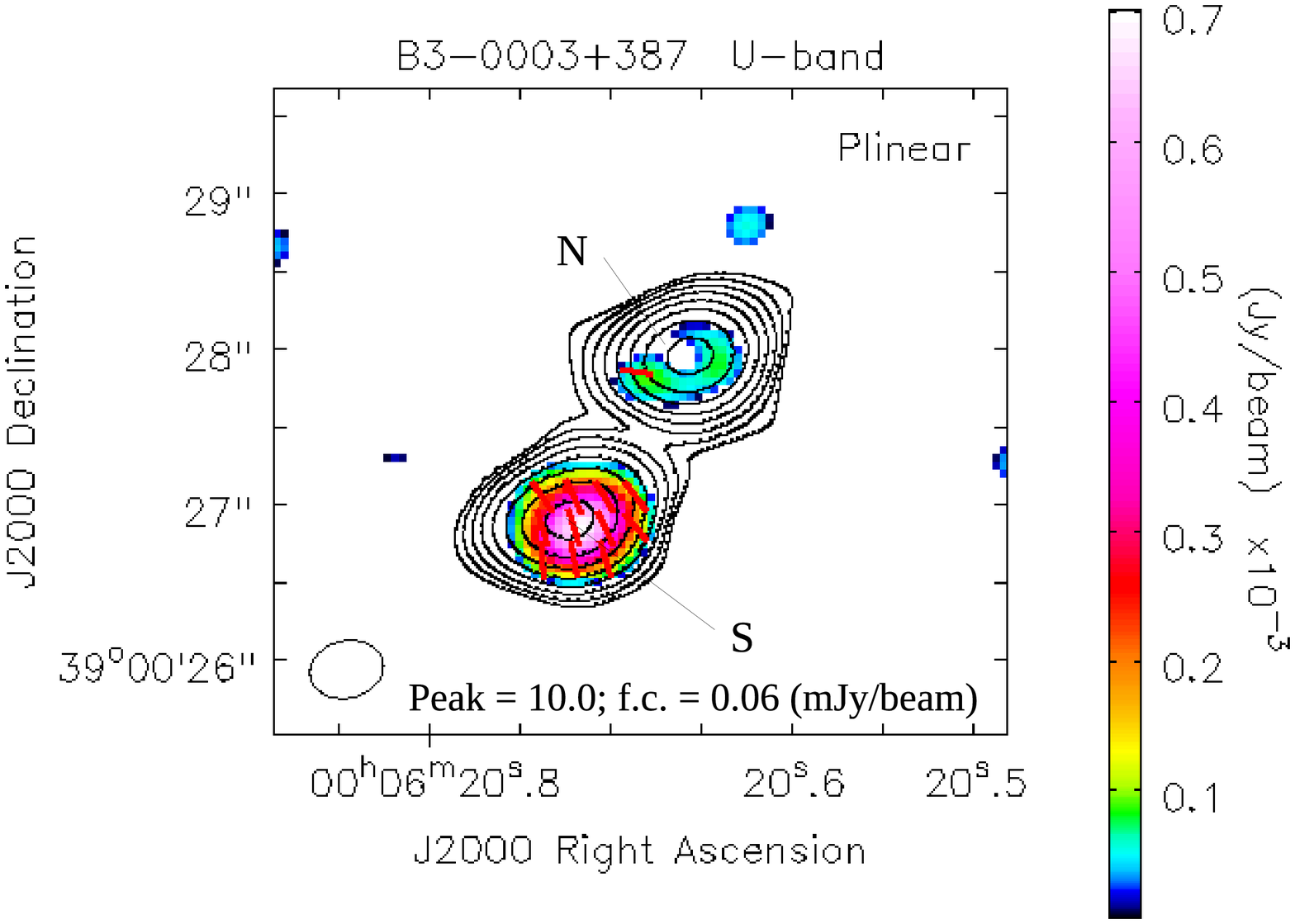}
\includegraphics{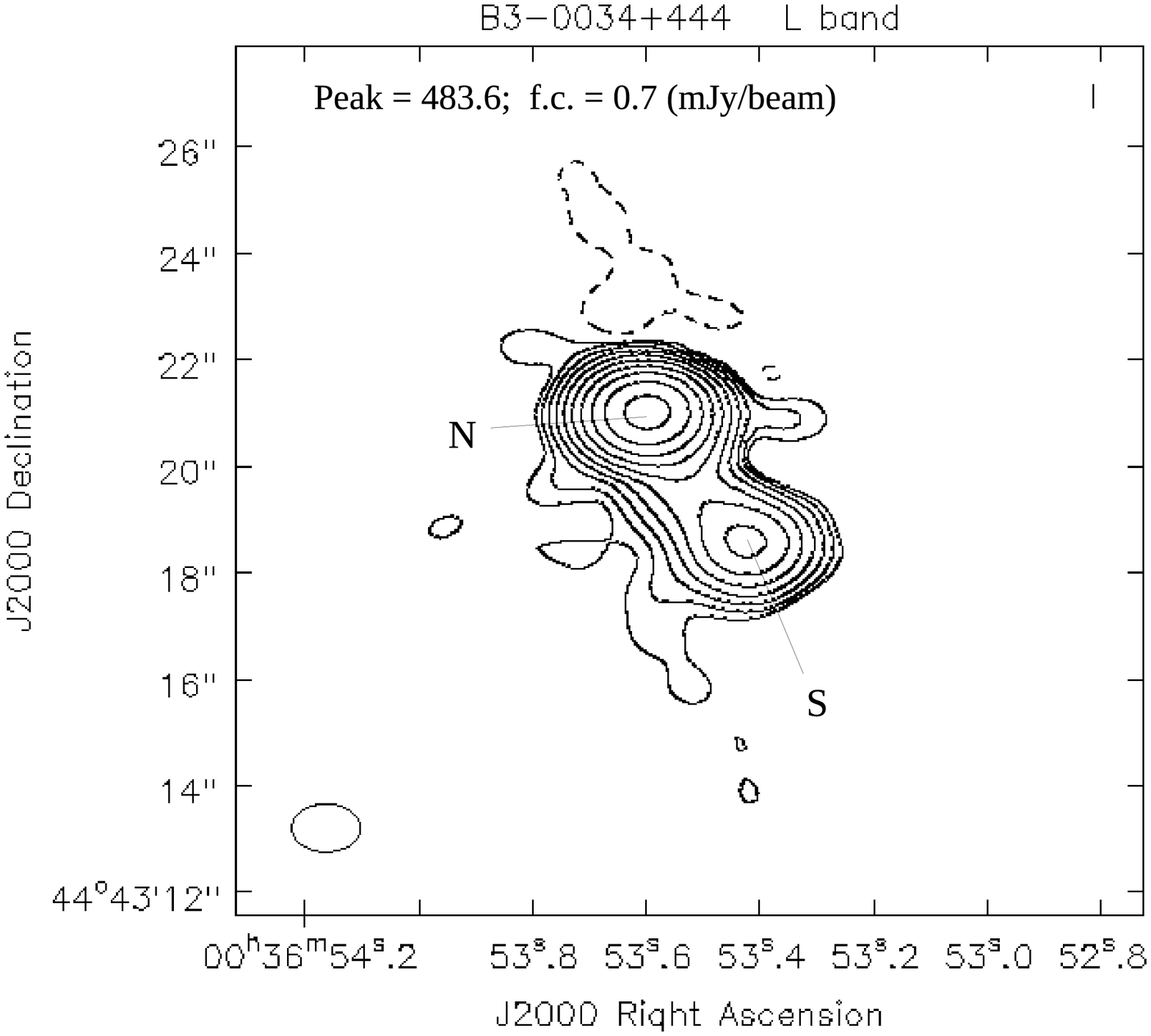}
\includegraphics{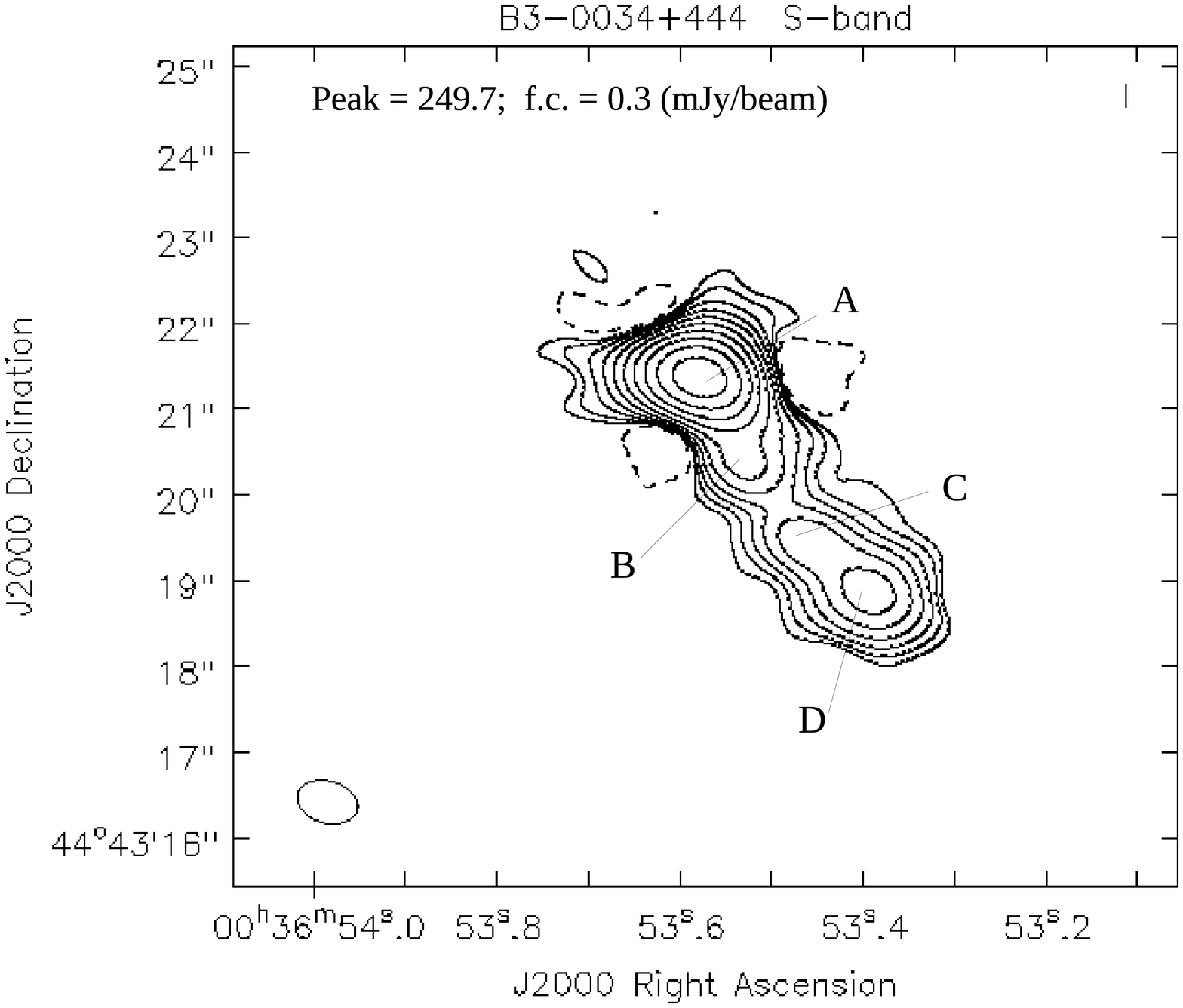}
\includegraphics{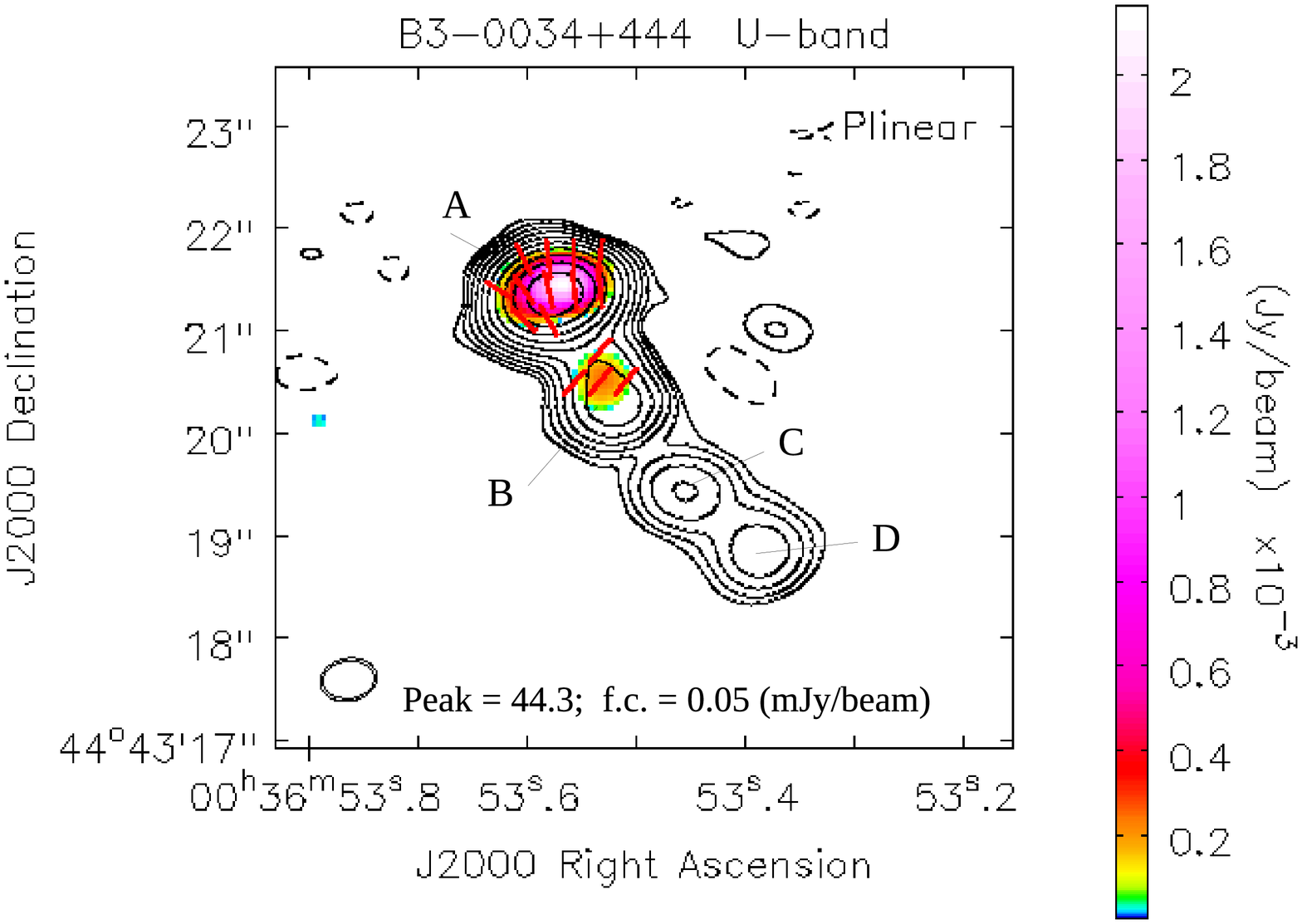}
\vspace{21cm}
\caption{VLA images in L, S and U bands
  of the CSS sources with LAS $>$ 1 arcsec. On each
image, we provide the source name, the observing band, the peak
brightness (peak) and the first contour (f.c.), which is three times
the off-source noise level on the image plane. Contours increase by a
factor of 2. The beam is plotted in the bottom left-hand corner of
each image. For the data sets with polarization information, the
color-scale represents the polarization intensity, while vectors
represent the EVPA.}
\label{vla-figure}
\end{center}
\end{figure*}

\addtocounter{figure}{-1}
\begin{figure*}
\begin{center}
\includegraphics{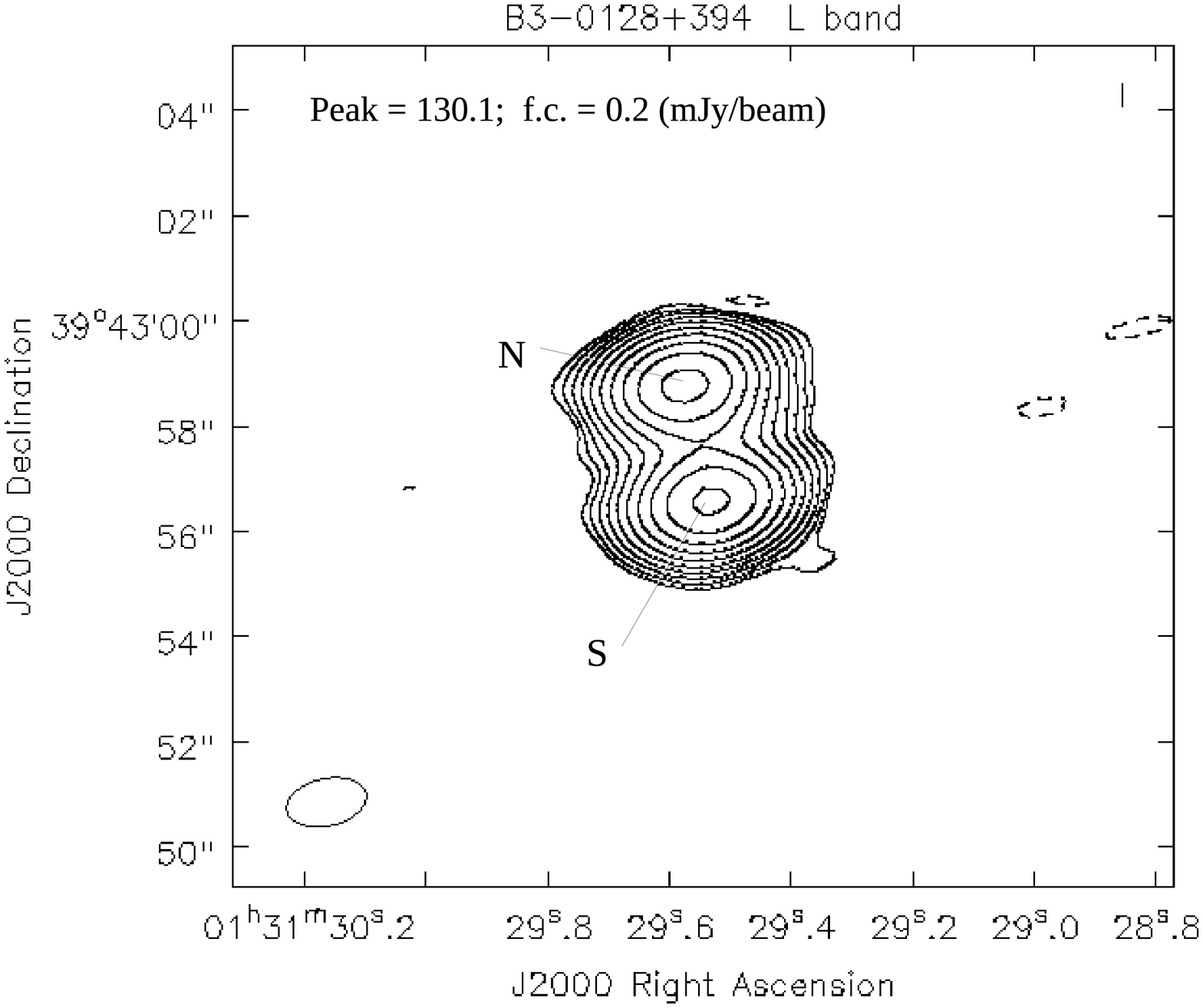}
\includegraphics{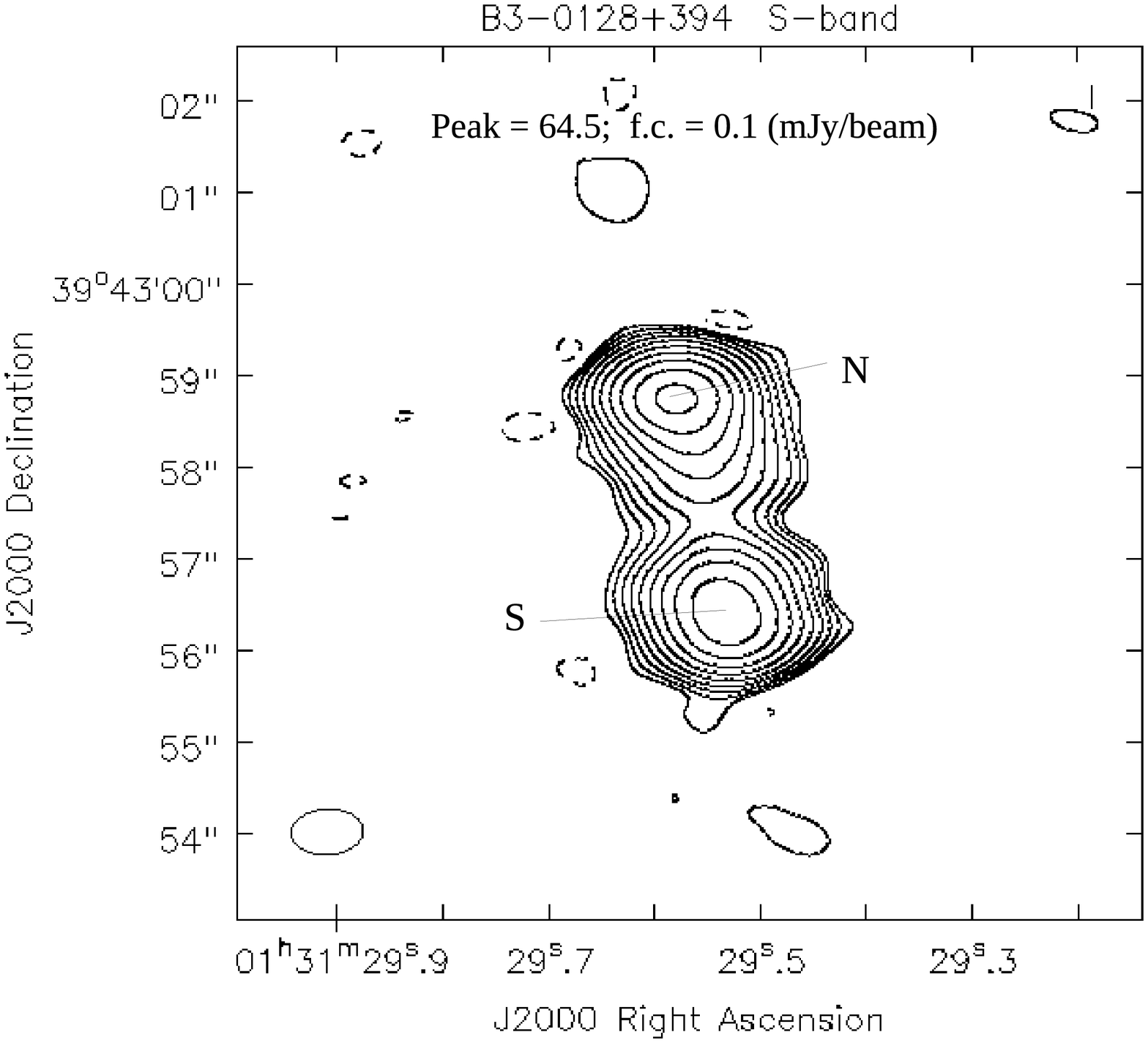}
\includegraphics{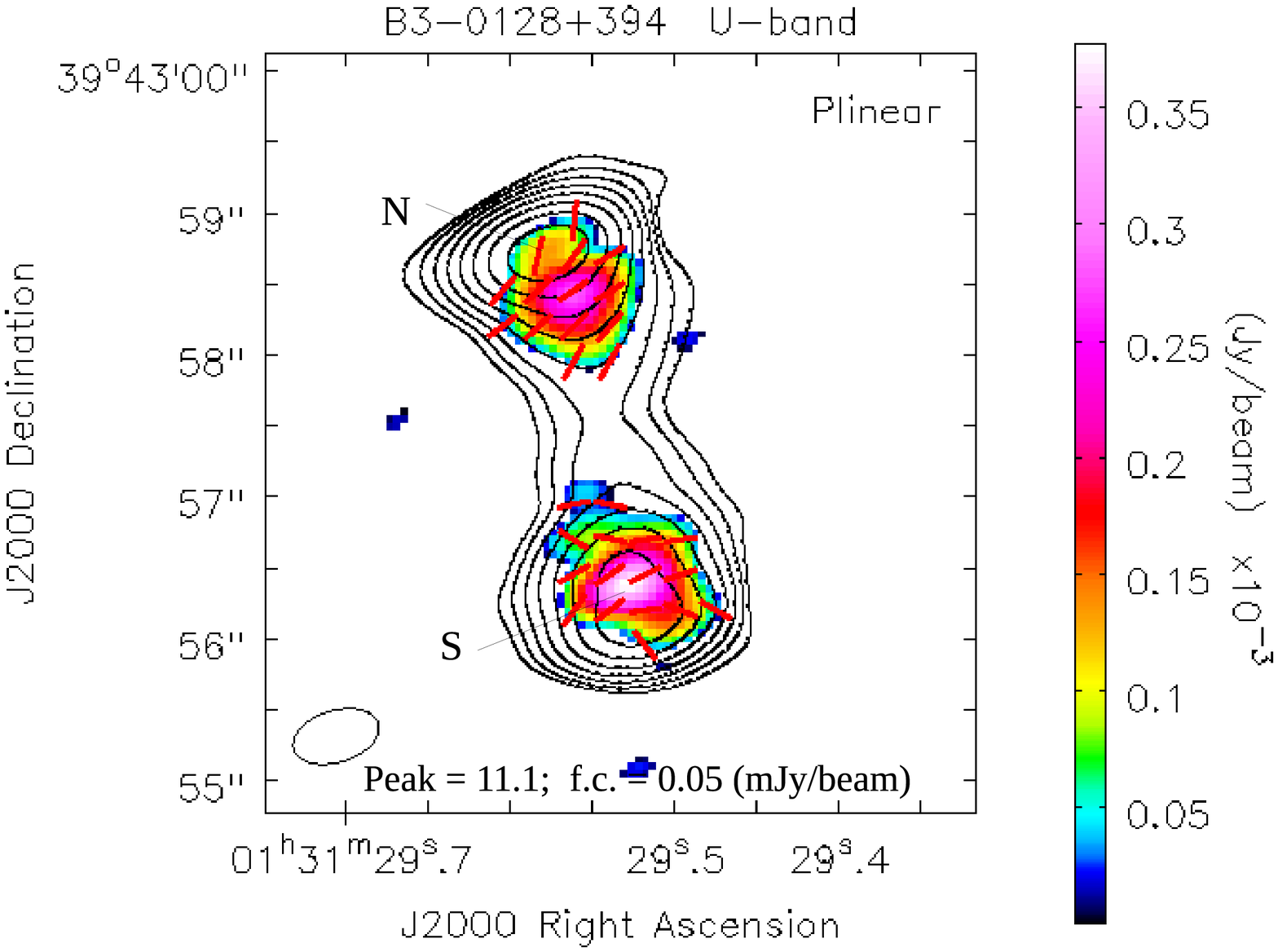}
\includegraphics{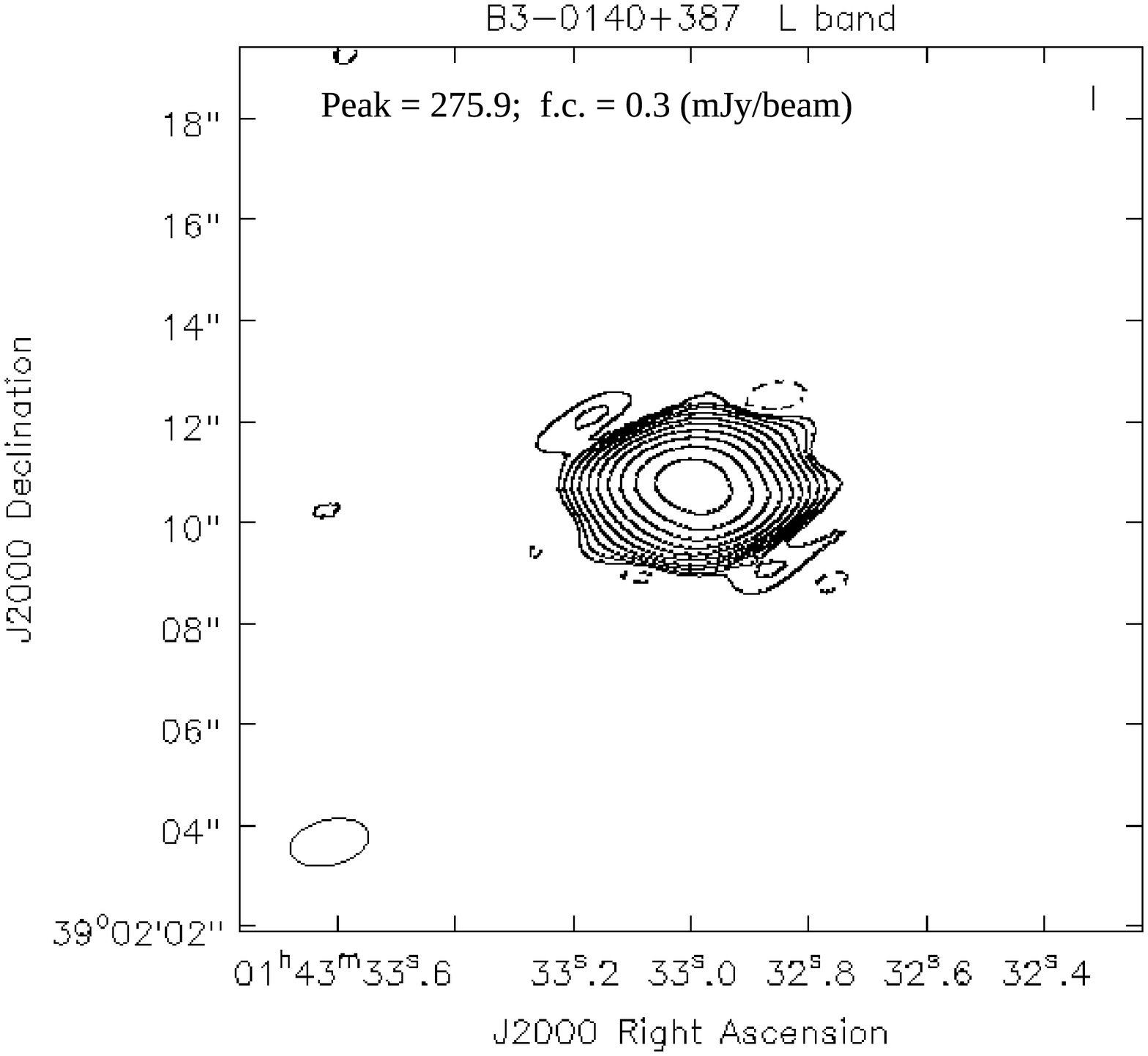}
\includegraphics{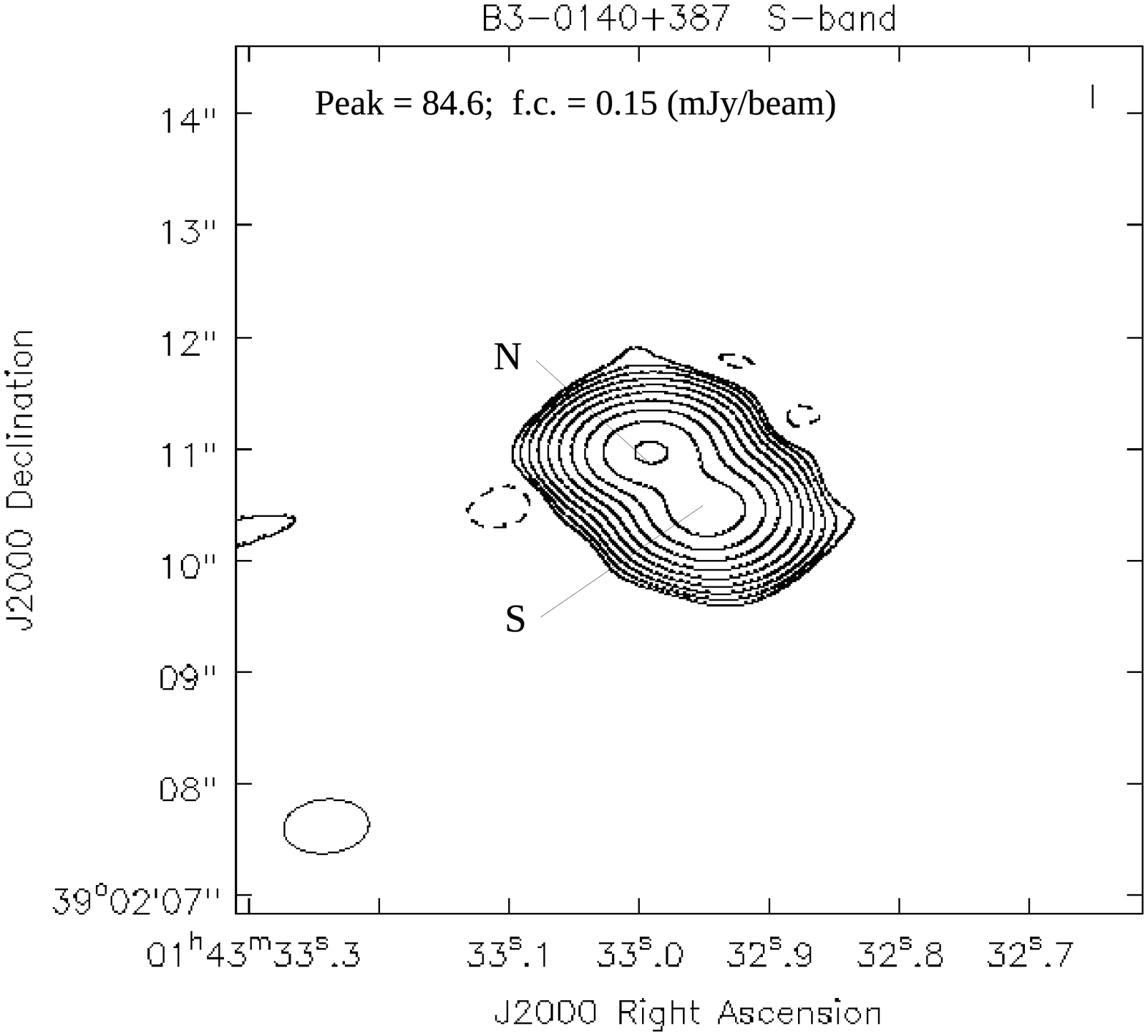}
\includegraphics{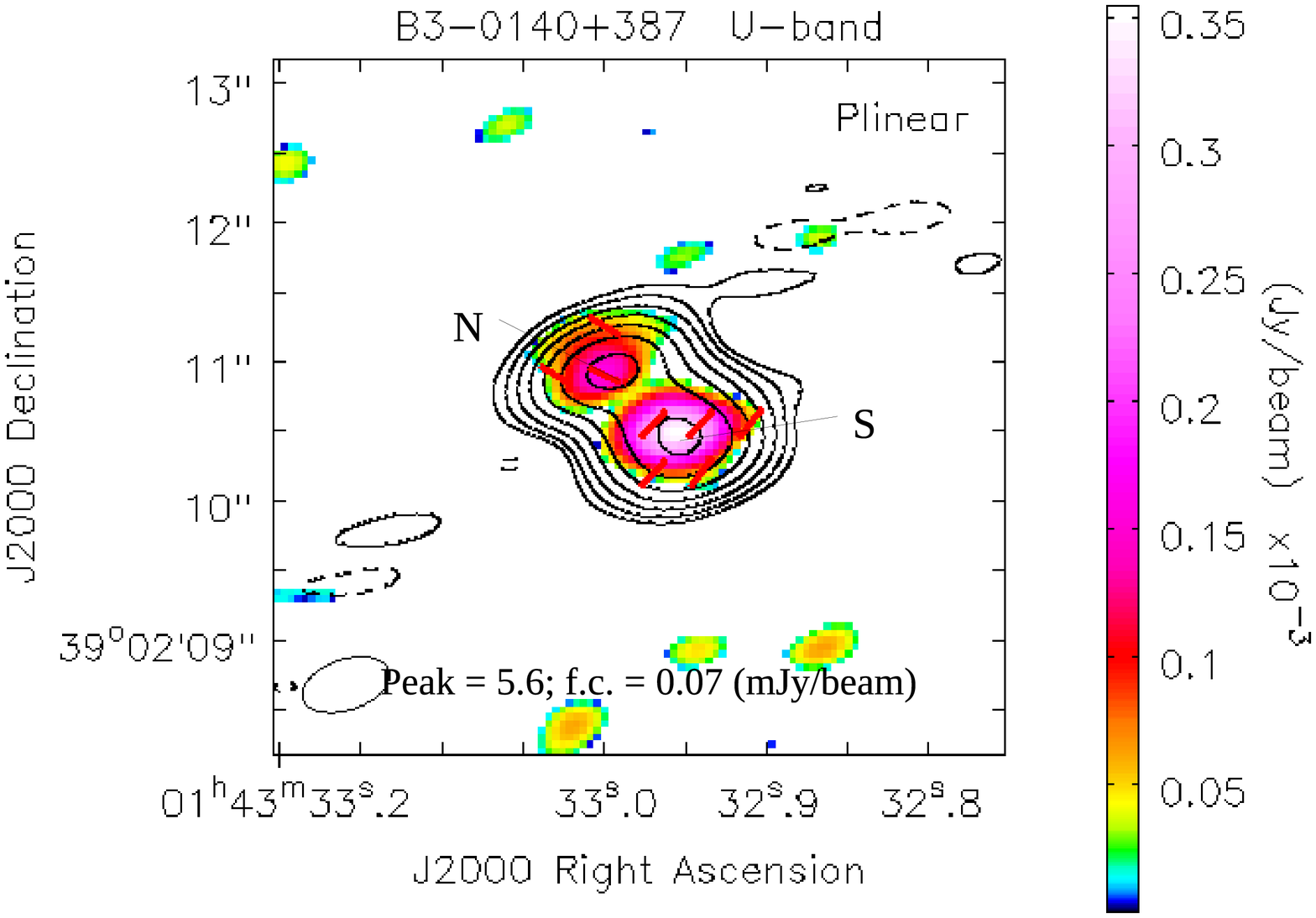}
\vspace{21cm}
\caption{Continued.}
\end{center}
\end{figure*}

\addtocounter{figure}{-1}
\begin{figure*}
\begin{center}
\includegraphics{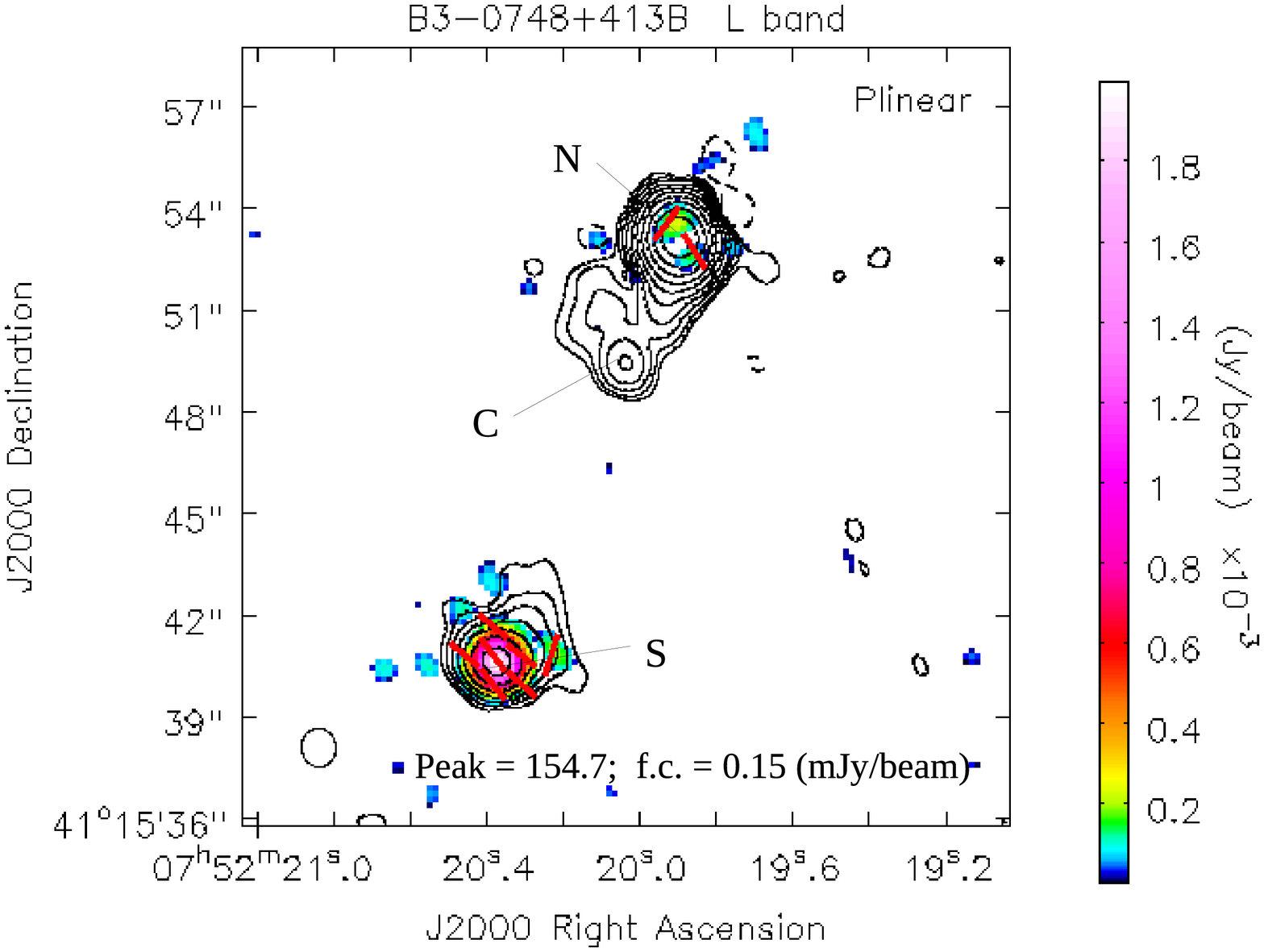}
\includegraphics{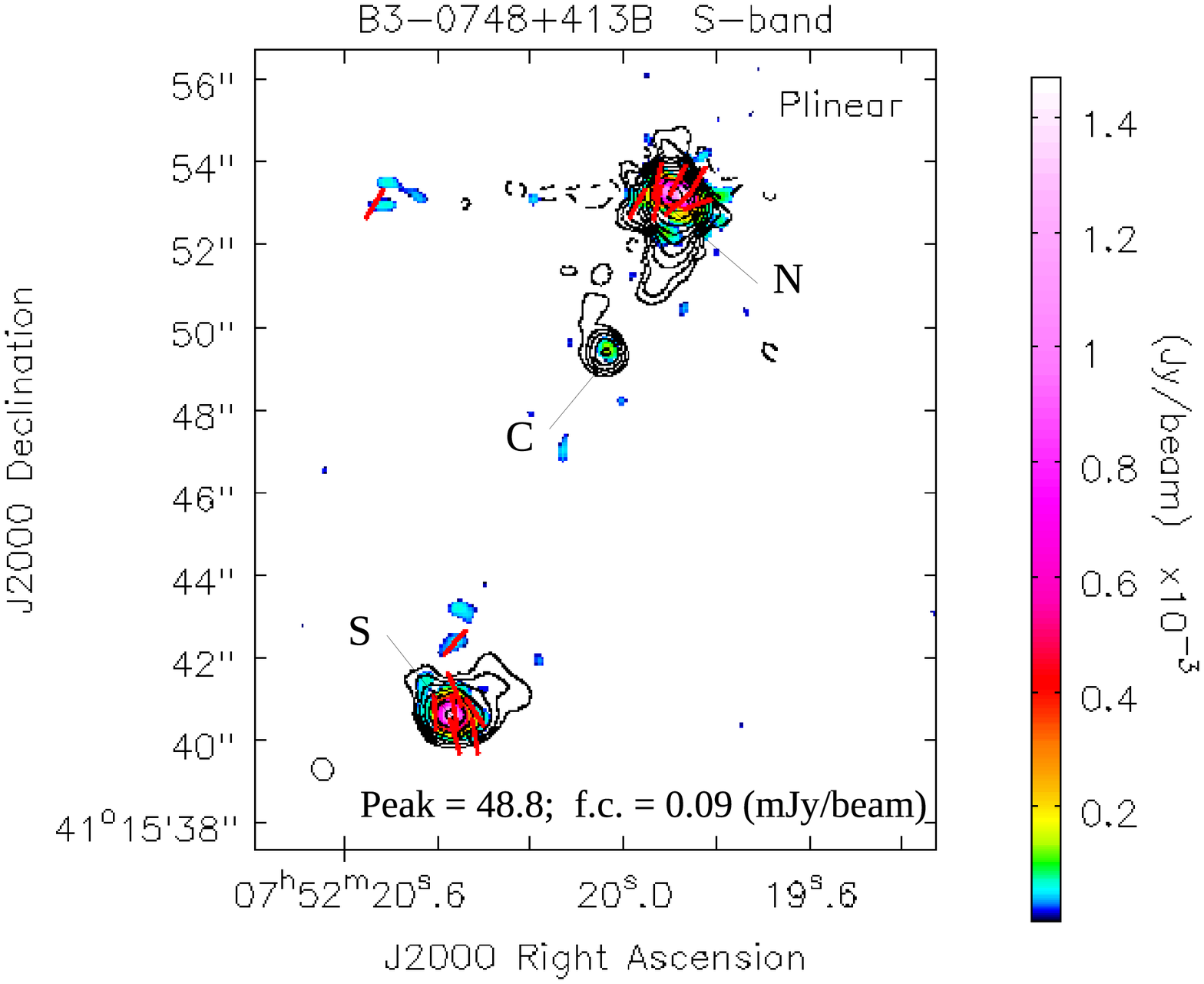}
\includegraphics{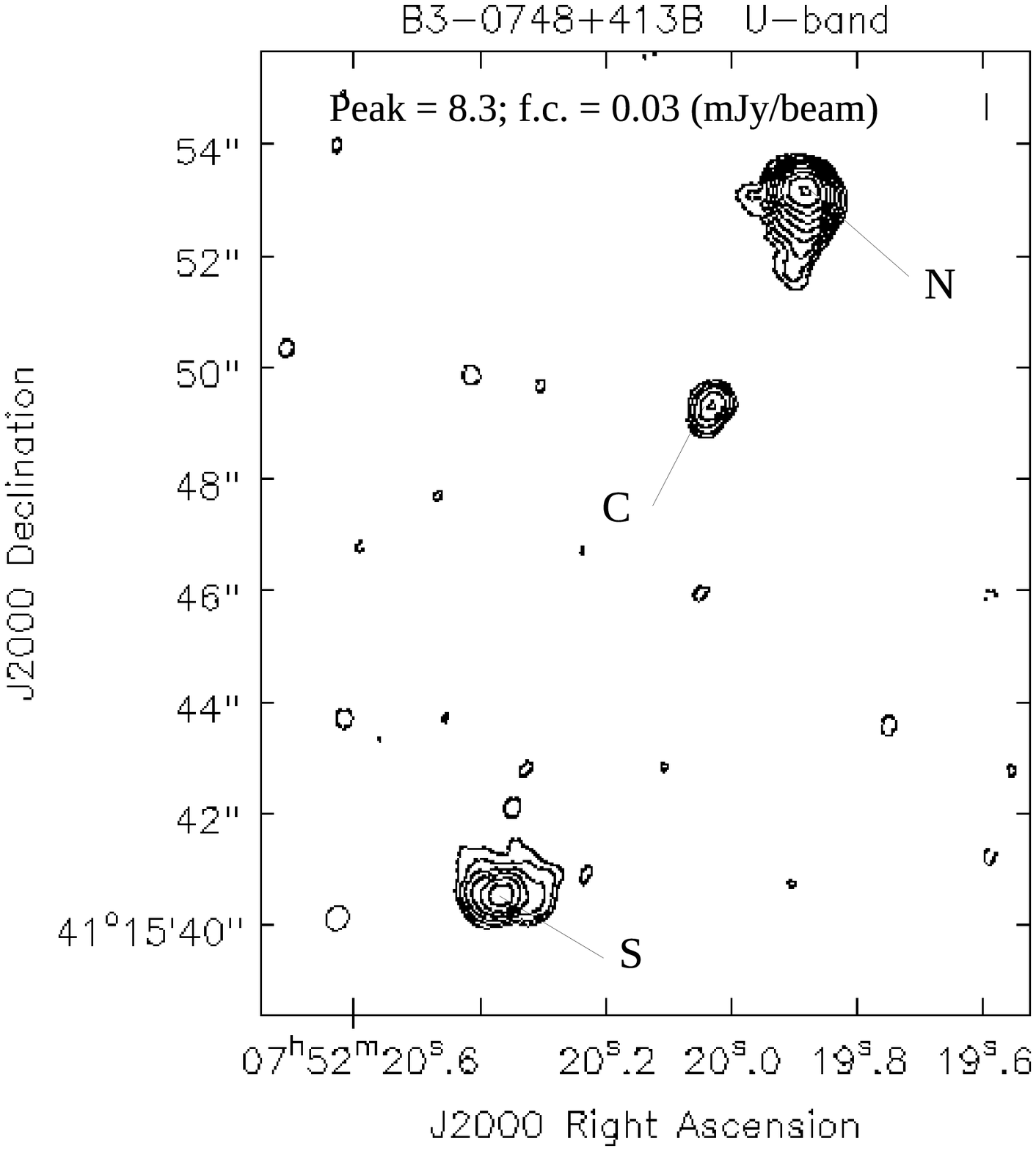}
\includegraphics{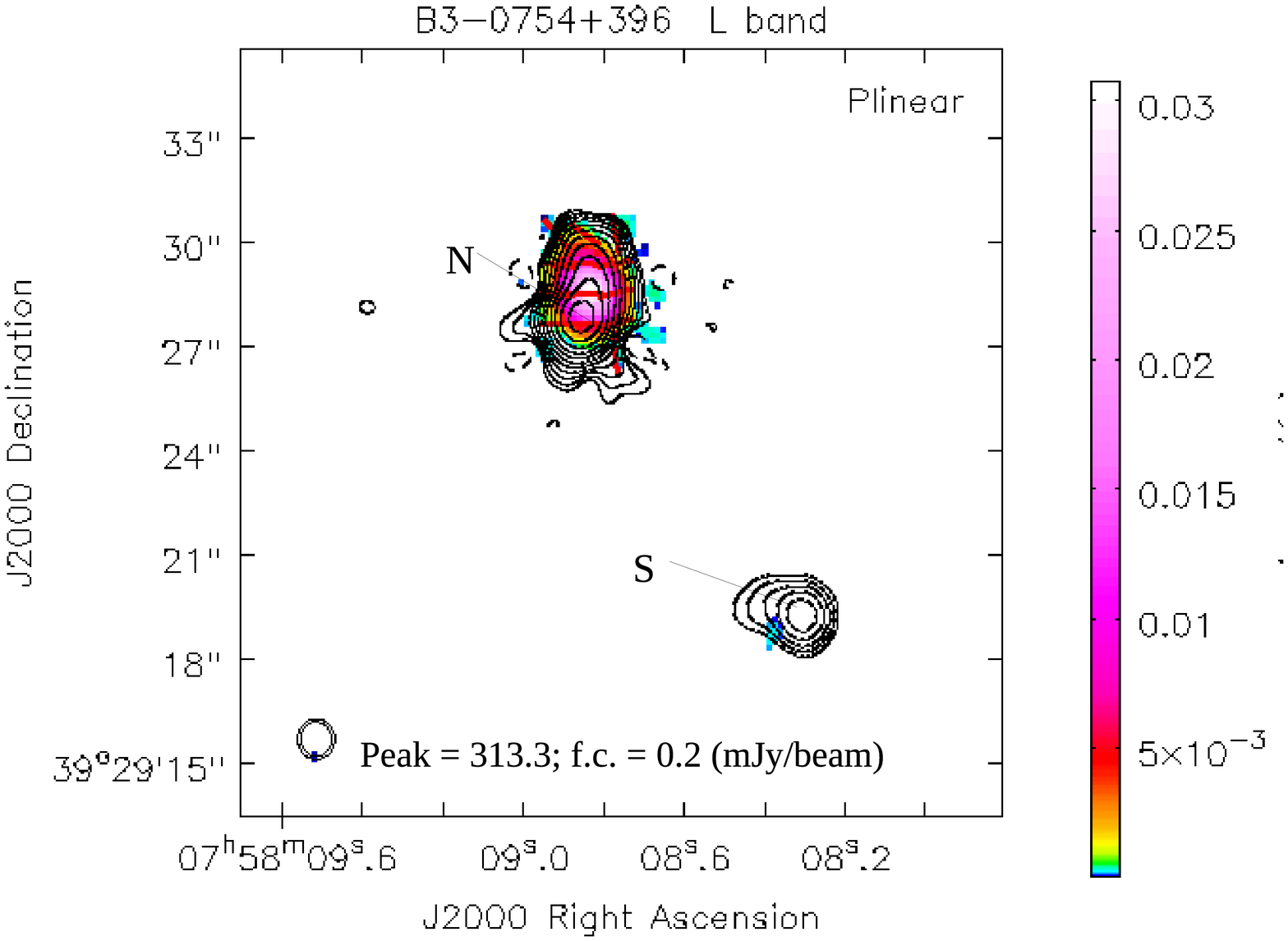}
\includegraphics{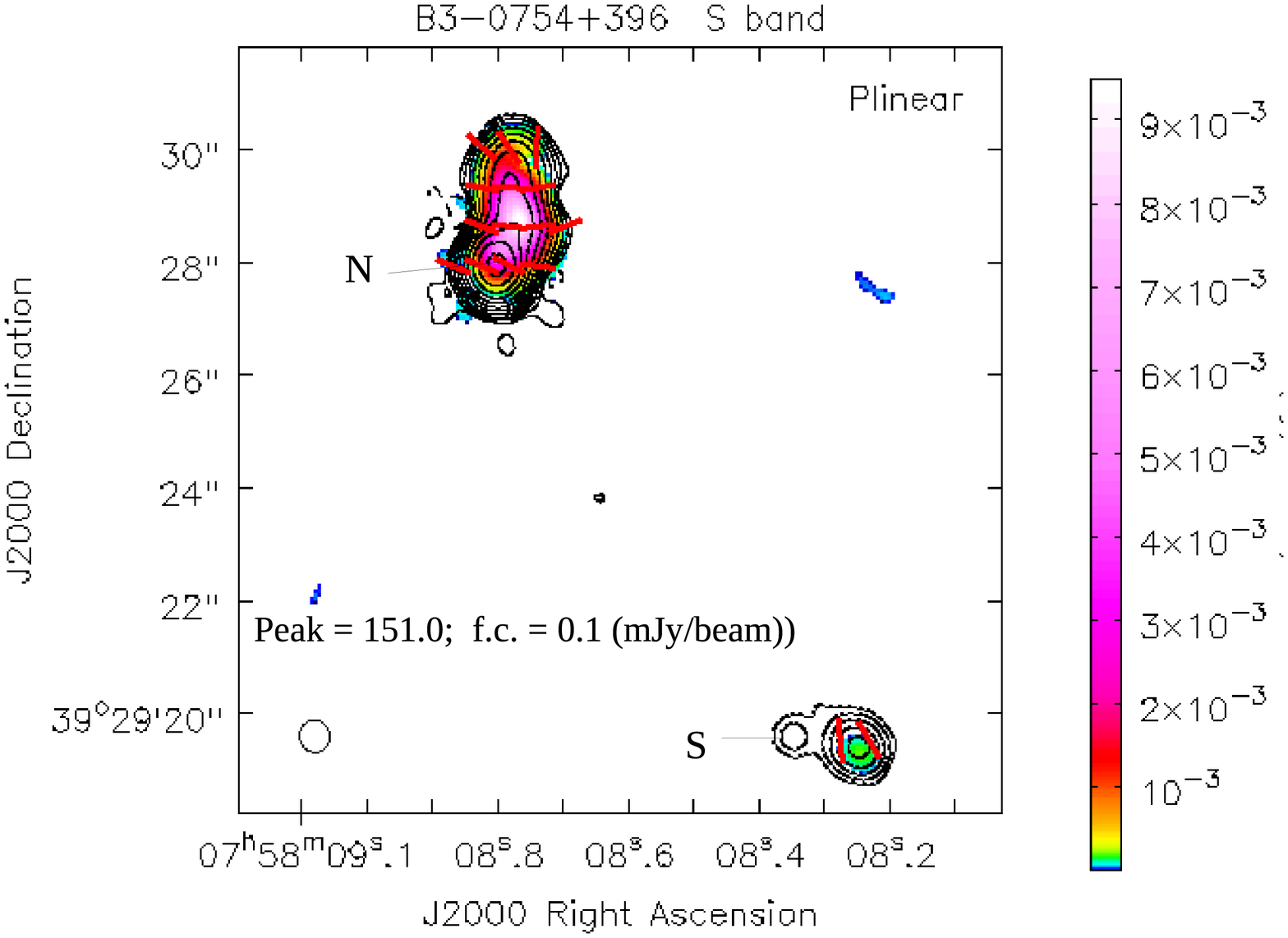}
\includegraphics{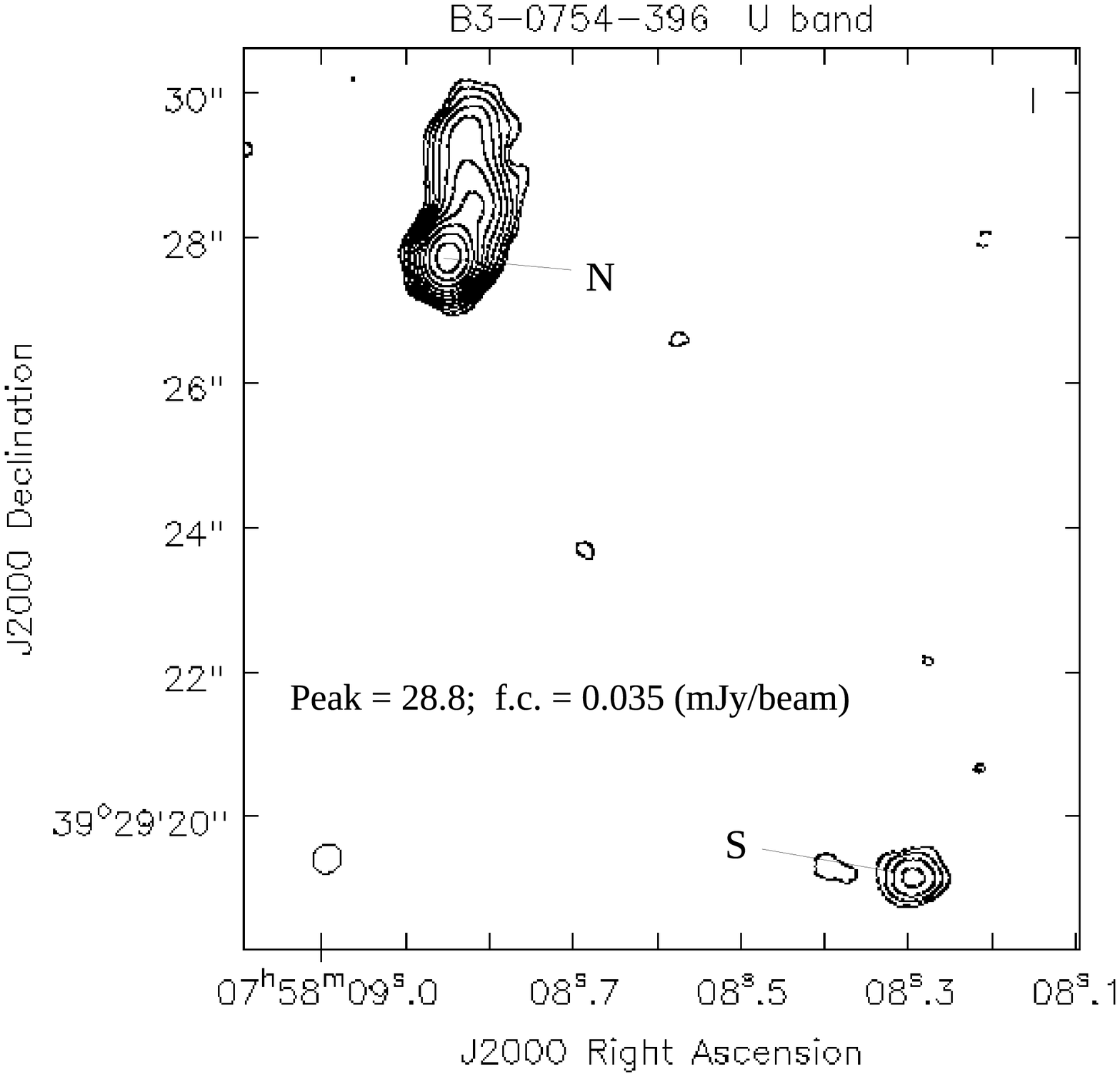}
\vspace{21cm}
\caption{Continued.}
\end{center}
\end{figure*}

\addtocounter{figure}{-1}
\begin{figure*}
\begin{center}
\includegraphics{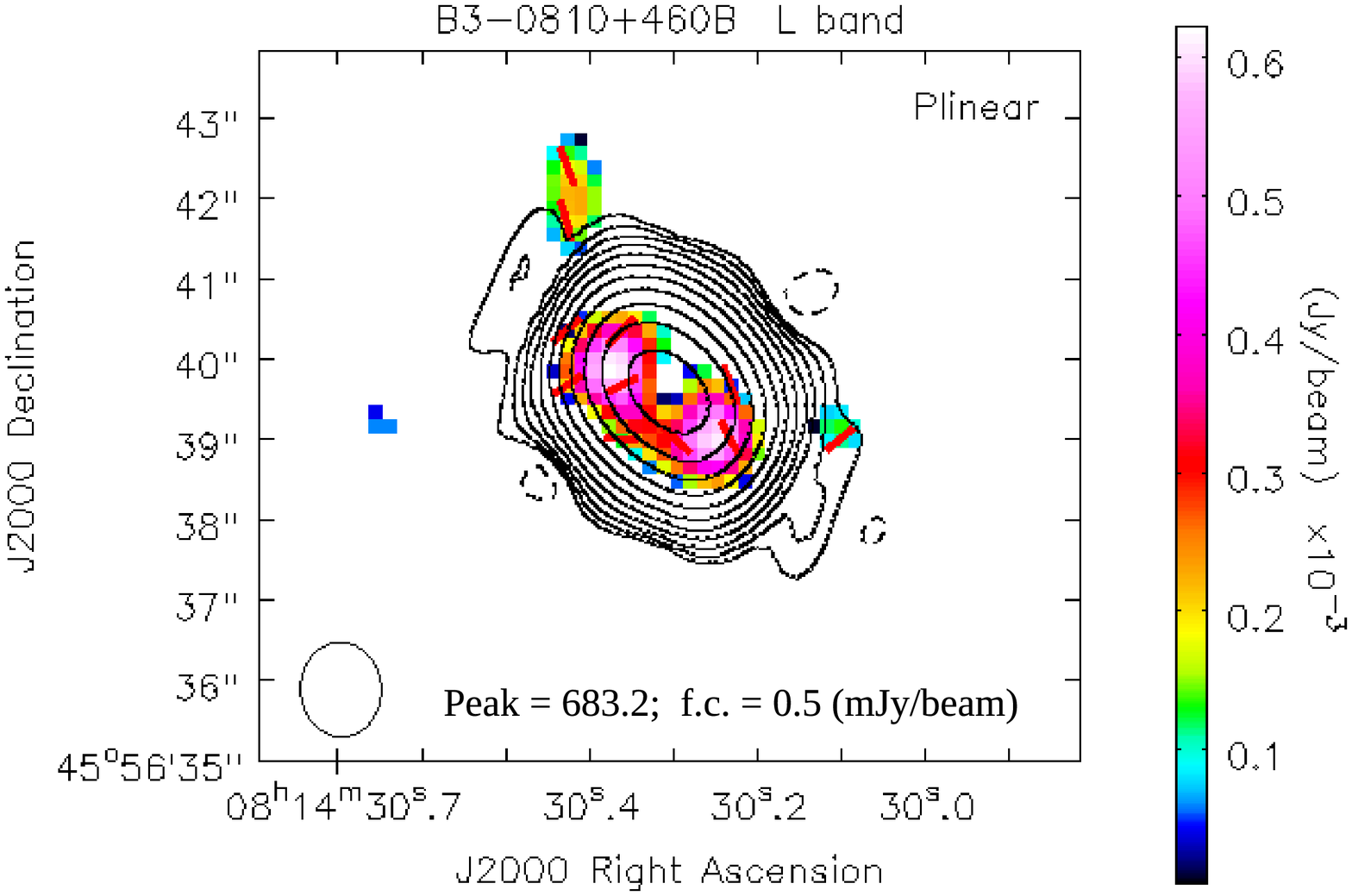}
\includegraphics{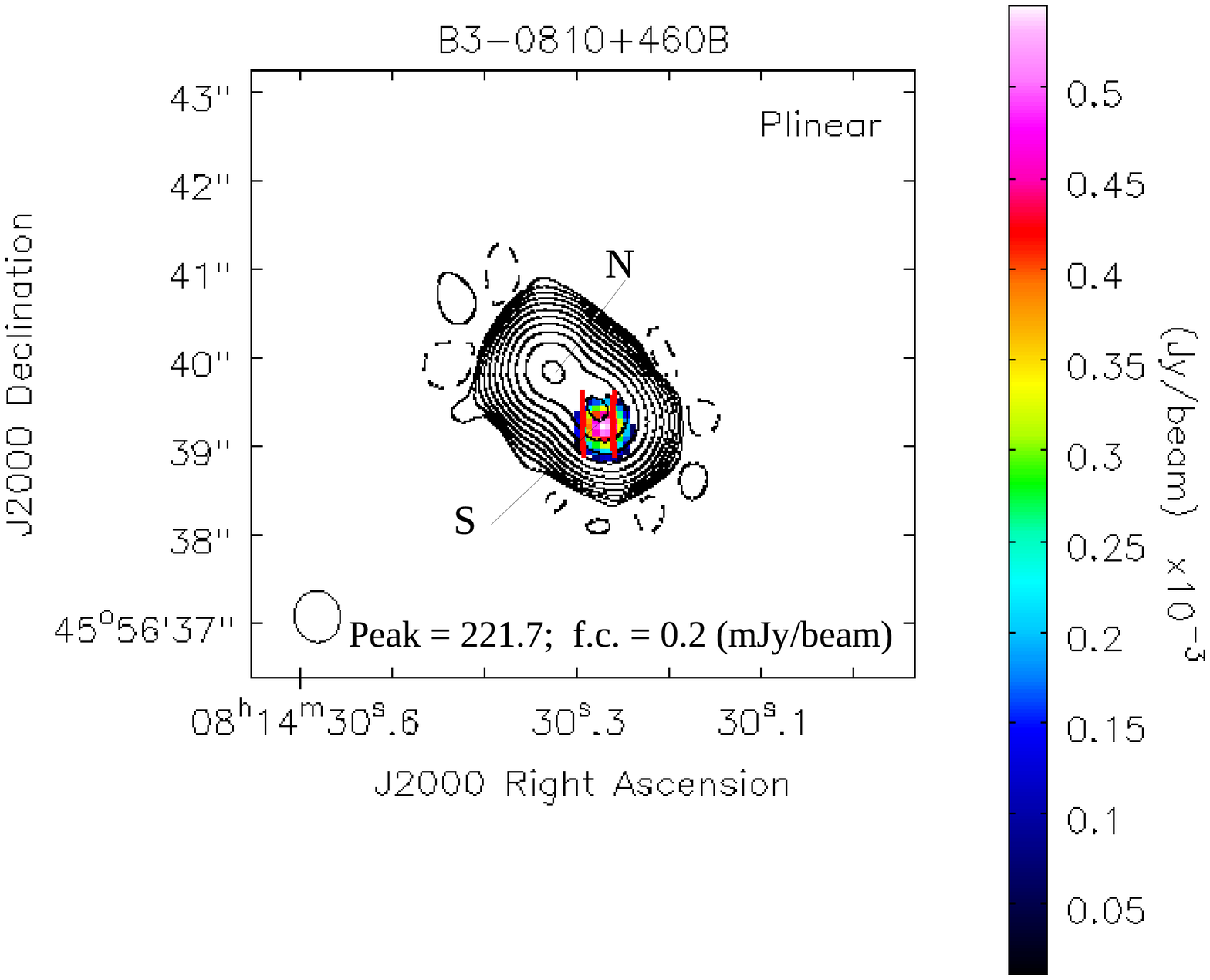}
\includegraphics{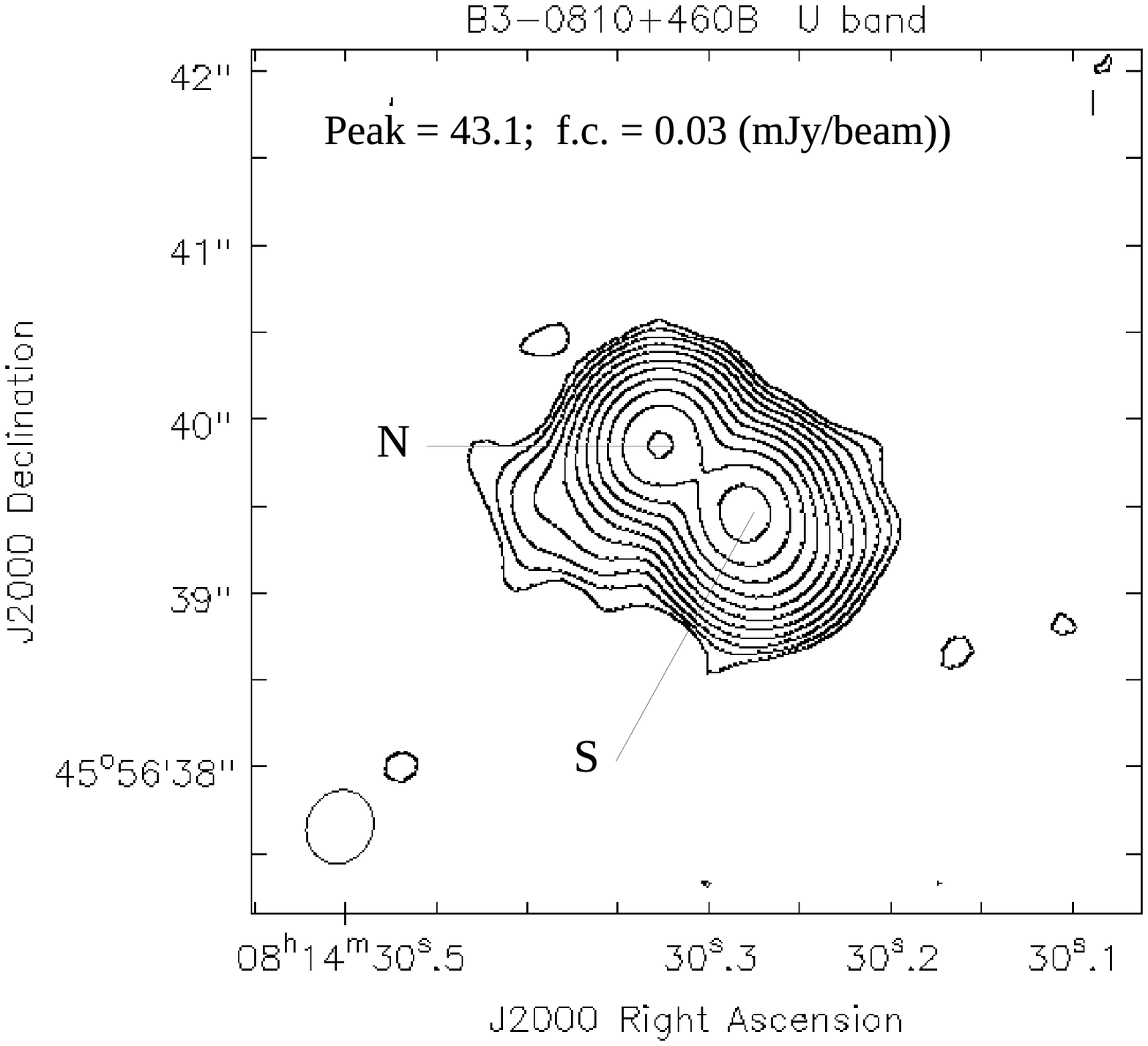}
\includegraphics{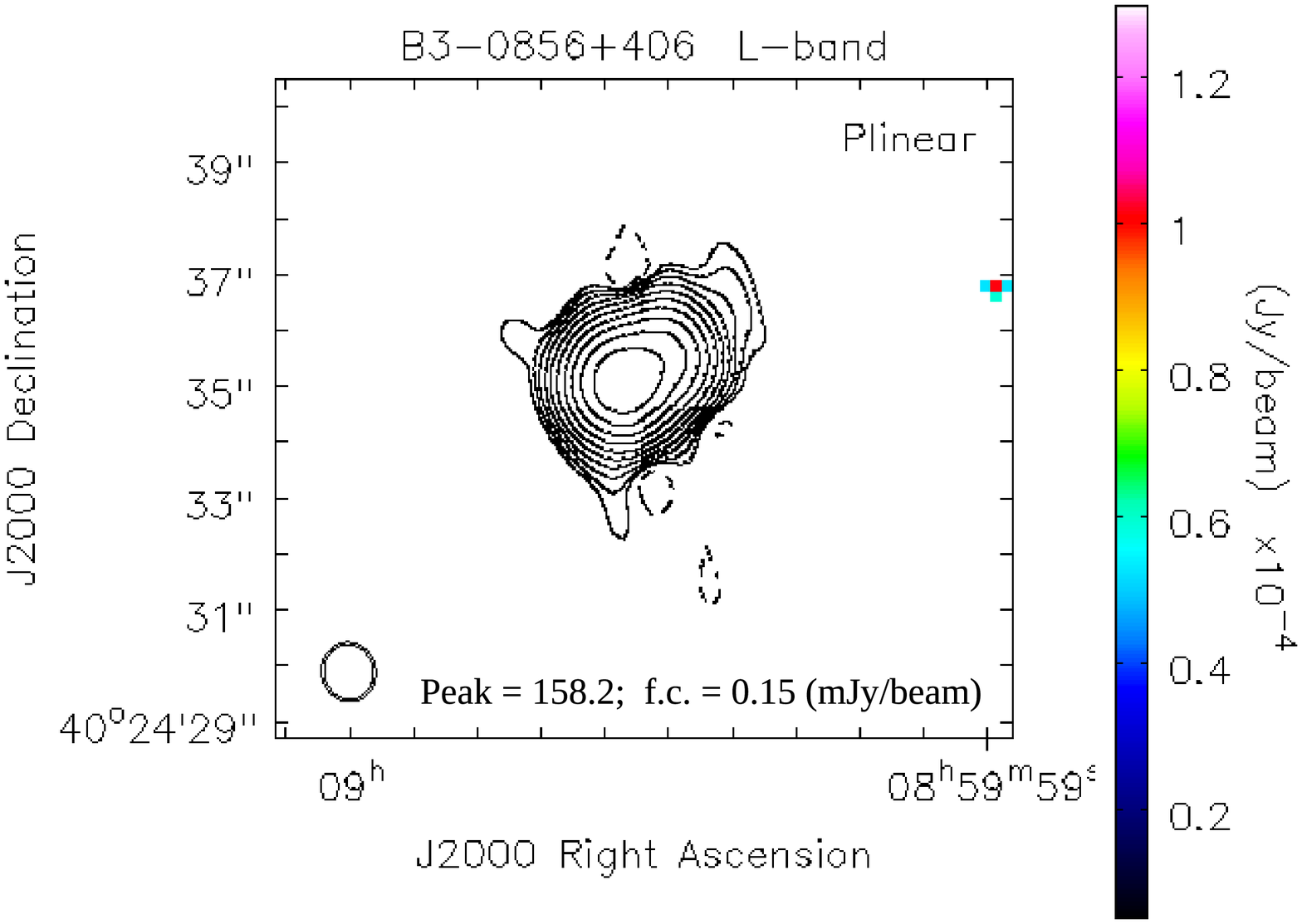}
\includegraphics{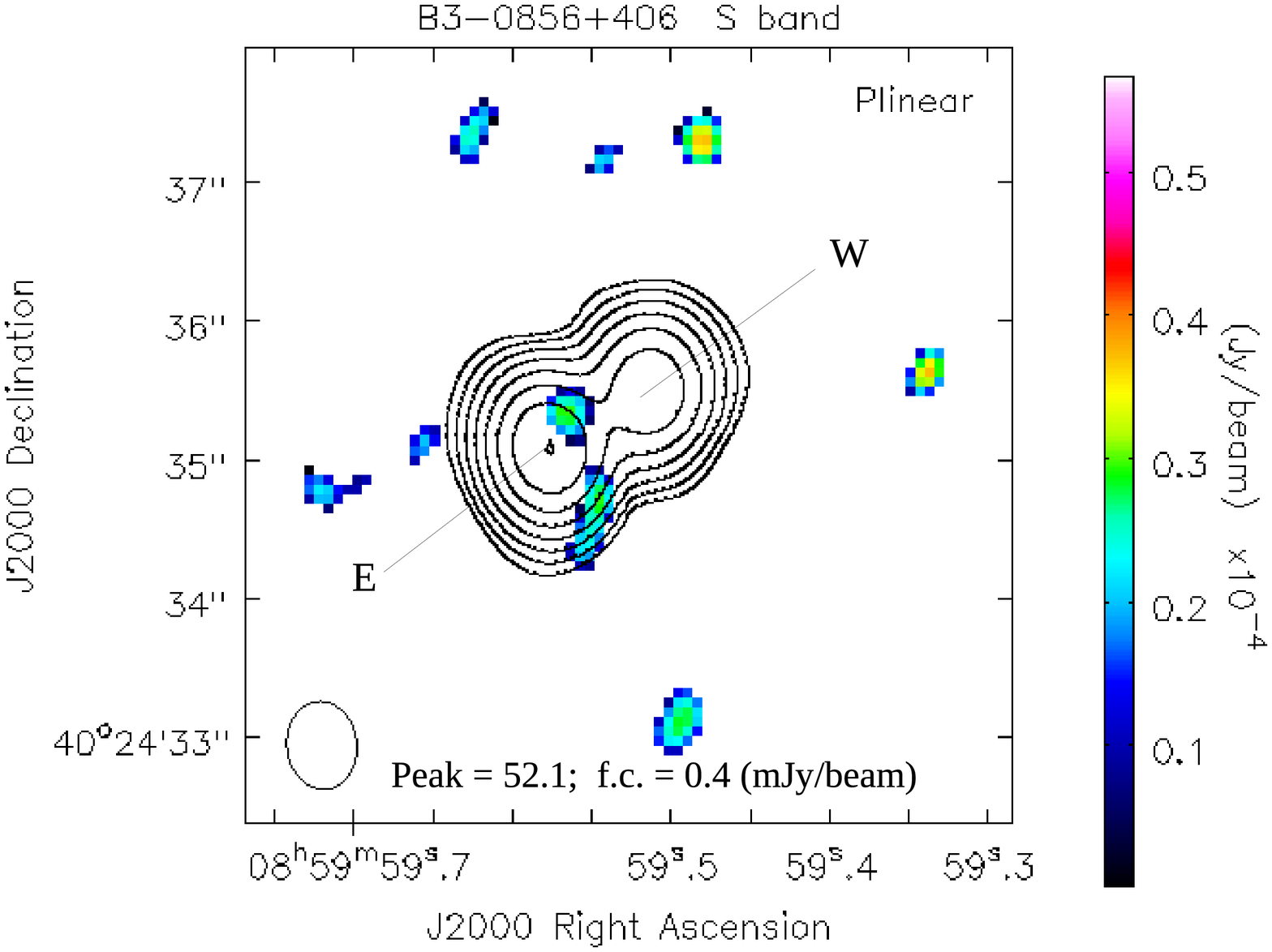}
\includegraphics{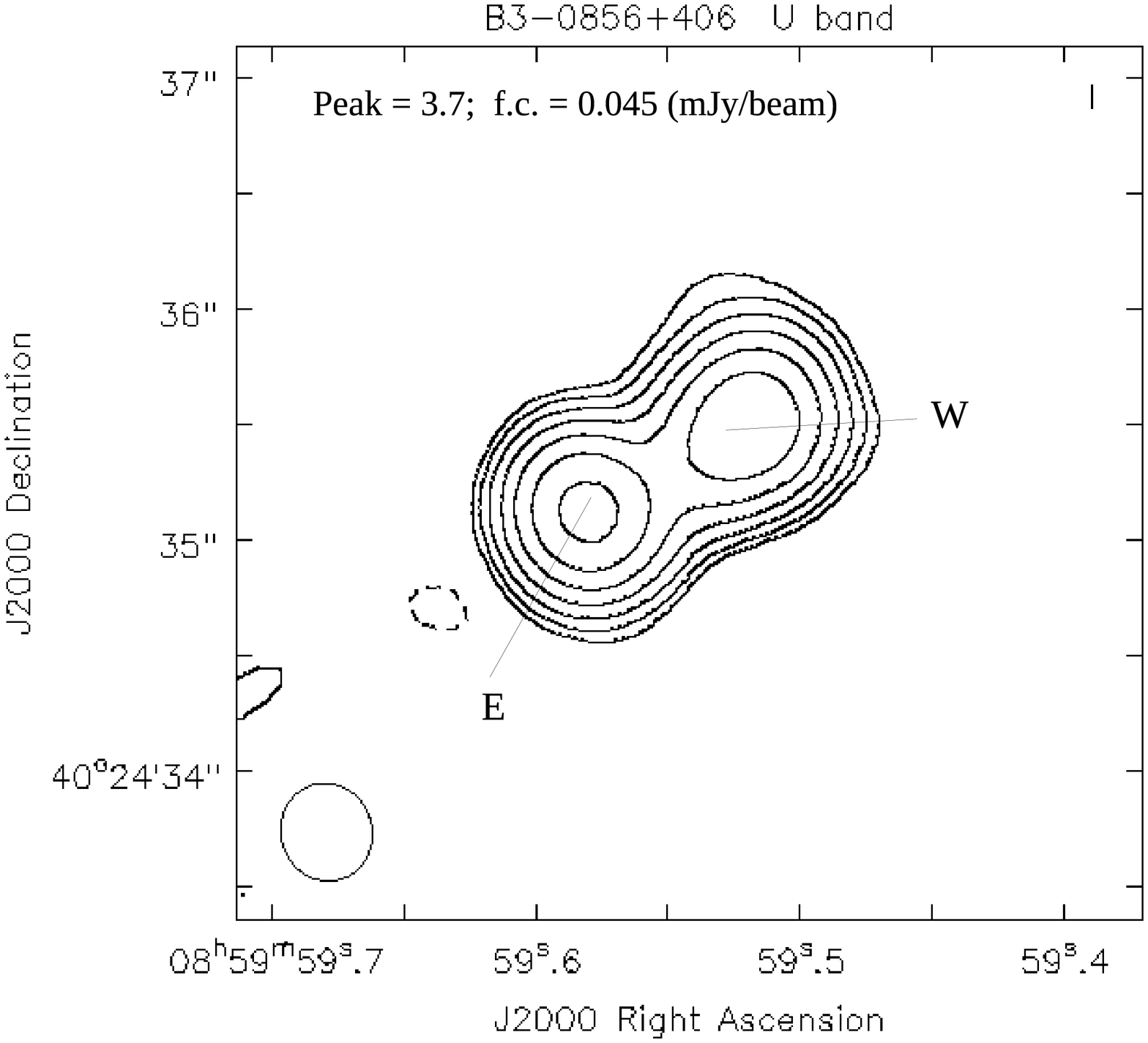}
\vspace{21cm}
\caption{Continued.}
\end{center}
\end{figure*}

\addtocounter{figure}{-1}
\begin{figure*}
\begin{center}
\includegraphics{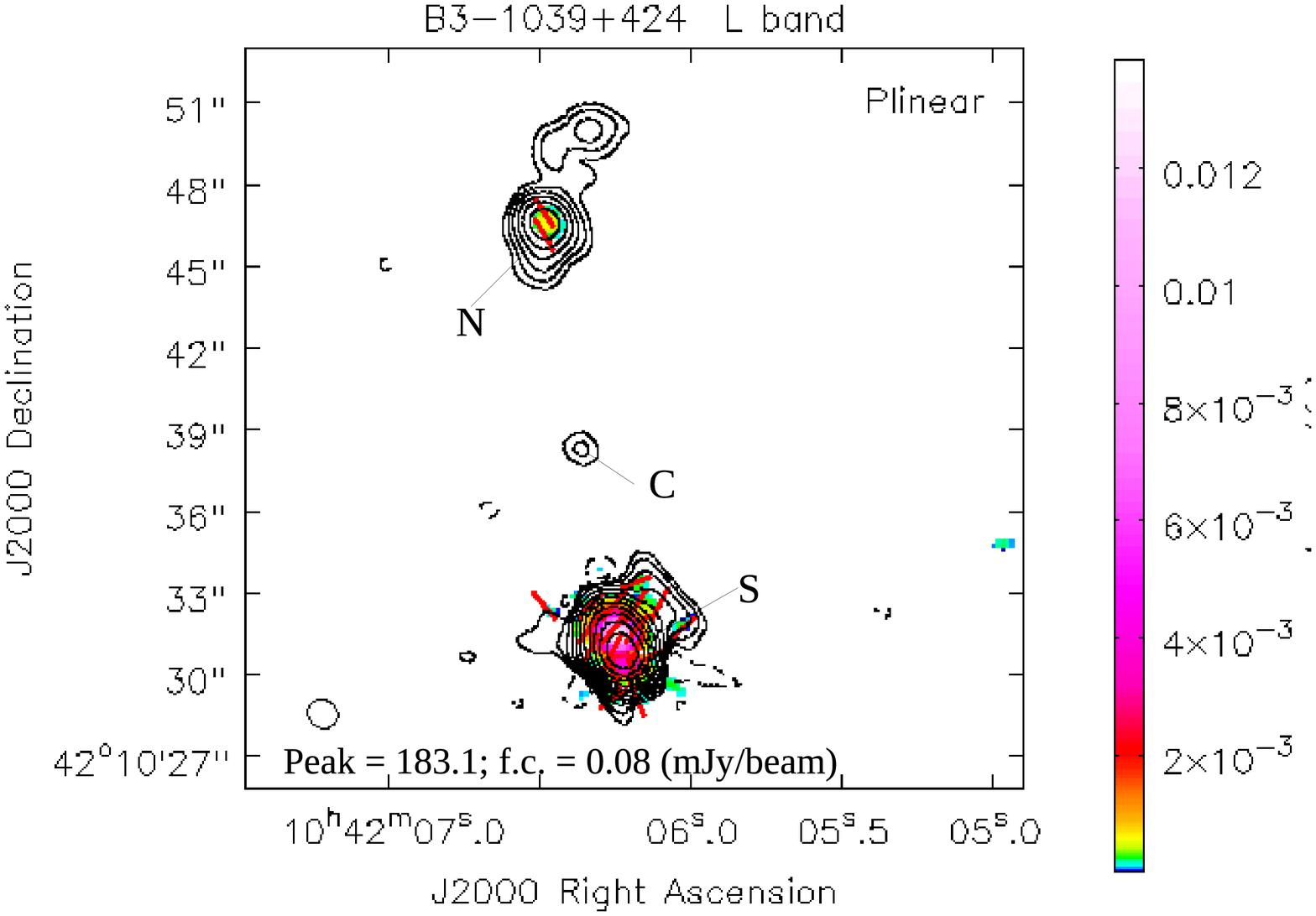}
\includegraphics{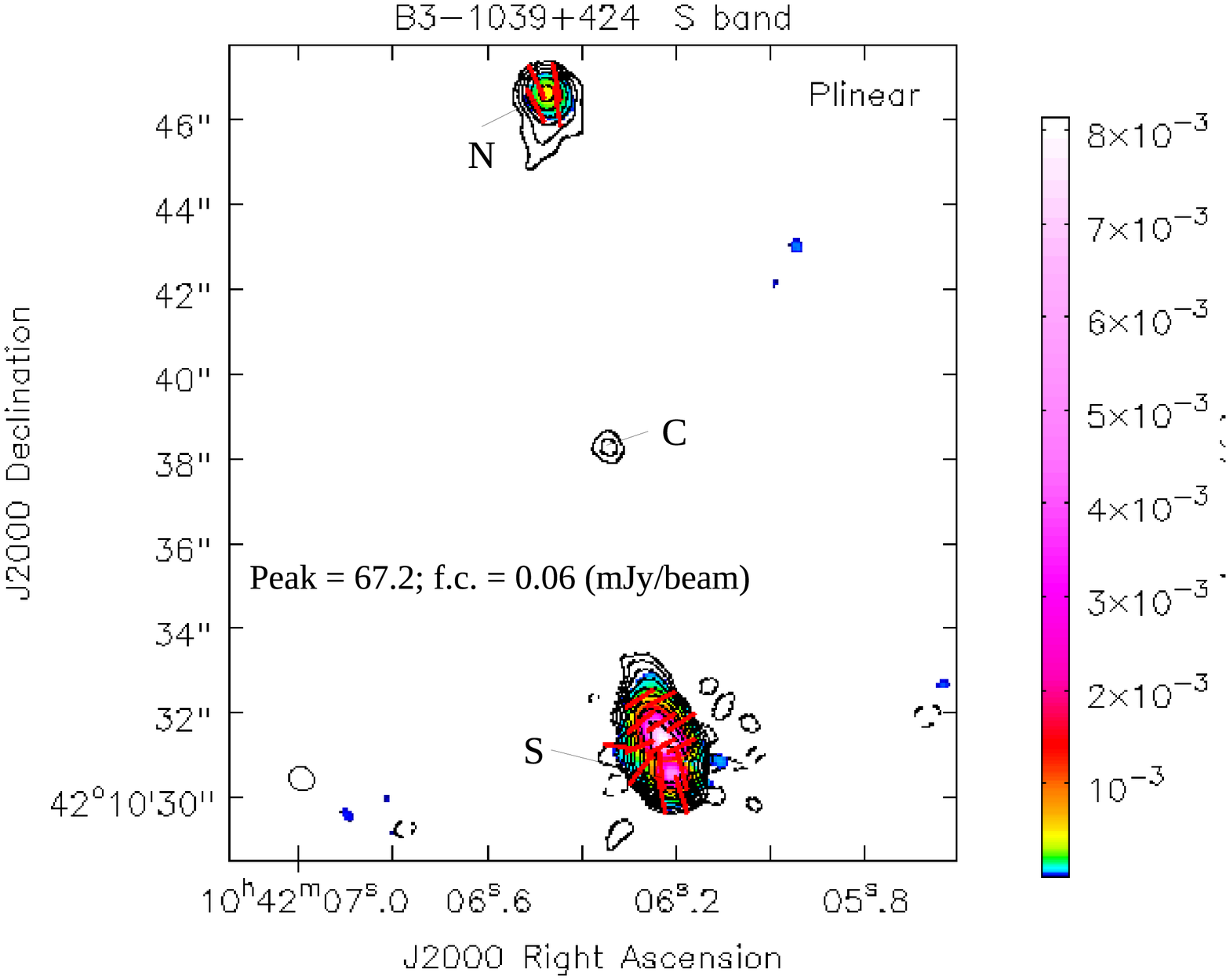}
\includegraphics{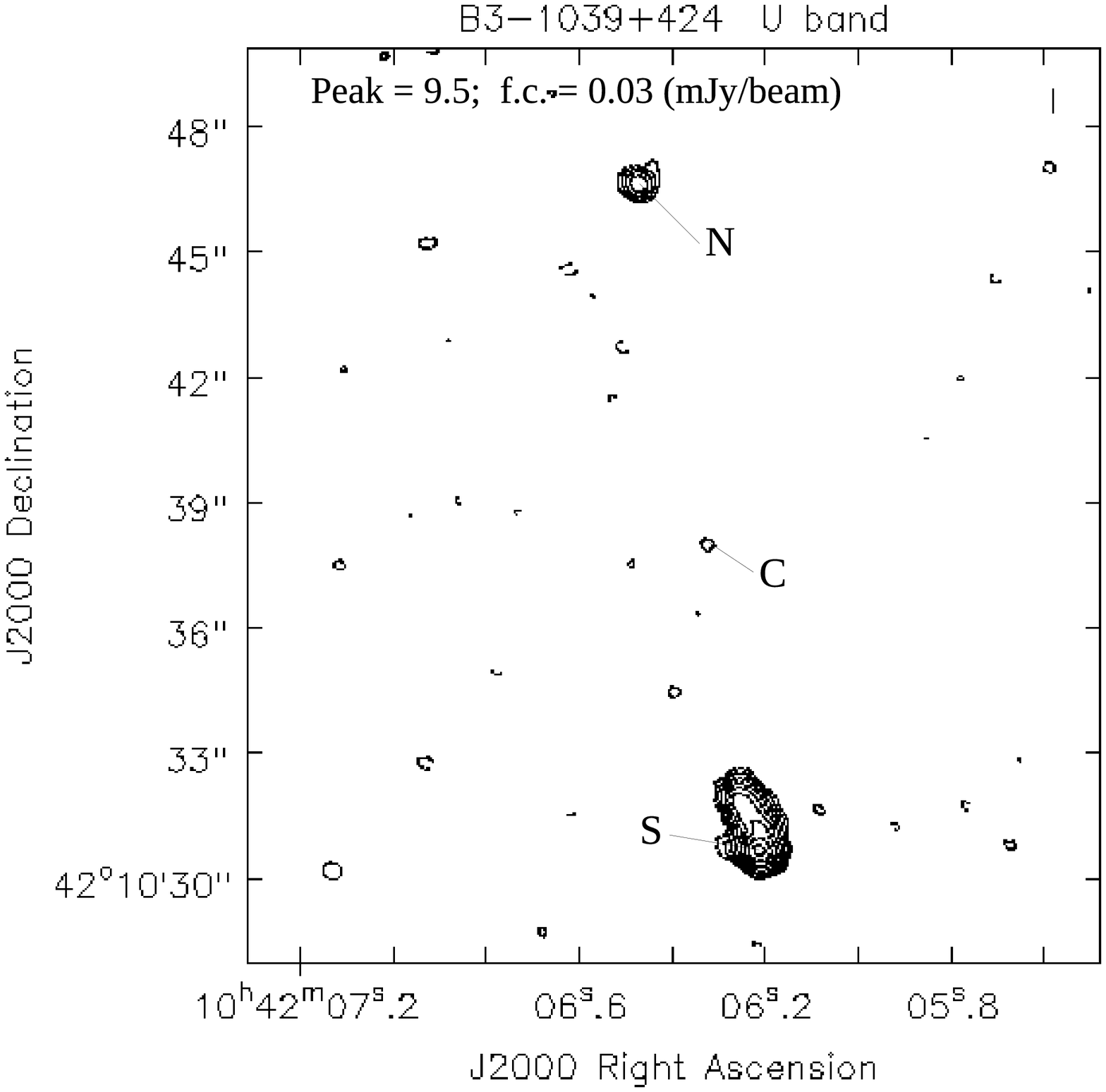}
\includegraphics{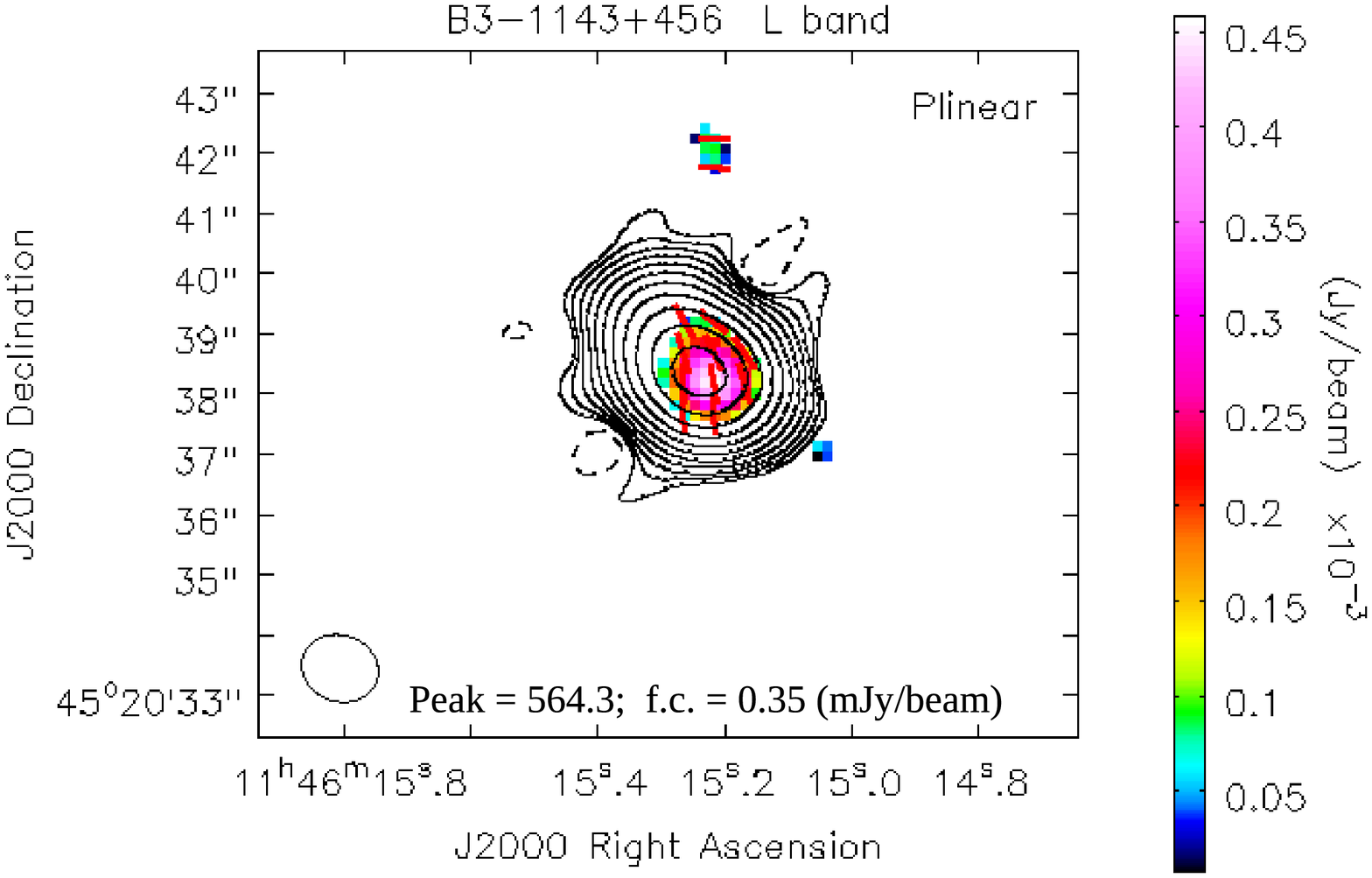}
\includegraphics{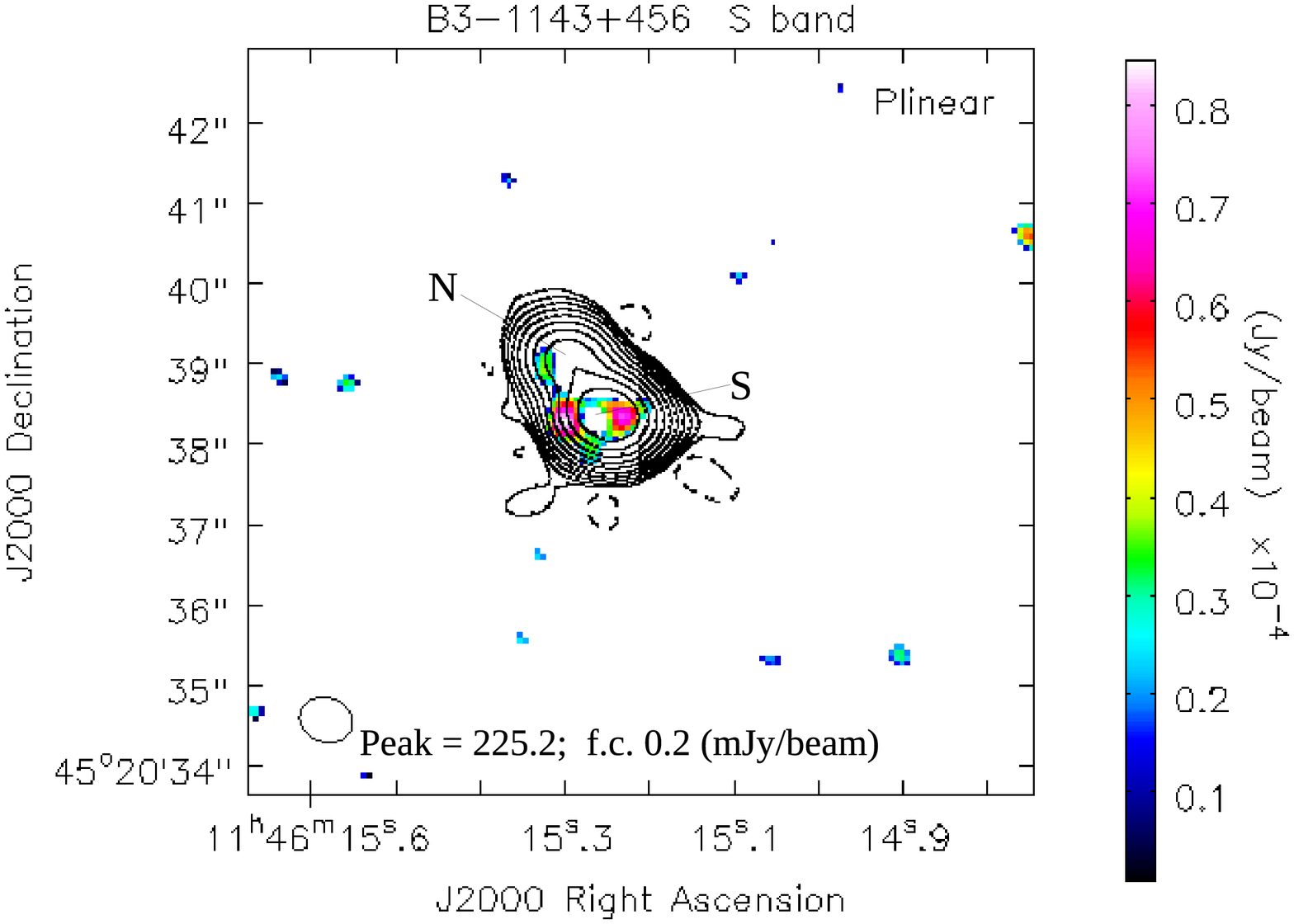}
\includegraphics{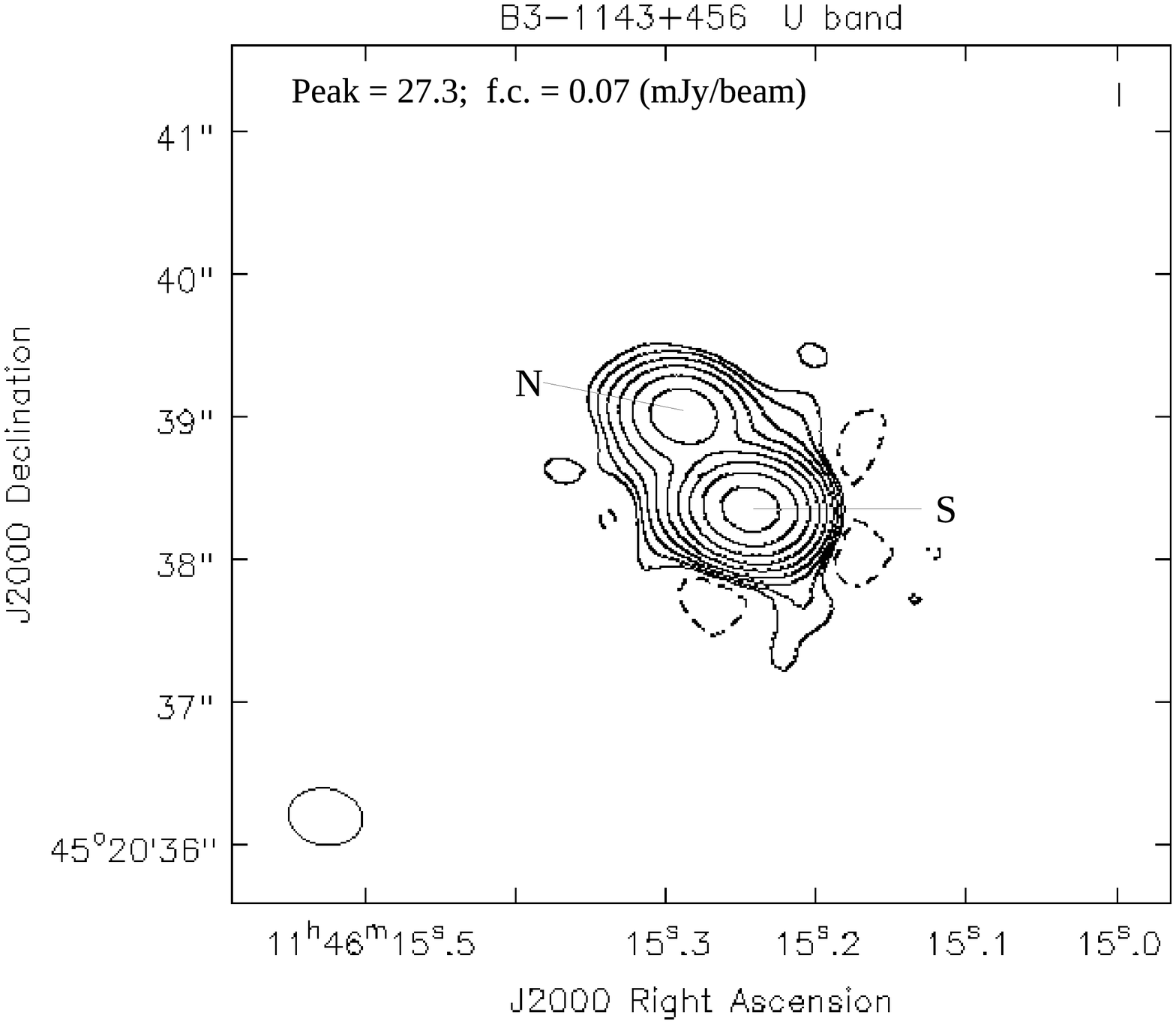}
\vspace{21cm}
\caption{Continued.}
\end{center}
\end{figure*}

\addtocounter{figure}{-1}
\begin{figure*}
\begin{center}
\includegraphics{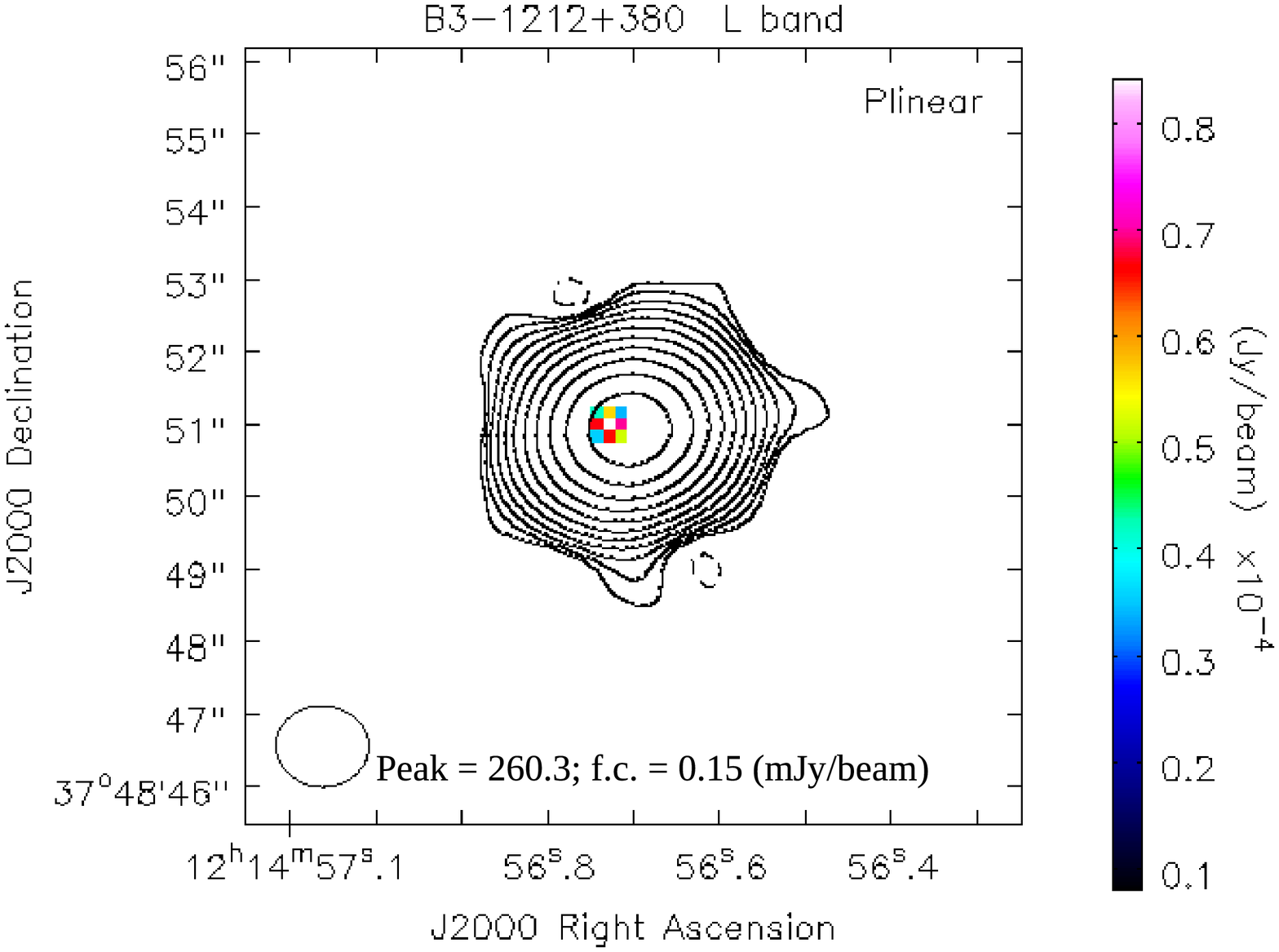}
\includegraphics{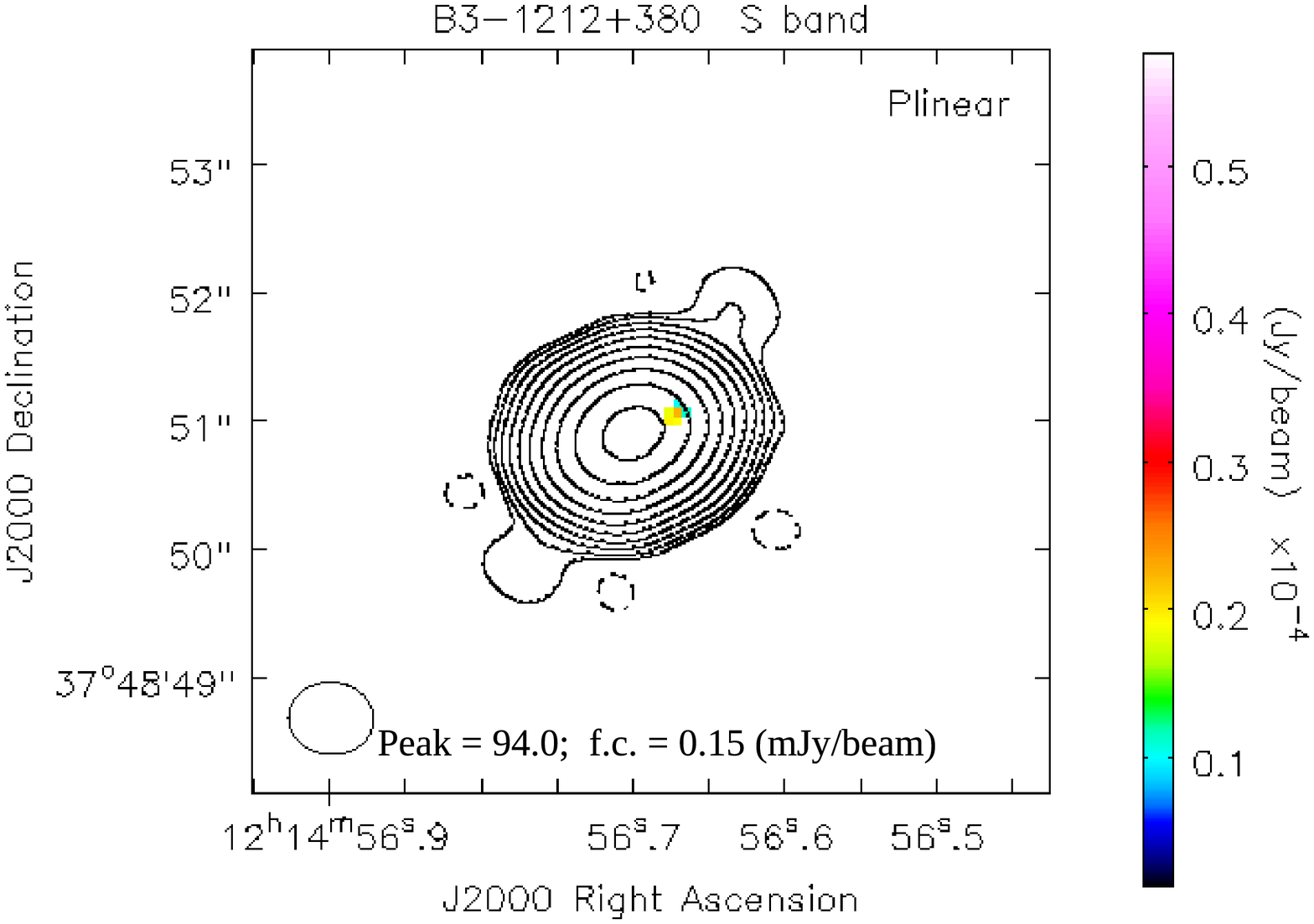}
\includegraphics{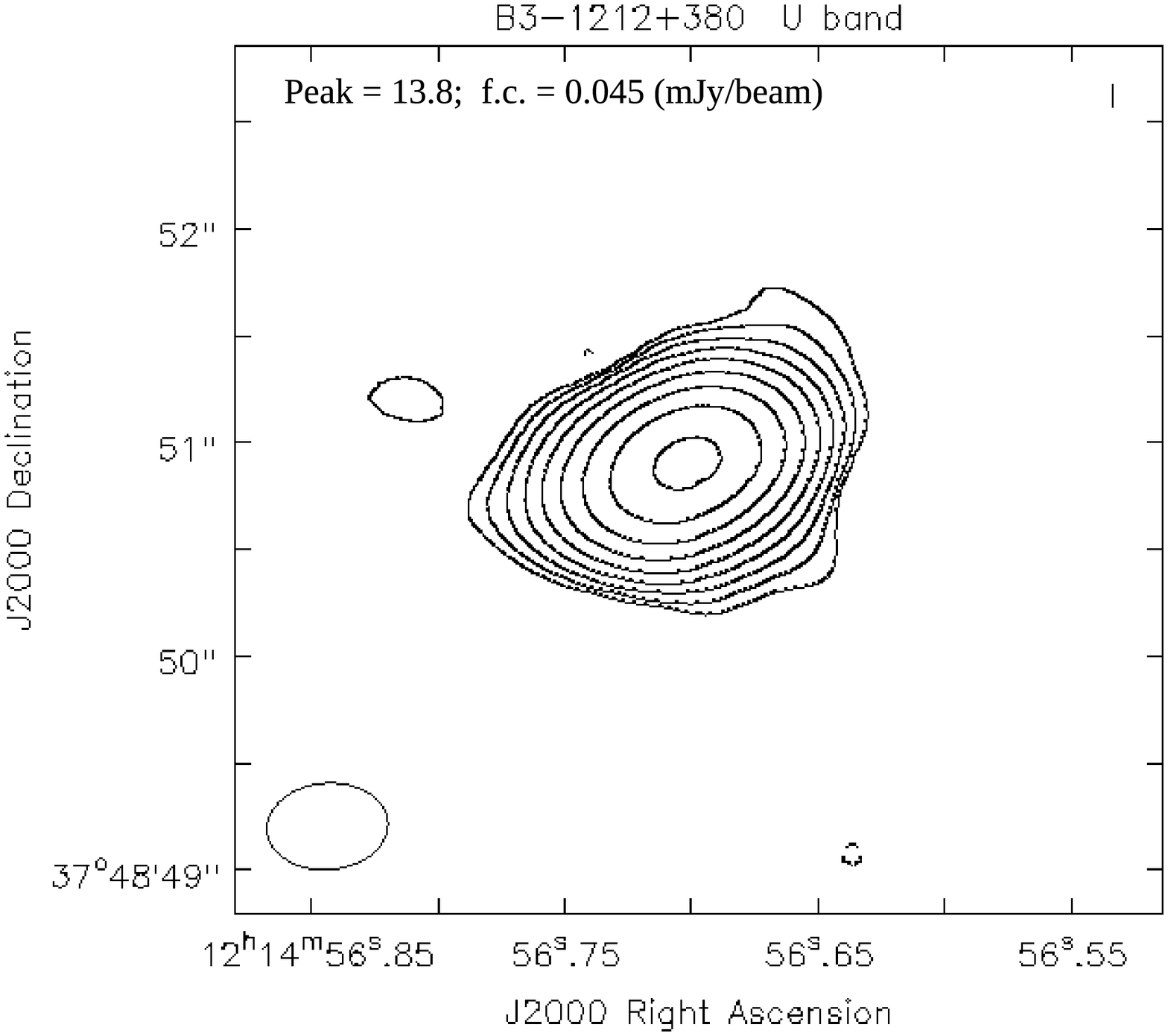}
\includegraphics{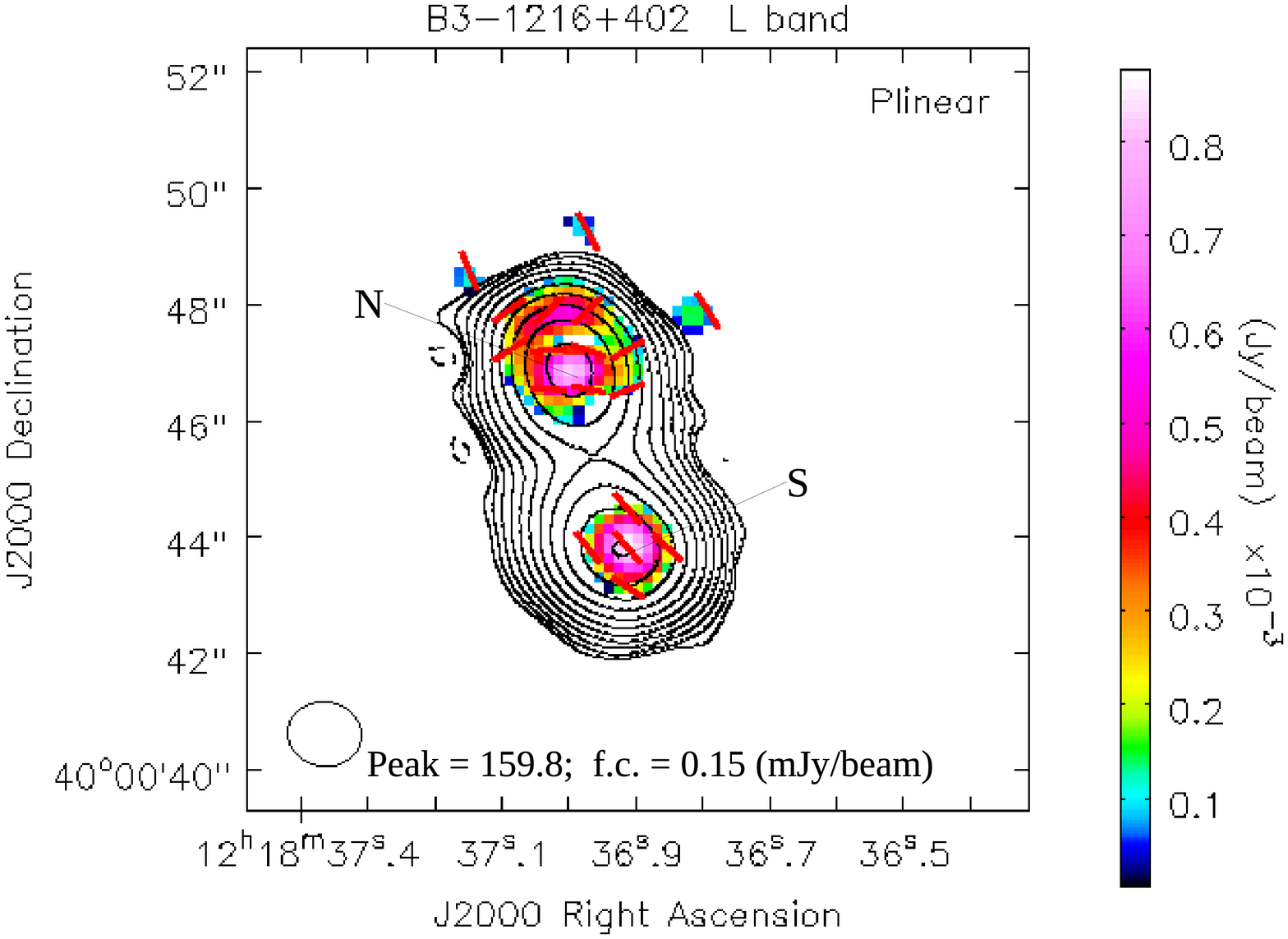}
\includegraphics{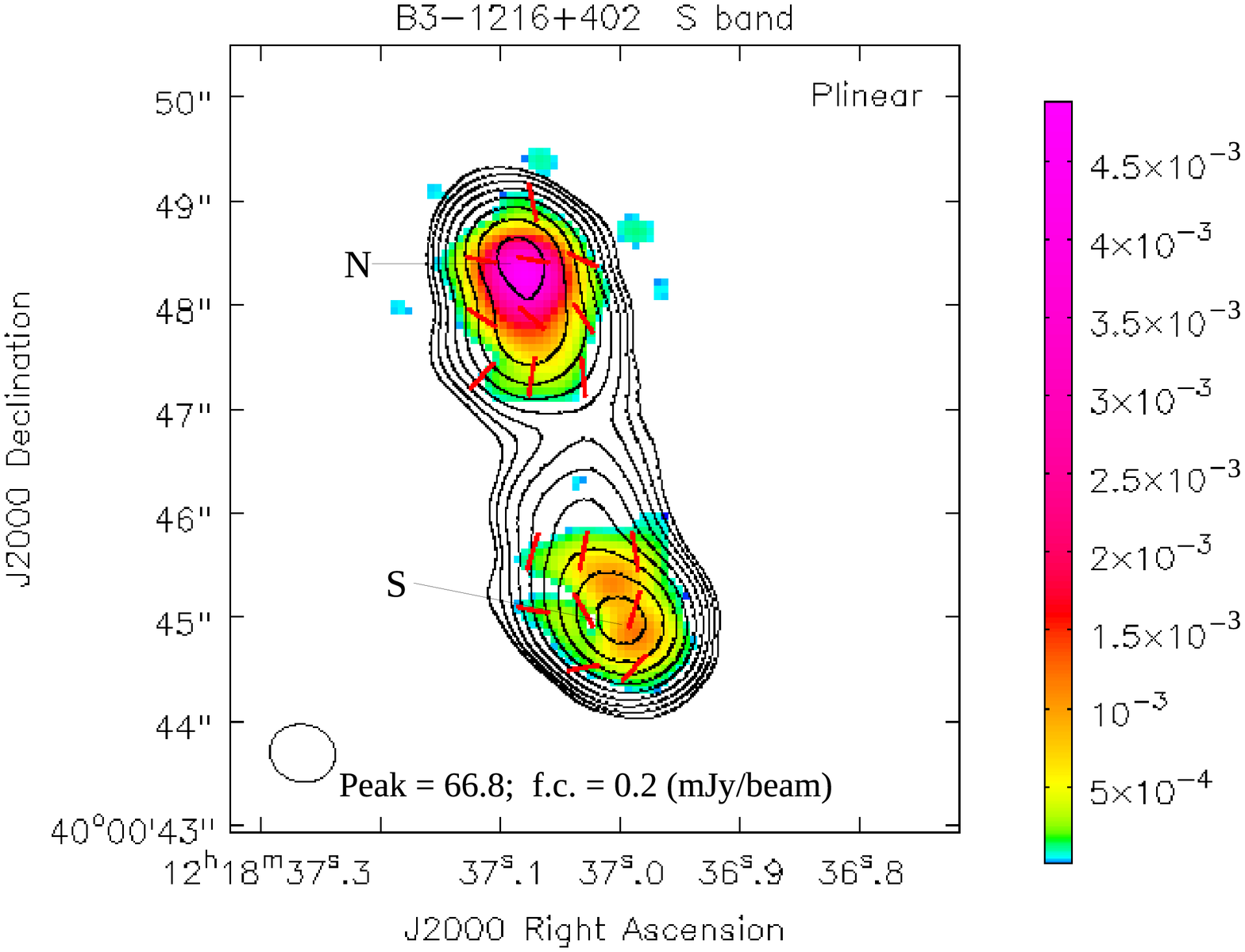}
\includegraphics{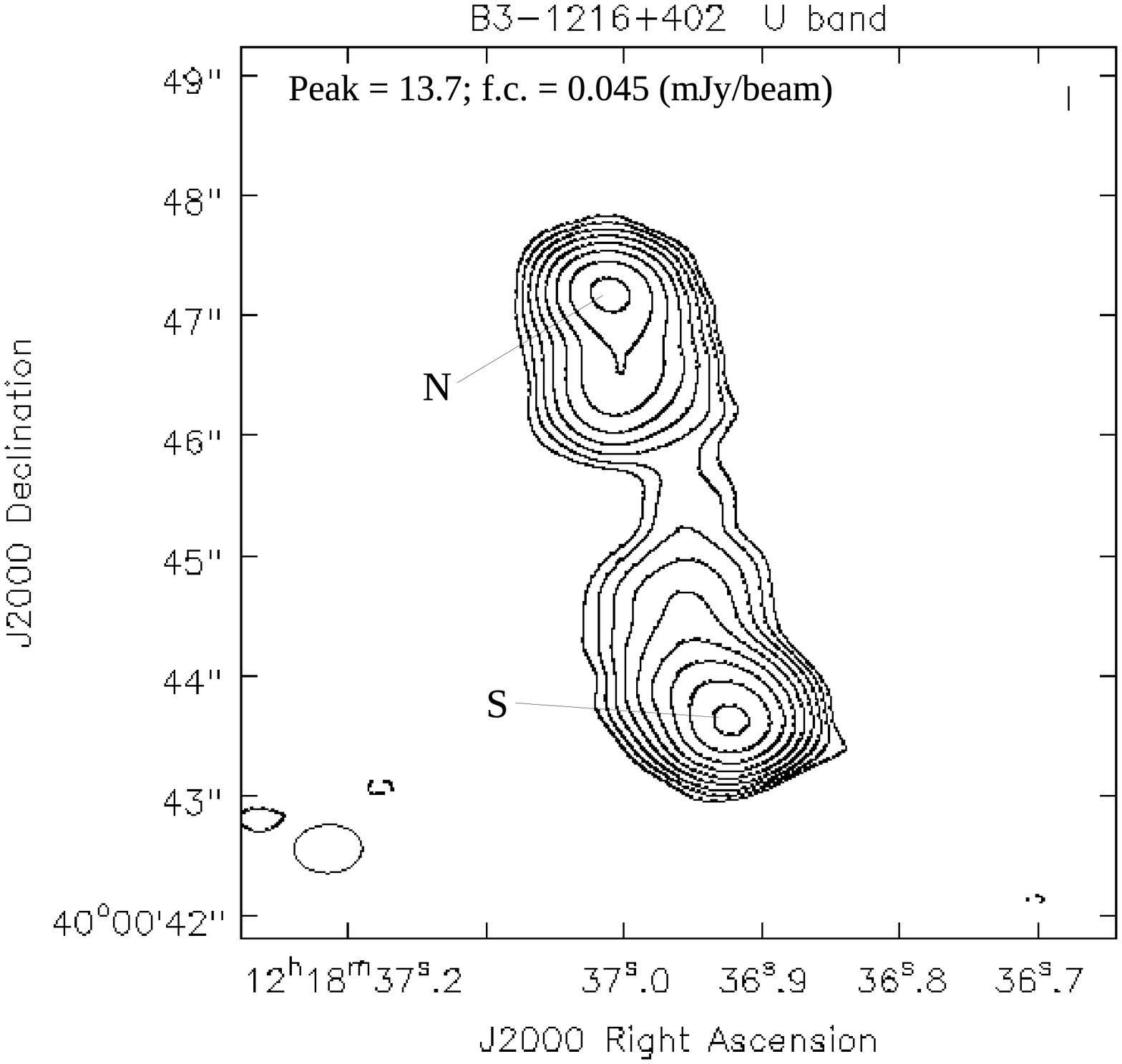}
\vspace{21cm}
\caption{Continued.}
\end{center}
\end{figure*}

\begin{figure*}
\begin{center}
\includegraphics{1016_C-AIPS.PS}
\includegraphics{1016_X-AIPS.PS}
\includegraphics{1049_X-AIPS.PS}
\includegraphics{1225_X-AIPS.PS}
\includegraphics{1340_C-AIPS.PS}
\includegraphics{1340_X-AIPS.PS}
\vspace{23cm}
\caption{VLBA images of the CSS sources with LAS $<$ 1 arcsec. On each
image, we provide the source name, the observing band, the peak
brightness (peak) and the first contour (f.c.), which is three times
the off-source noise level on the image plane. Contours increase by a
factor of 2. The beam is plotted in the bottom left-hand corner of
each image.}
\label{vlba-figure}
\end{center}
\end{figure*}

\addtocounter{figure}{-1}
\begin{figure*}
\begin{center}
\includegraphics{1449_C-AIPS.PS}
\includegraphics{1449_X-AIPS.PS}
\vspace{9cm}
\caption{Continued.}
\label{vlba-figure}
\end{center}
\end{figure*}

\begin{figure*}
\begin{center}
\includegraphics{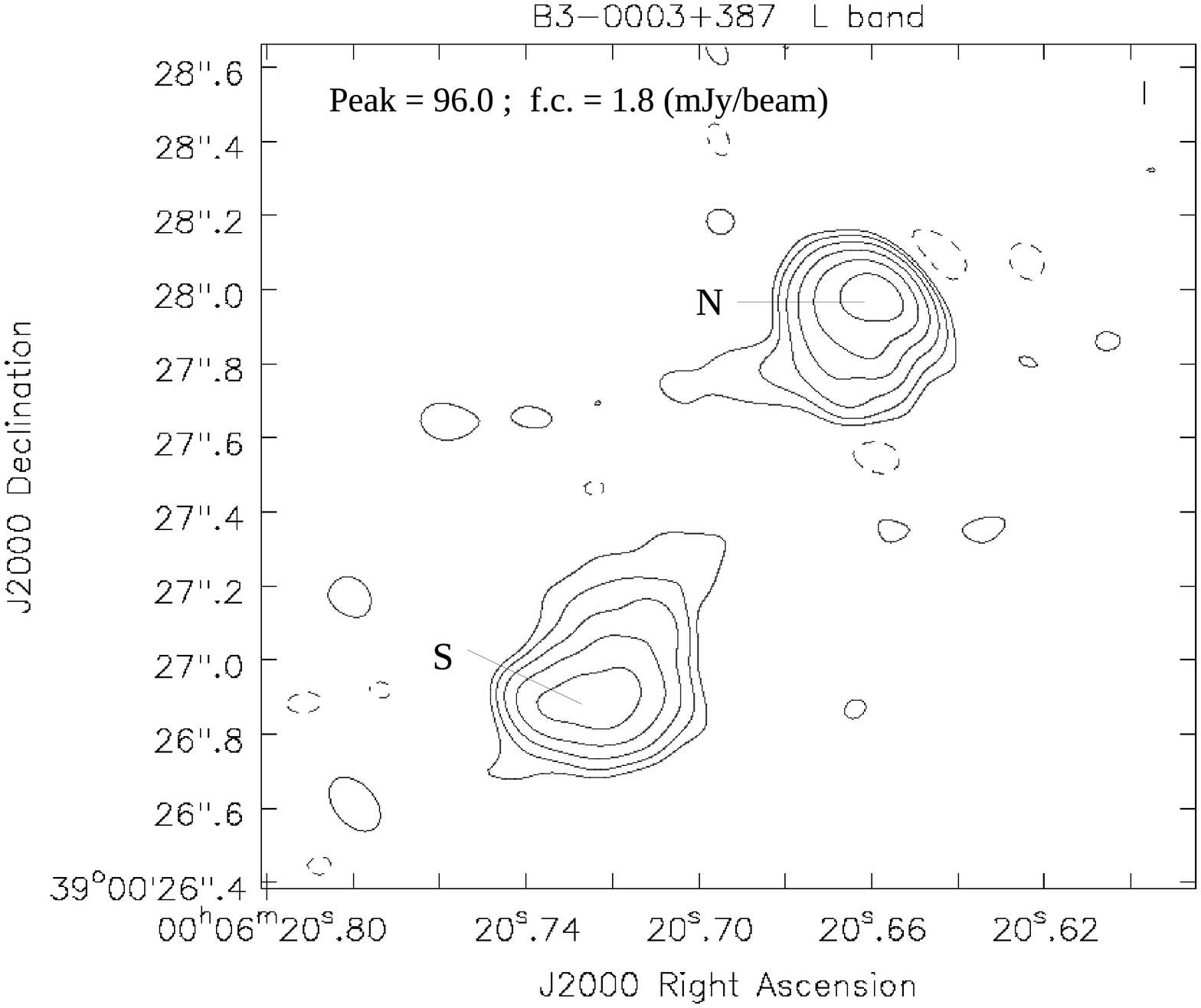}
\includegraphics{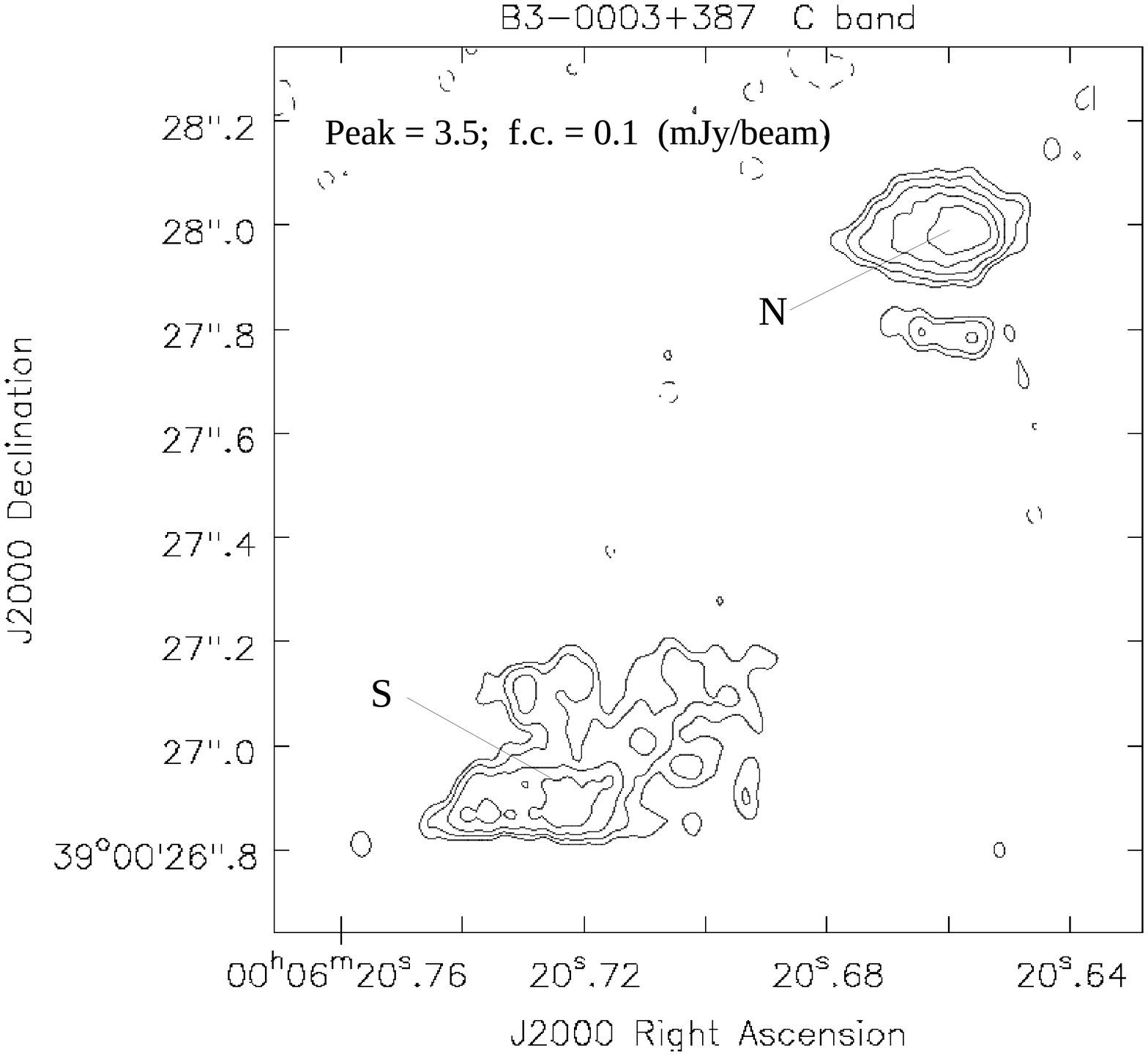}
\includegraphics{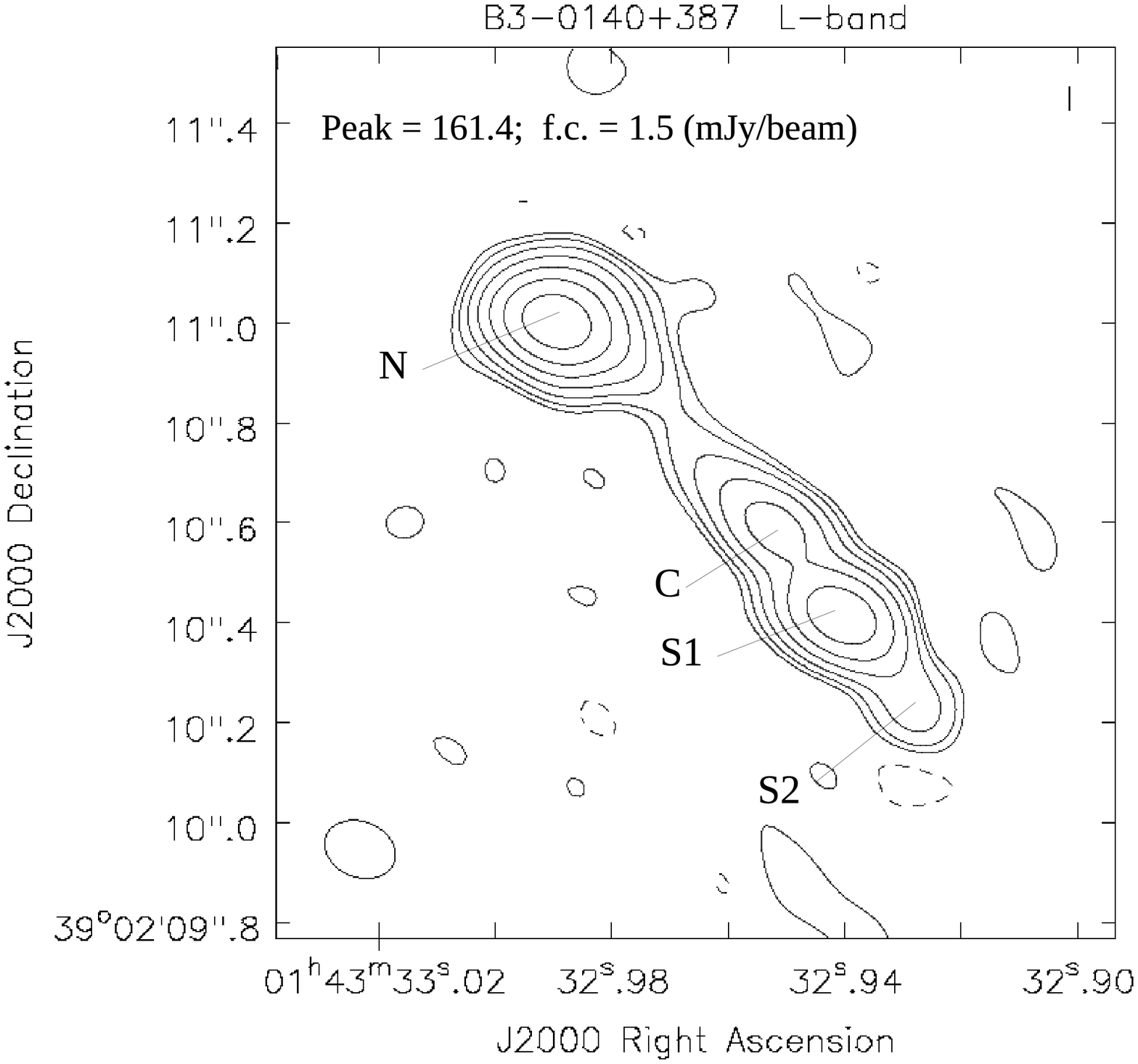}
\includegraphics{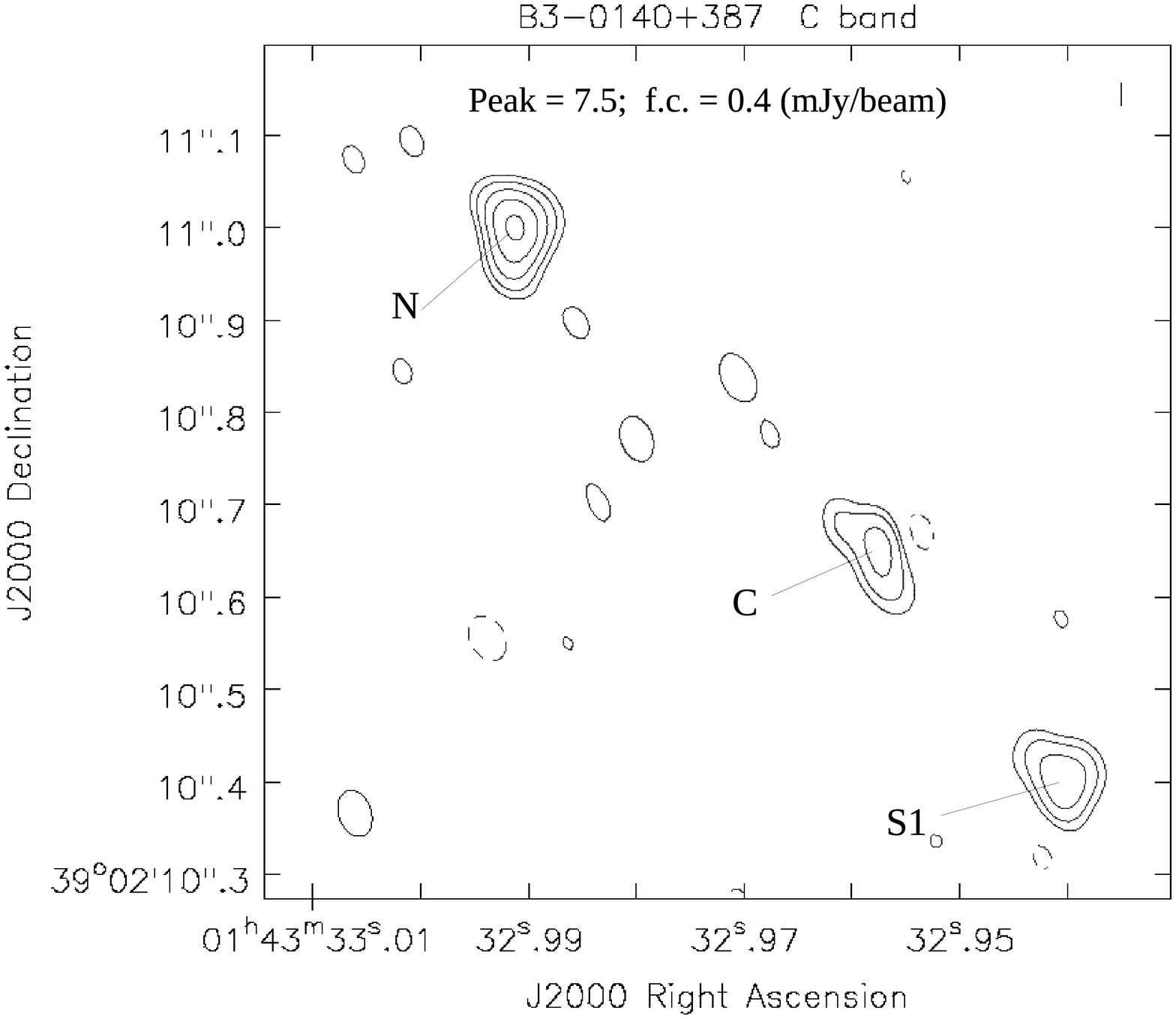}
\includegraphics{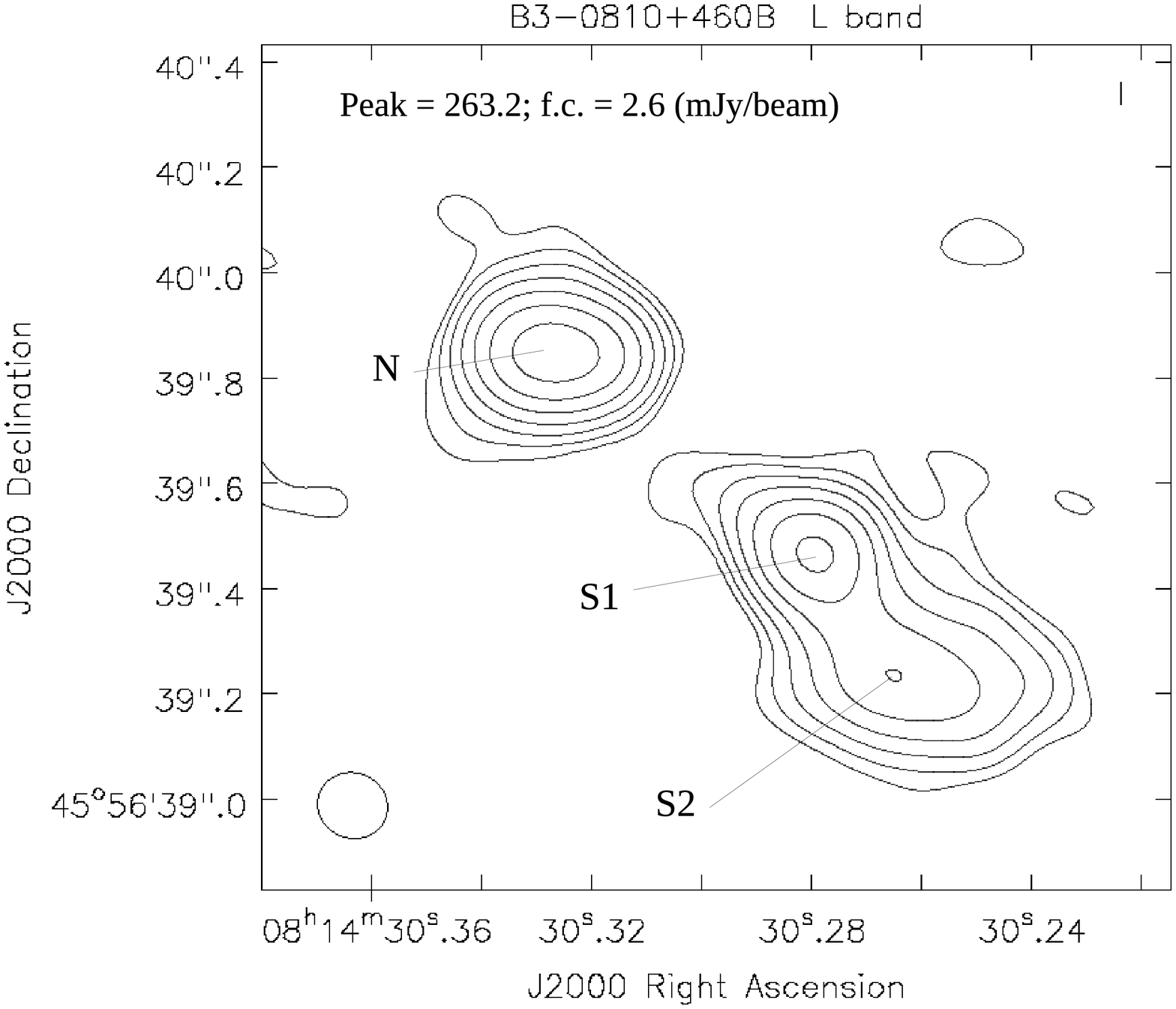}
\includegraphics{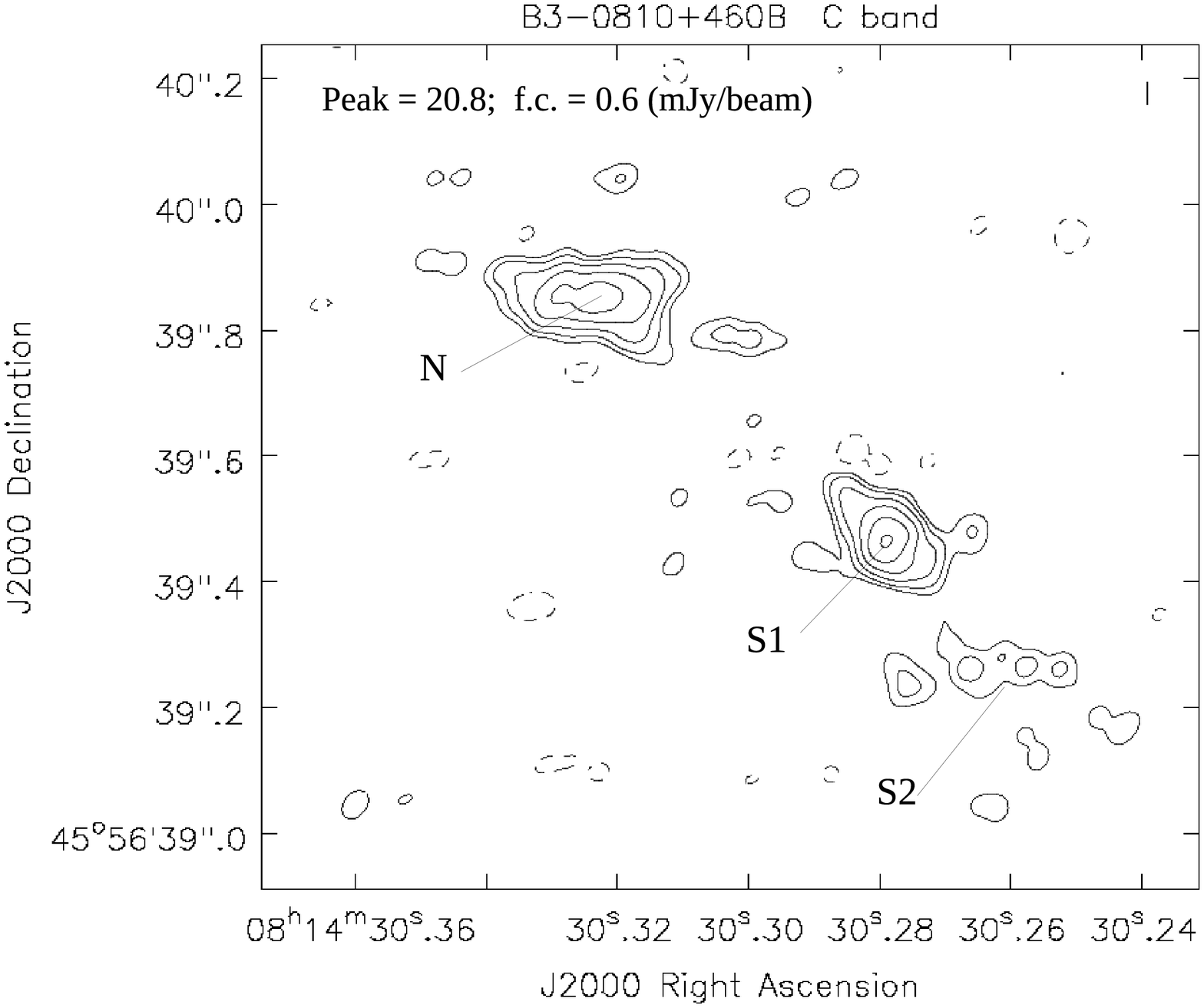}
\vspace{20cm}
\caption{eMERLIN images of the CSS sources. On each
image, we provide the source name, the observing band, the peak
brightness (peak) and the first contour (f.c.), which is three times
the off-source noise level on the image plane. Contours increase by a
factor of 2. The beam is plotted in the bottom left-hand corner of
each image.}
\label{merlin-figure}
\end{center}
\end{figure*}

\addtocounter{figure}{-1}
\begin{figure*}
\begin{center}
\includegraphics{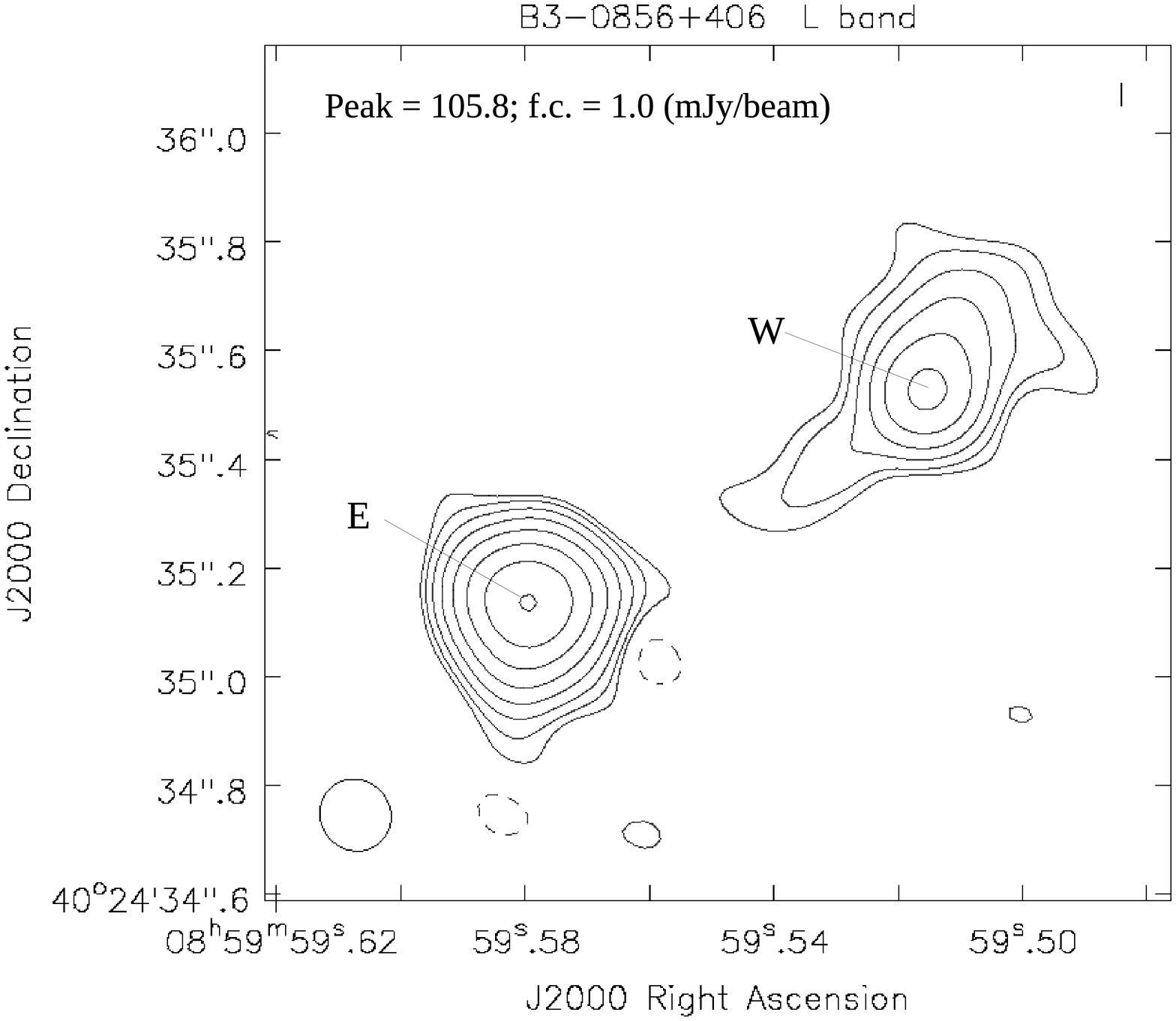}
\includegraphics{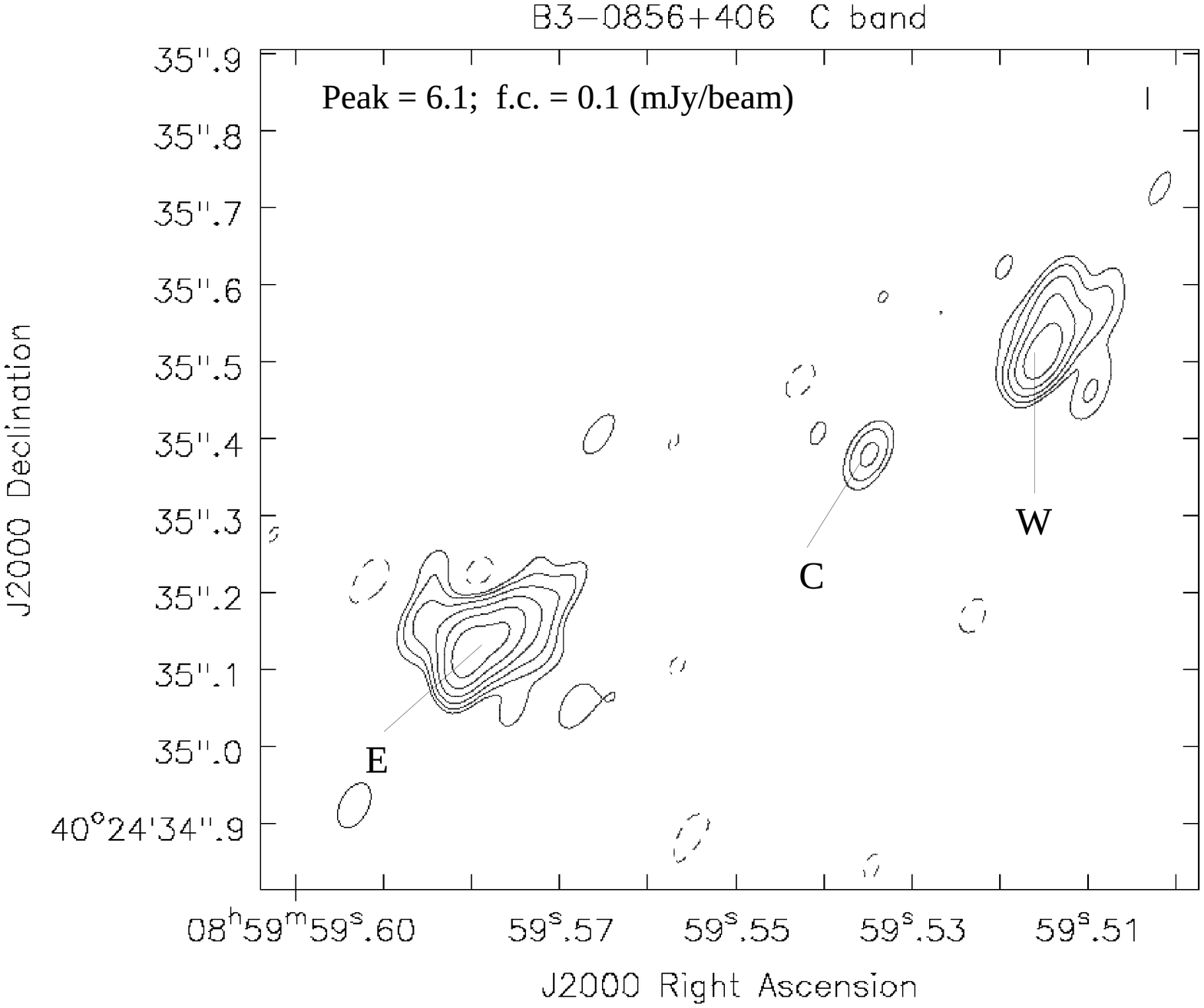}
\includegraphics{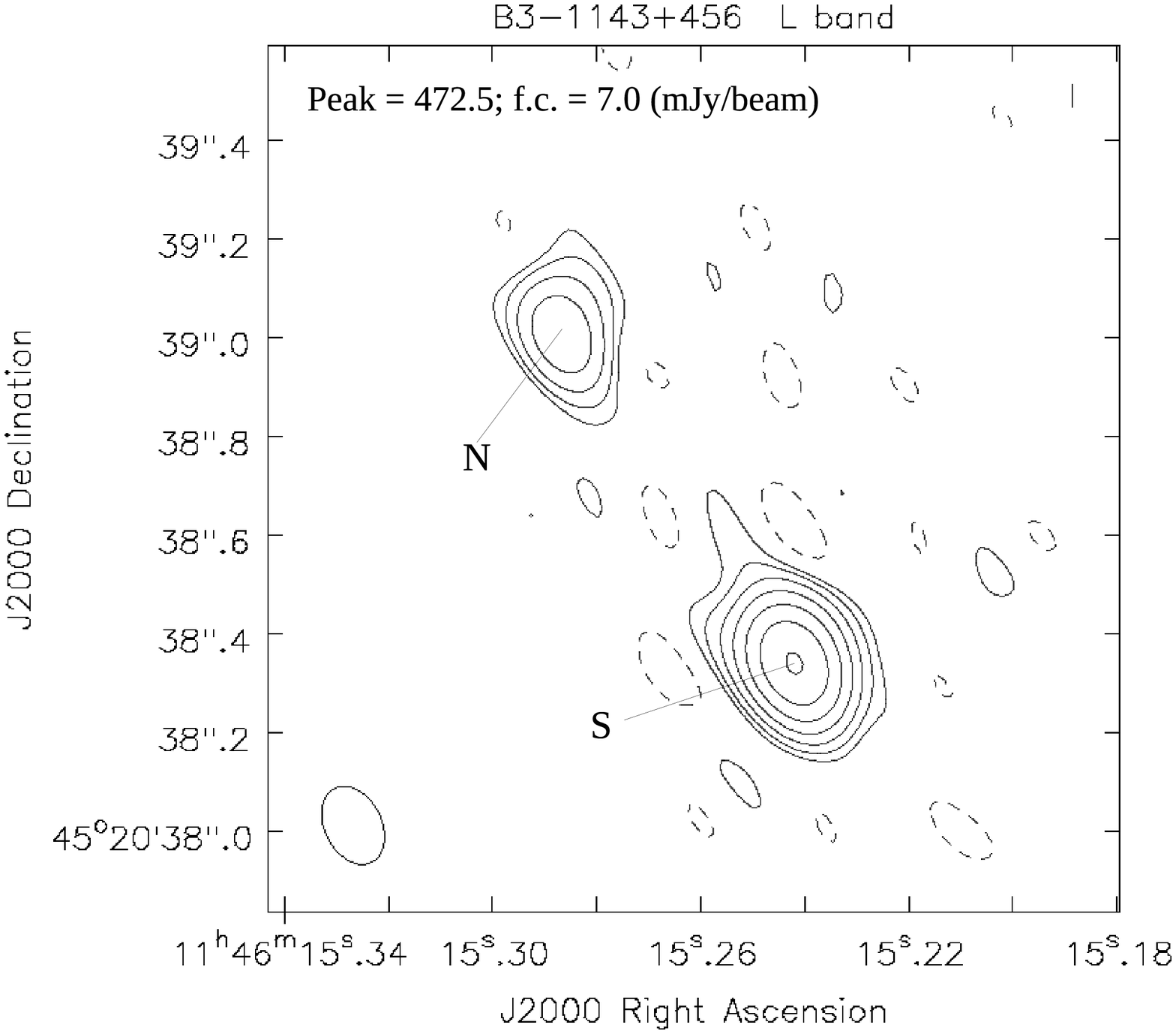}
\includegraphics{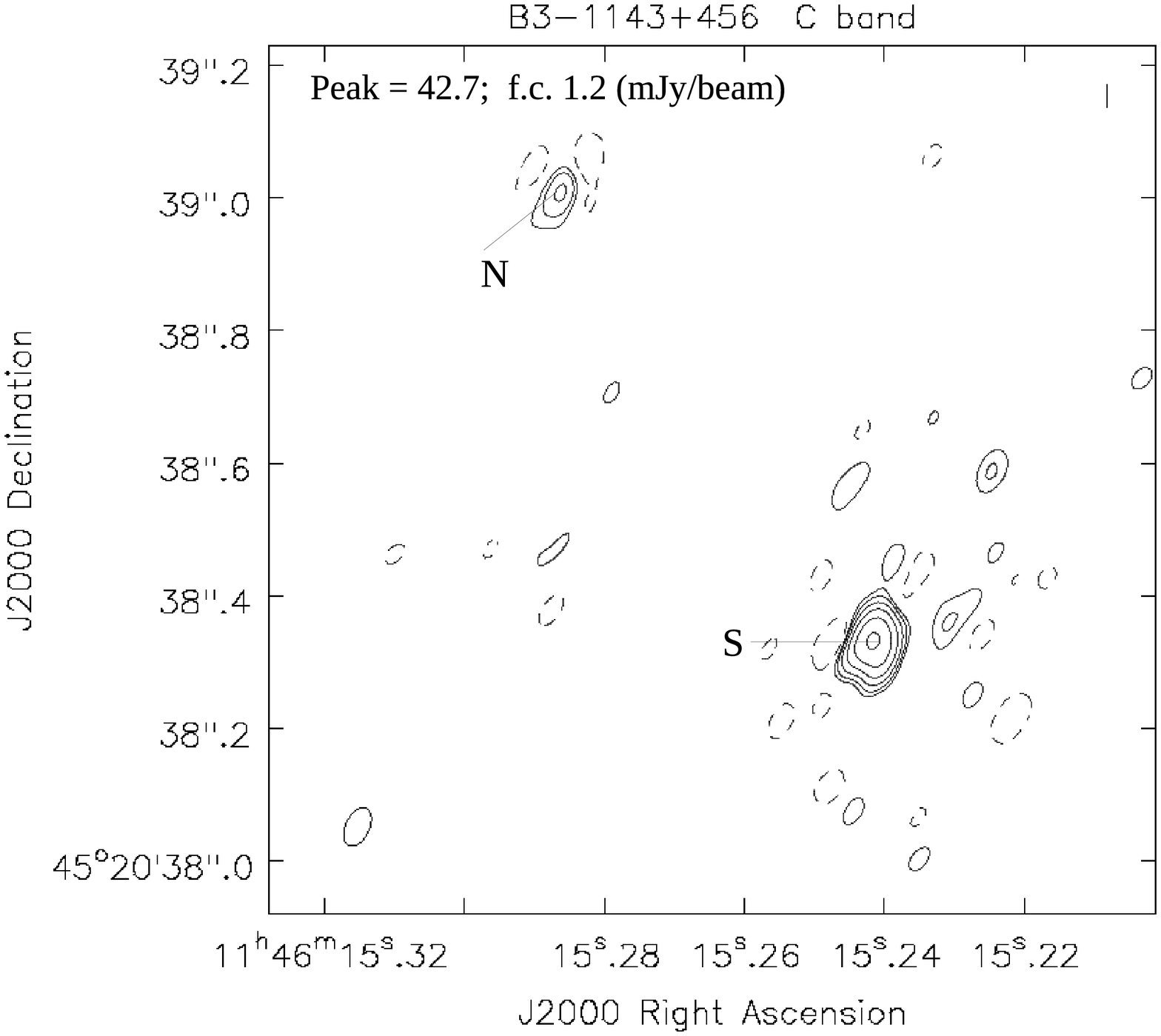}
\includegraphics{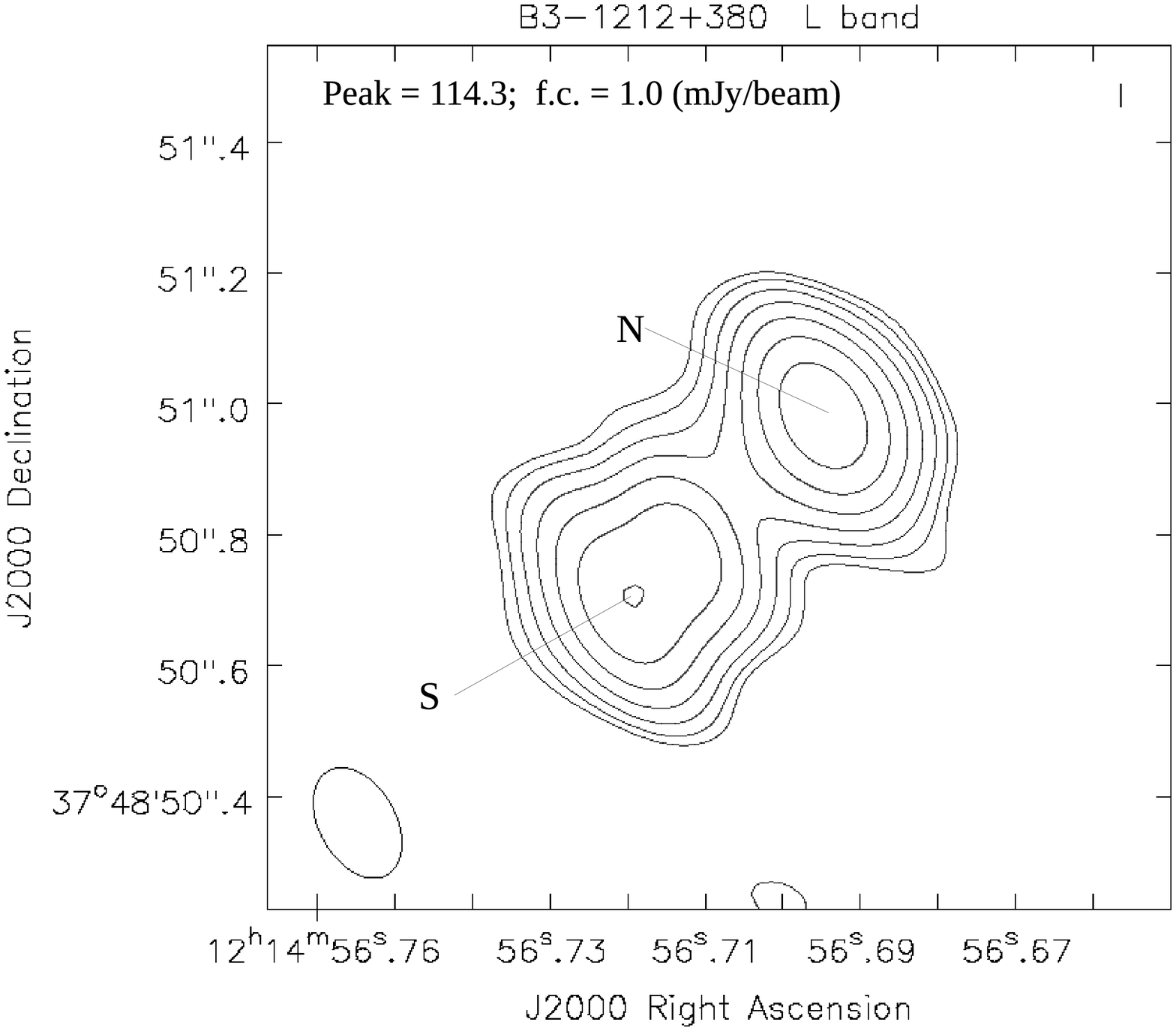}
\includegraphics{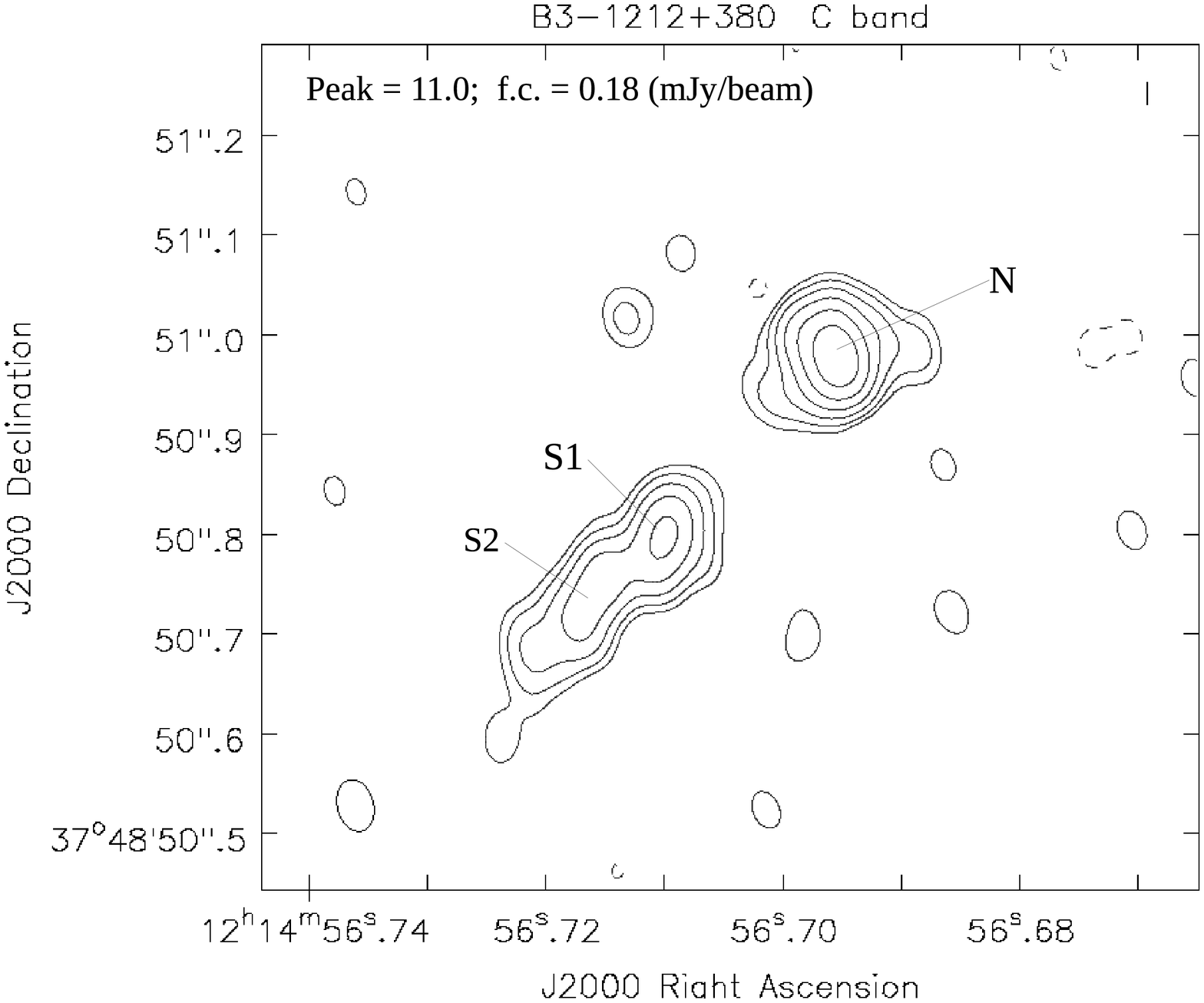}
\vspace{20cm}
\caption{Continued.}
\end{center}
\end{figure*}

\section{Field sources}
\label{appendix_out}

\begin{figure*}
\begin{center} 
\includegraphics{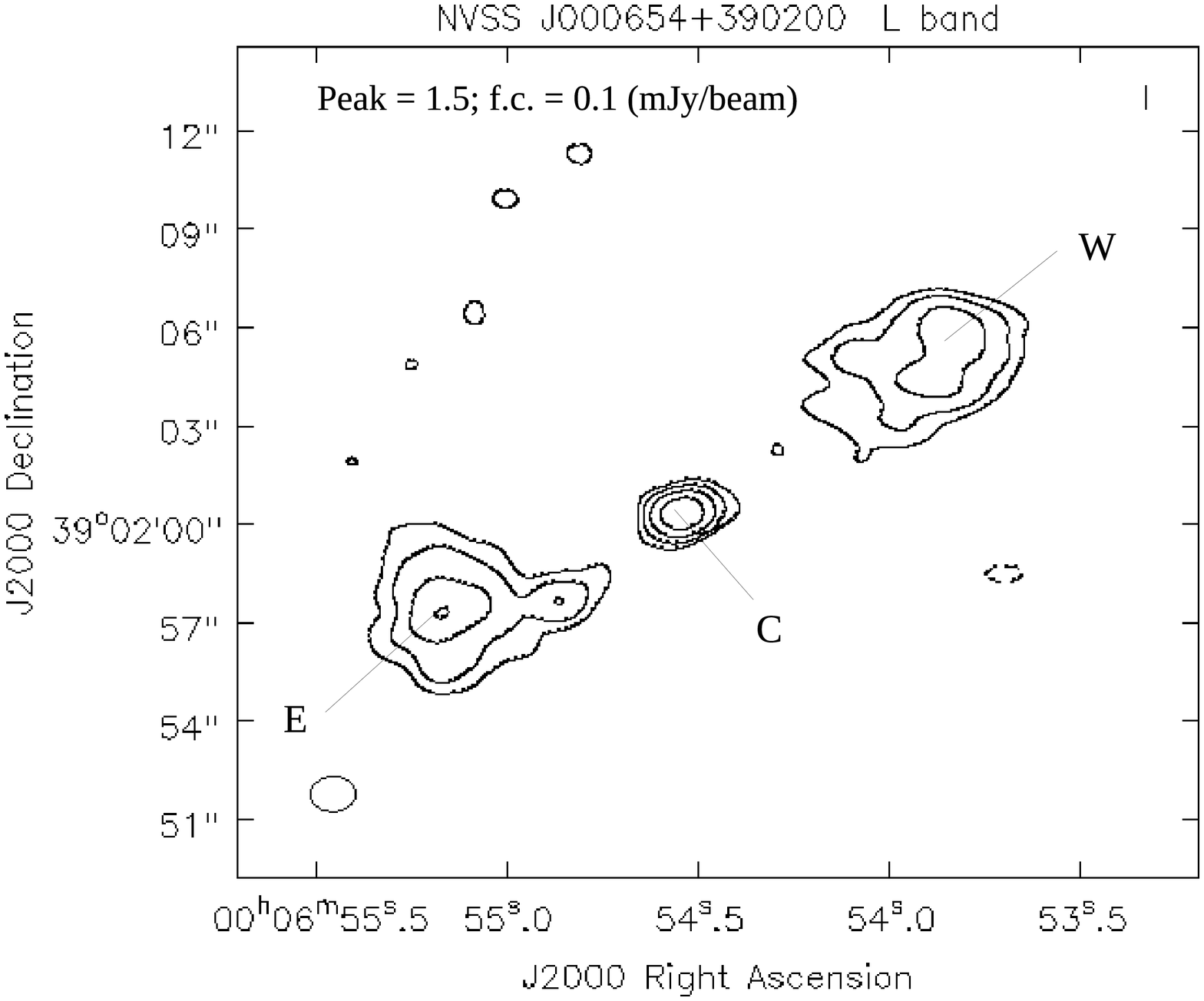}
\includegraphics{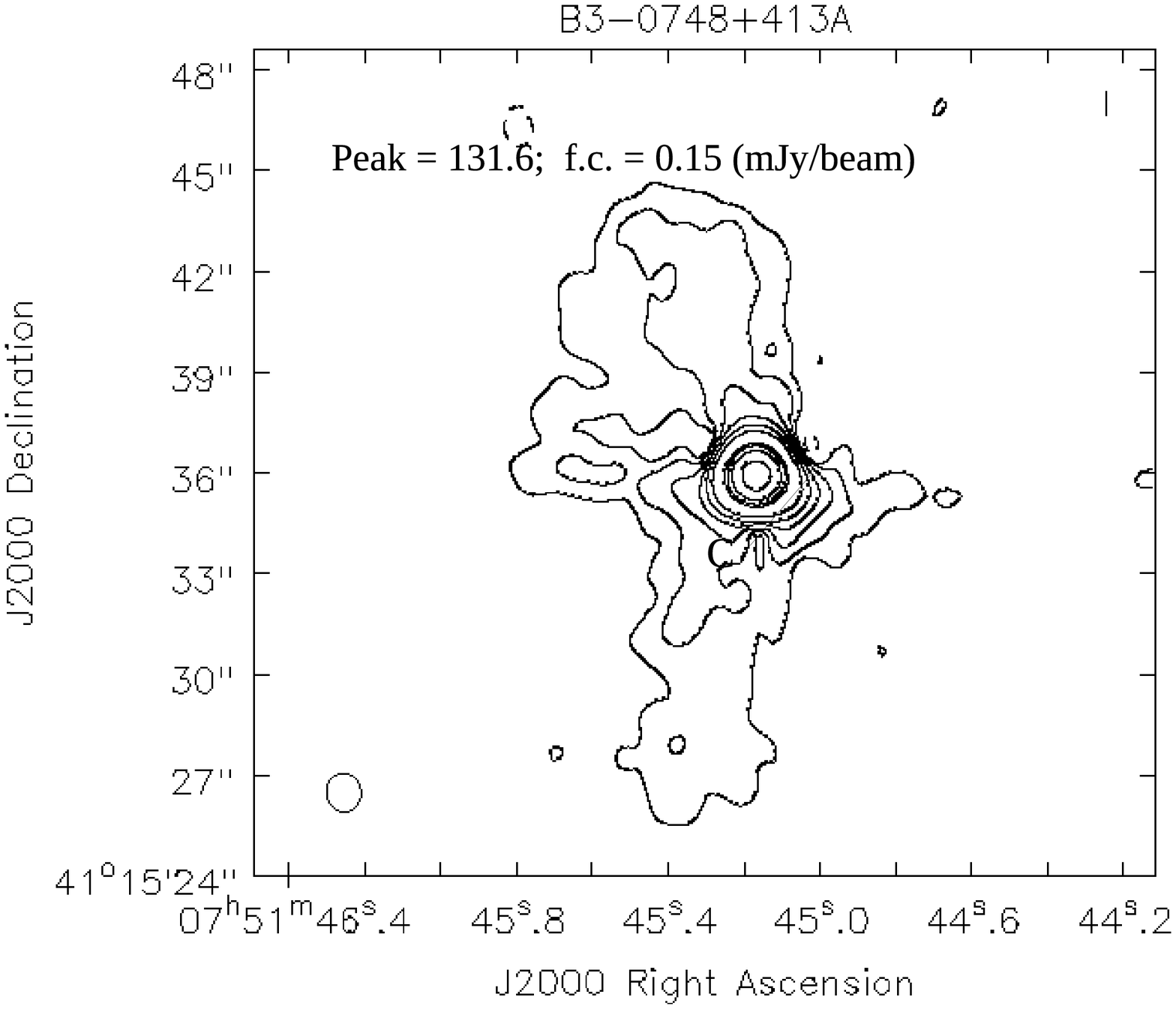}
\includegraphics{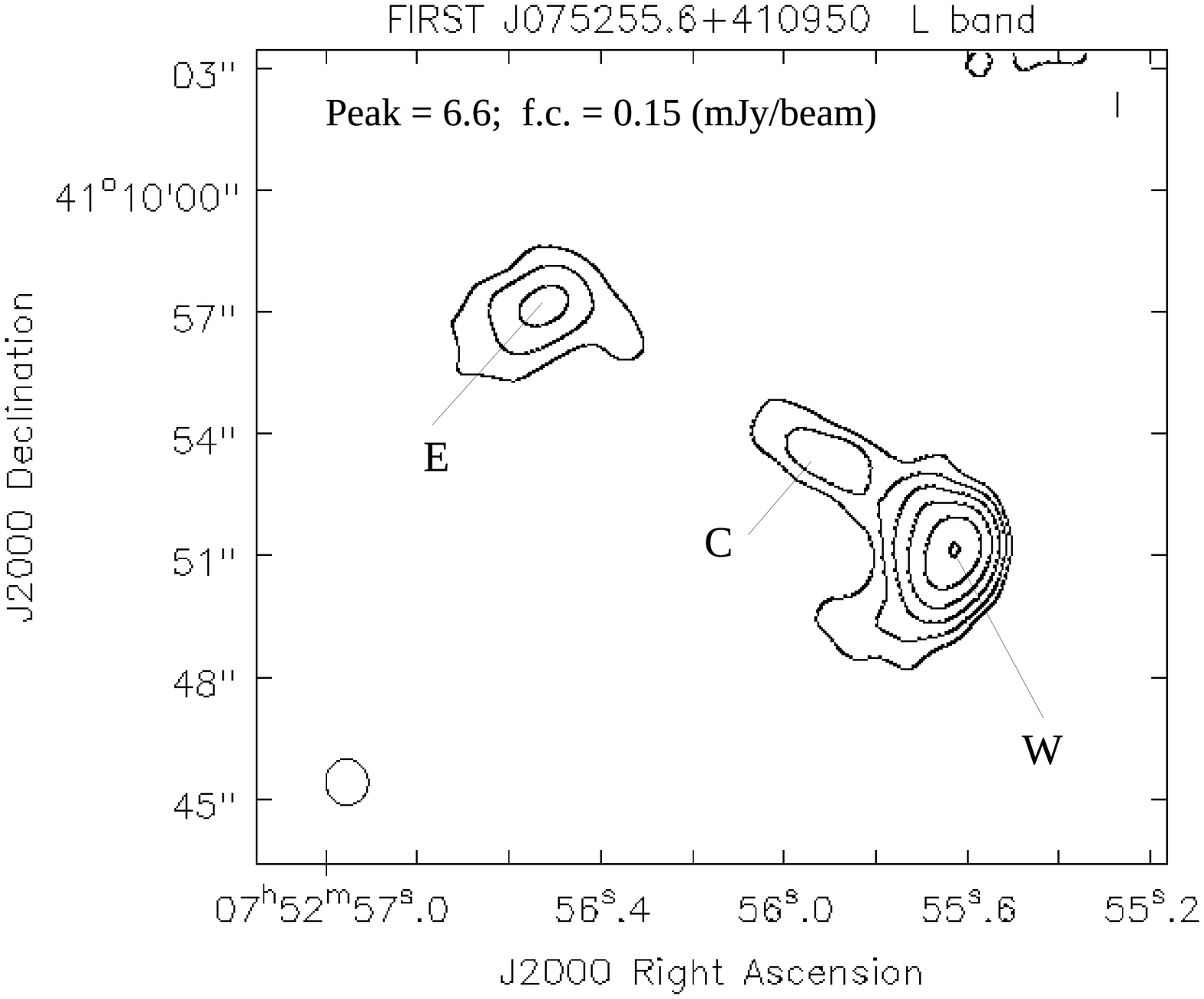}
\includegraphics{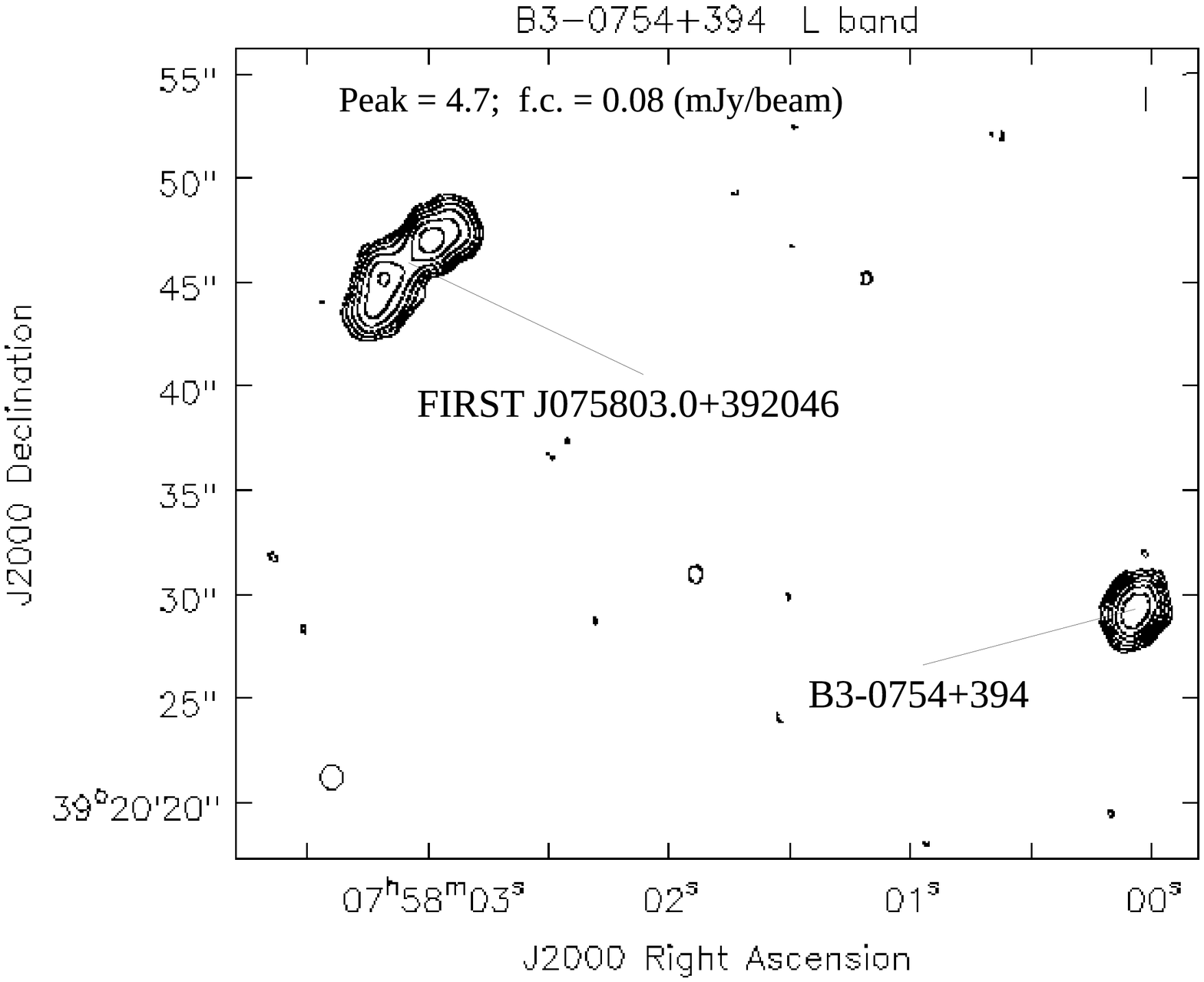}
\includegraphics{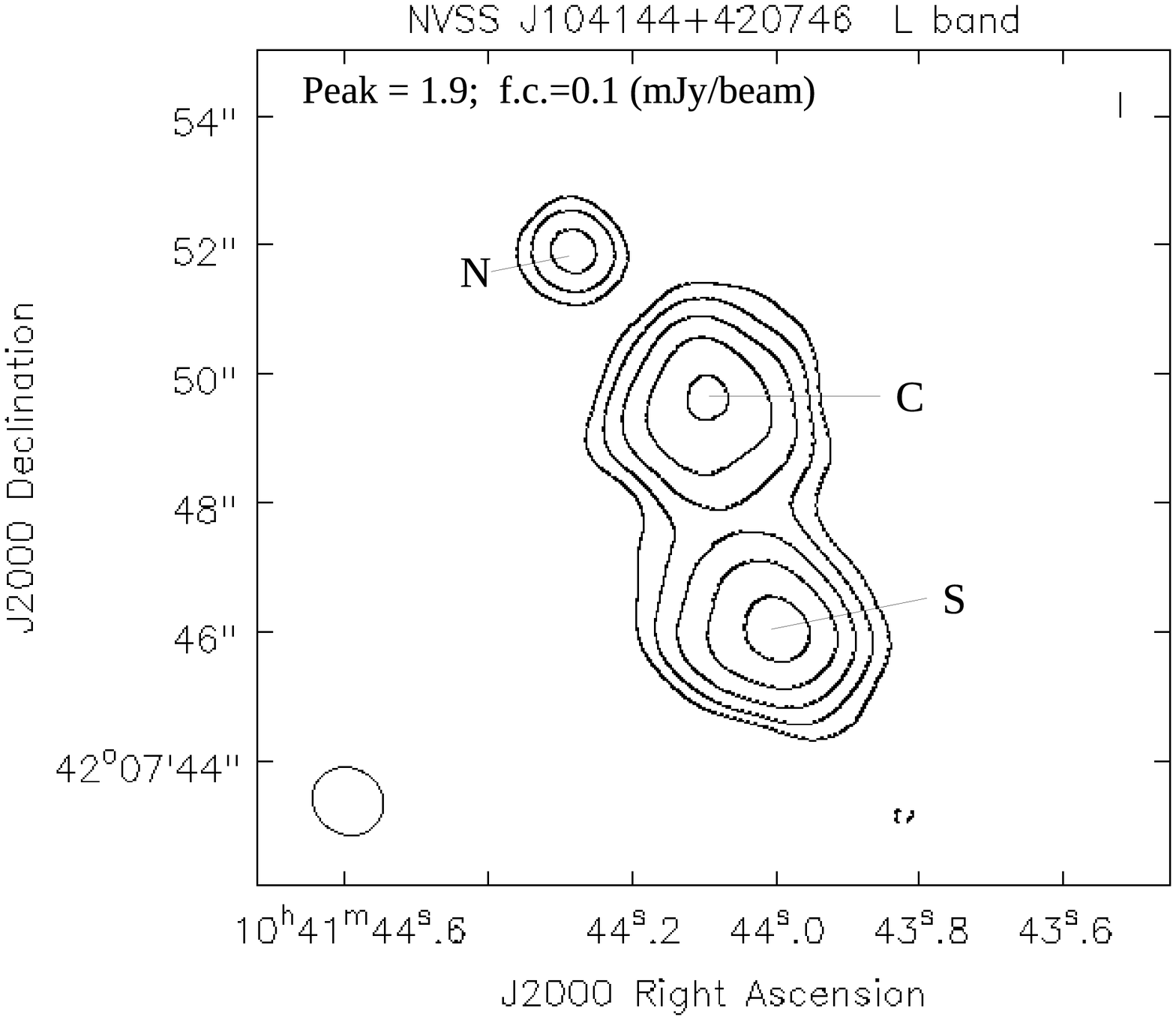}
\vspace{19cm}
\caption{VLA images in L band of the objects with a resolved radio
  structure that fall within the field of view of the observations. On each
image, we provide the source name, the observing band, the peak
brightness (peak) and the first contour (f.c.), which is three times
the off-source noise level on the image plane. Contours increase by a
factor of 2. The beam is plotted in the bottom left-hand corner of
each image.}
\label{fig-out}
\end{center}
\end{figure*}

In L band we had to image several objects that fall within the
primary beam of the array. For no high-angular
resolution observations are available in the literature, here we
provide information on the radio structure and flux density at 1.5 GHz
based on images with a resolution of about 1 - 1.5 arcsec (Table
\ref{table-out}). Flux density is measured on images corrected for the
response of the primary beam.
Images are presented in Fig. \ref{fig-out}. \\

\begin{table*}
\caption{Flux density and morphology of the objects in the field of
  the target sources. Column 1: source name; columns 2 and 3: right
  ascension (RA) and declination (Dec); column 4: source component; column 5: flux density
at 1.5 GHz; column 6: radio morphology. Cx: complex structure; CJ: core-jet, FRI: Fanaroff-Riley type-I
radio source, FRII: Fanaroff-Riley type-II radio source, SR: slightly
resolved, T: triple, U: unresolved.}
\begin{center}
\begin{tabular}{cccccc}
\hline
Source name & RA (J2000) & Dec (J2000) & Component & Flux density & Morph.\\
           &              &           &            &  mJy & \\
\hline
NVSS\,J000654+390200& 00:06:54.54 & 39:02:00.34 & Tot &  9.7$\pm$0.4& FRII \\ 
                    &             &             & W   &  4.2$\pm$0.2&
\\
                    &             &             & C   &  1.5$\pm$0.1 &
\\
                    &             &             & E   &  4.0$\pm$0.2 & \\
B3-0004+387         & 00:07:30.34 & 39:03:47.0  & Tot &  42.5$\pm$1.3
& U\\
NVSS\,J014352+385702& 01:43:52.46 & 38:57:02.38 & Tot & 29.4$\pm$0.9&
U\\
B3-0748+413A        & 07:51:45.16 & 41:15:35.93 & Tot & 180.3$\pm$5.5
& Cx \\
FIRST\,J075255.6+410950 & 07:52:55:.63 & 41:09:51.07 & Tot &
20.7$\pm$0.7 & CJ? \\
                        &    &    &  W  & 15.3$\pm$0.2 & \\
                        &    &    &  C  &  1.7$\pm$0.2 & \\
                        &    &    &  E  &  3.3$\pm$0.2 & \\
B3-0754+394 & 07:58:00.06  & 39:20:29.20 & Tot & 11.3$\pm$0.4 & U\\
FIRST\,J075803.0+392046 & 07:58:02.97 & 39:20:47.1 &Tot & 20.6$\pm$0.7&
FRI?\\
NVSS\,J085935+401826 & 08:59:35.96 & 40:18:26.13 & & 22.6$\pm$0.7 & U \\
NVSS\,J104144+420746 & 10:41:44.00 & 42:07:46.08 &Tot& 9.9$\pm$0.3& T
\\
                     &             &             & N & 0.5$\pm$0.1& \\
                     &             &             & C & 4.9$\pm$0.2& \\
                     &             &             & S & 4.5$\pm$0.2& \\
NVSS\,J104134+420916 & 10:41:34.78 & 42:09:16.43 &Tot&
8.1$\pm$0.2&SR\\
WISEA\,J114540.82+452230.7& 11:45:40.81 & 45:22:31.05&Tot&5.9$\pm$0.2&U\\
\hline
\end{tabular}
\end{center}
\label{table-out}
\end{table*}

\section{Spectral fits}
\label{synage-appendix}

In Fig. \ref{synage-fig} we present the integrated spectrum of the
sources along with the best-fit CI model and best-fit
CI OFF model. In Table \ref{synage_table} we report the parameters of
the fits.\\

\begin{figure*}
\begin{center}
\includegraphics{0003_tot_CI_spectrum.ps}
\includegraphics{0003_tot_CIOFF_spectrum.ps}
\includegraphics{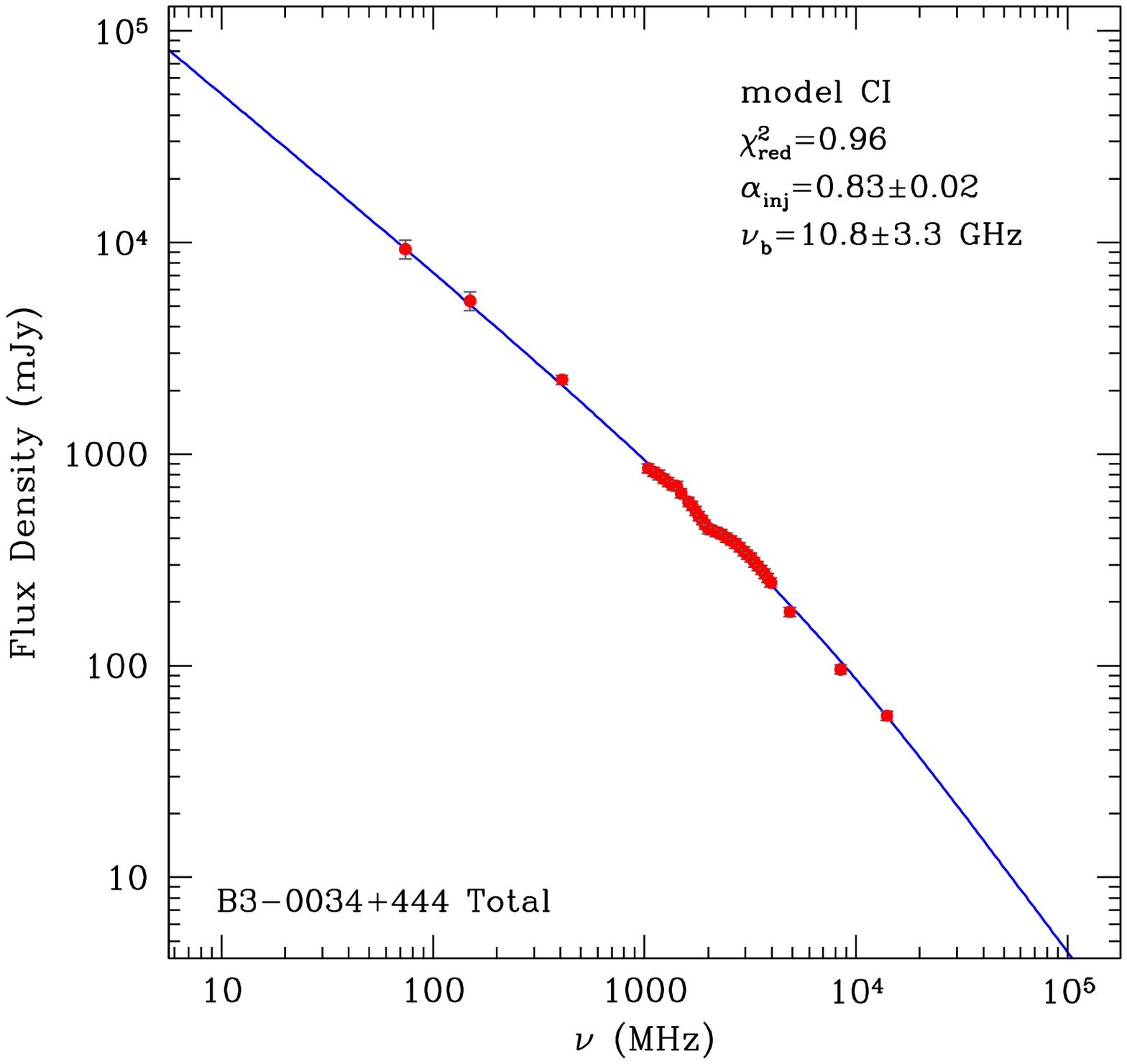}
\includegraphics{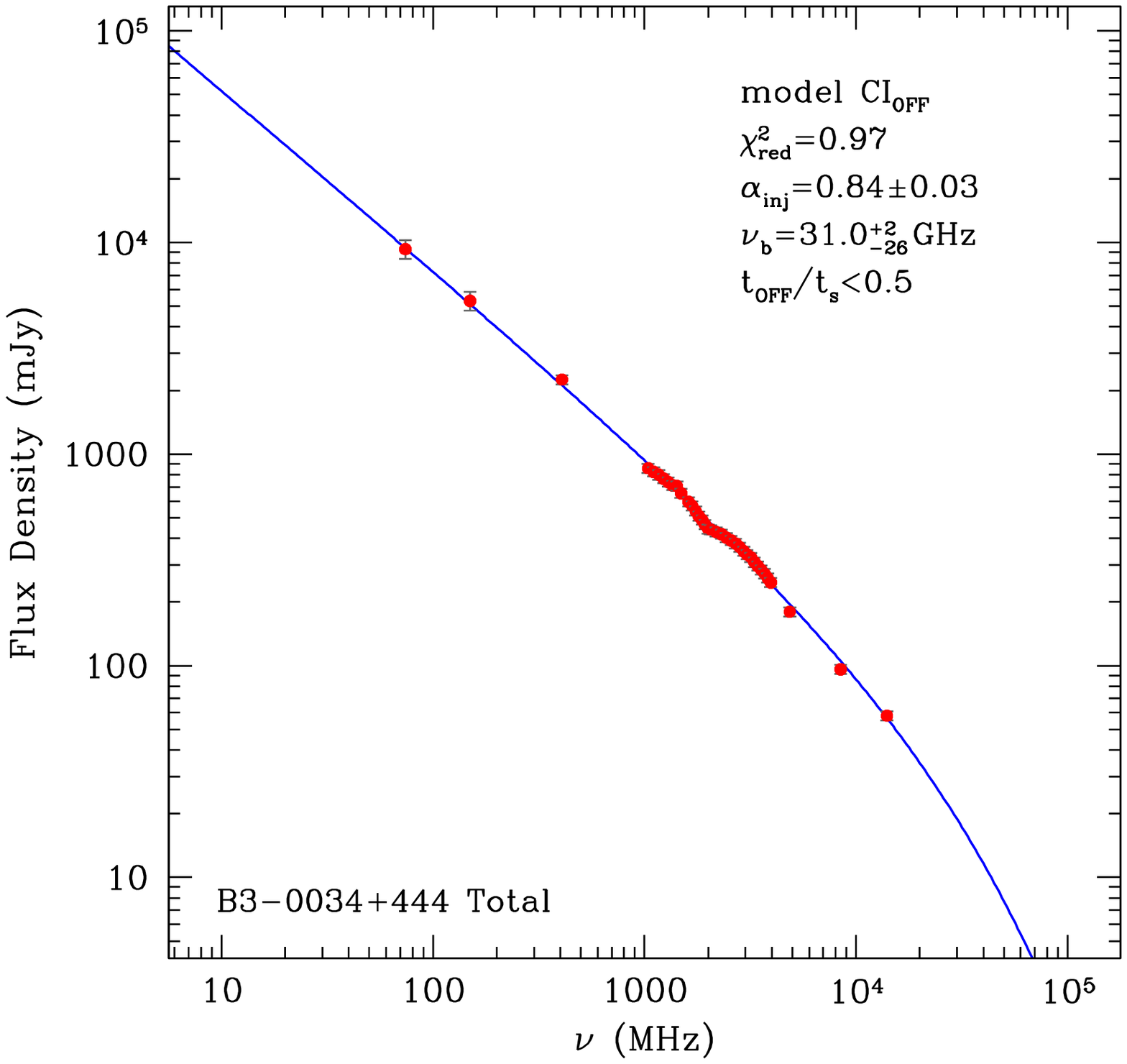}
\includegraphics{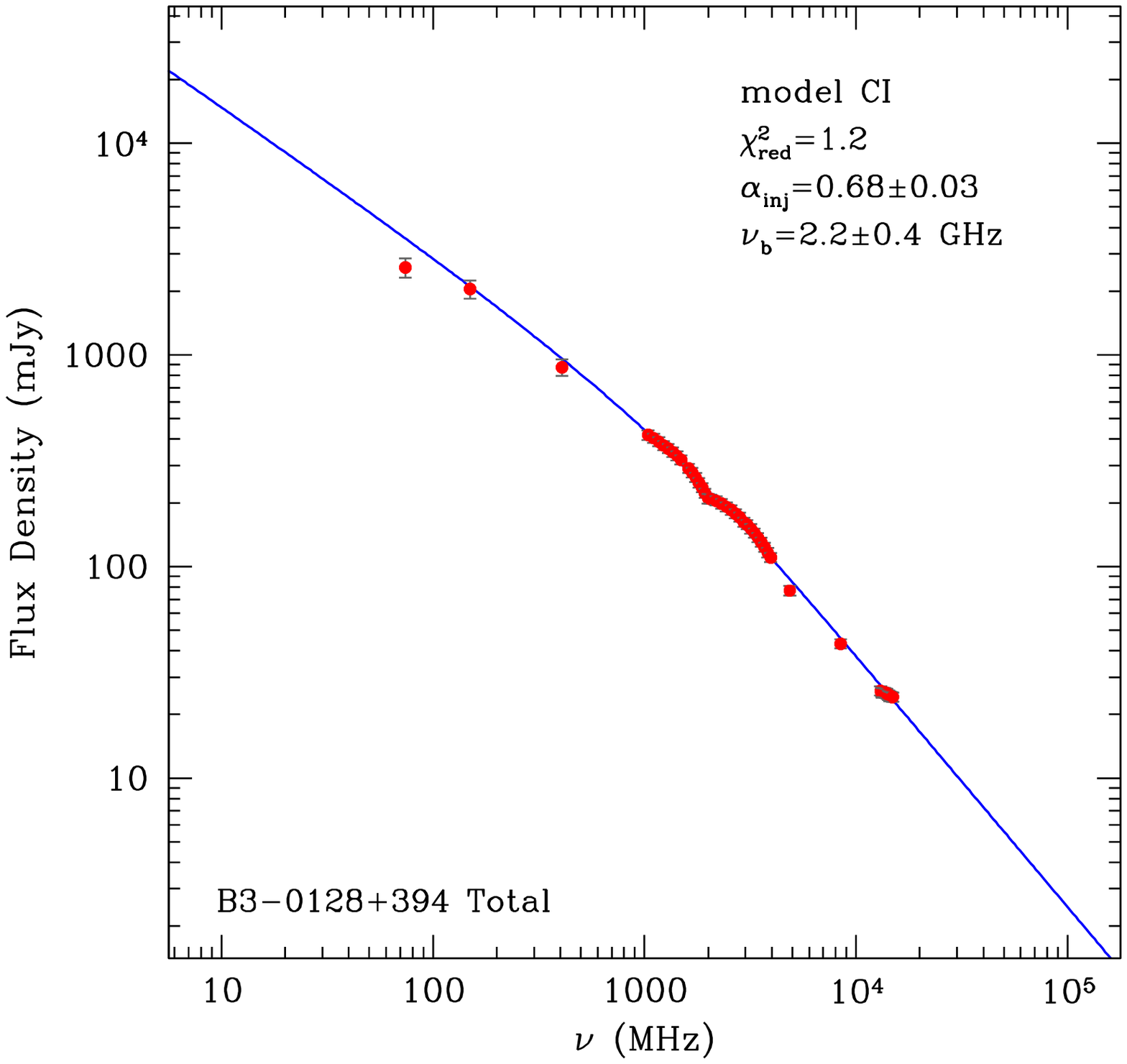}
\includegraphics{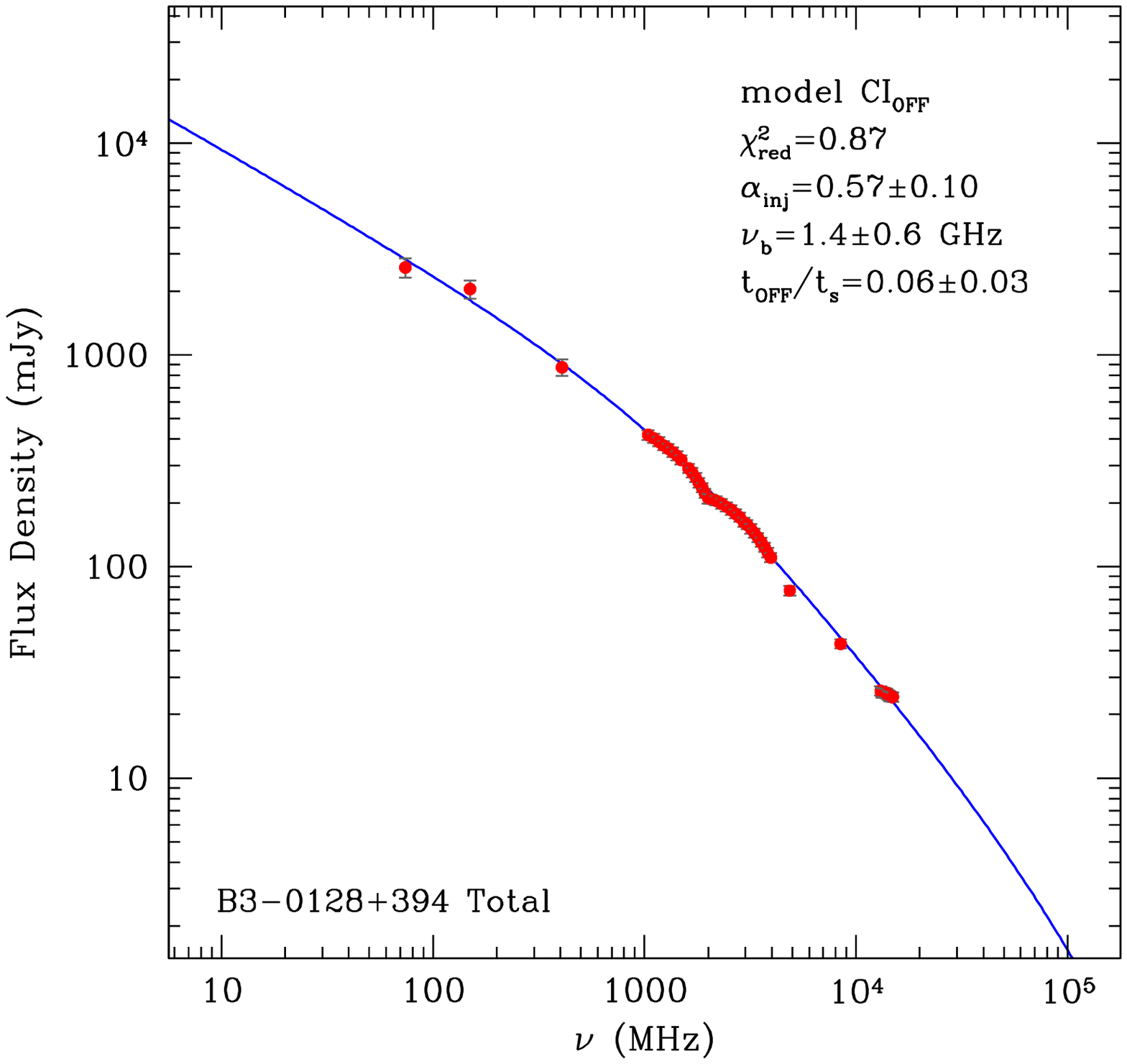}
\vspace{23cm}
\caption{Integrated spectrum of the sources along with the best-fit CI model
  ({\it left}) and best-fit CI OFF model ({\it right}).}
\label{synage-fig}
\end{center}
\end{figure*}

\addtocounter{figure}{-1}
\begin{figure*}
\begin{center}
\includegraphics{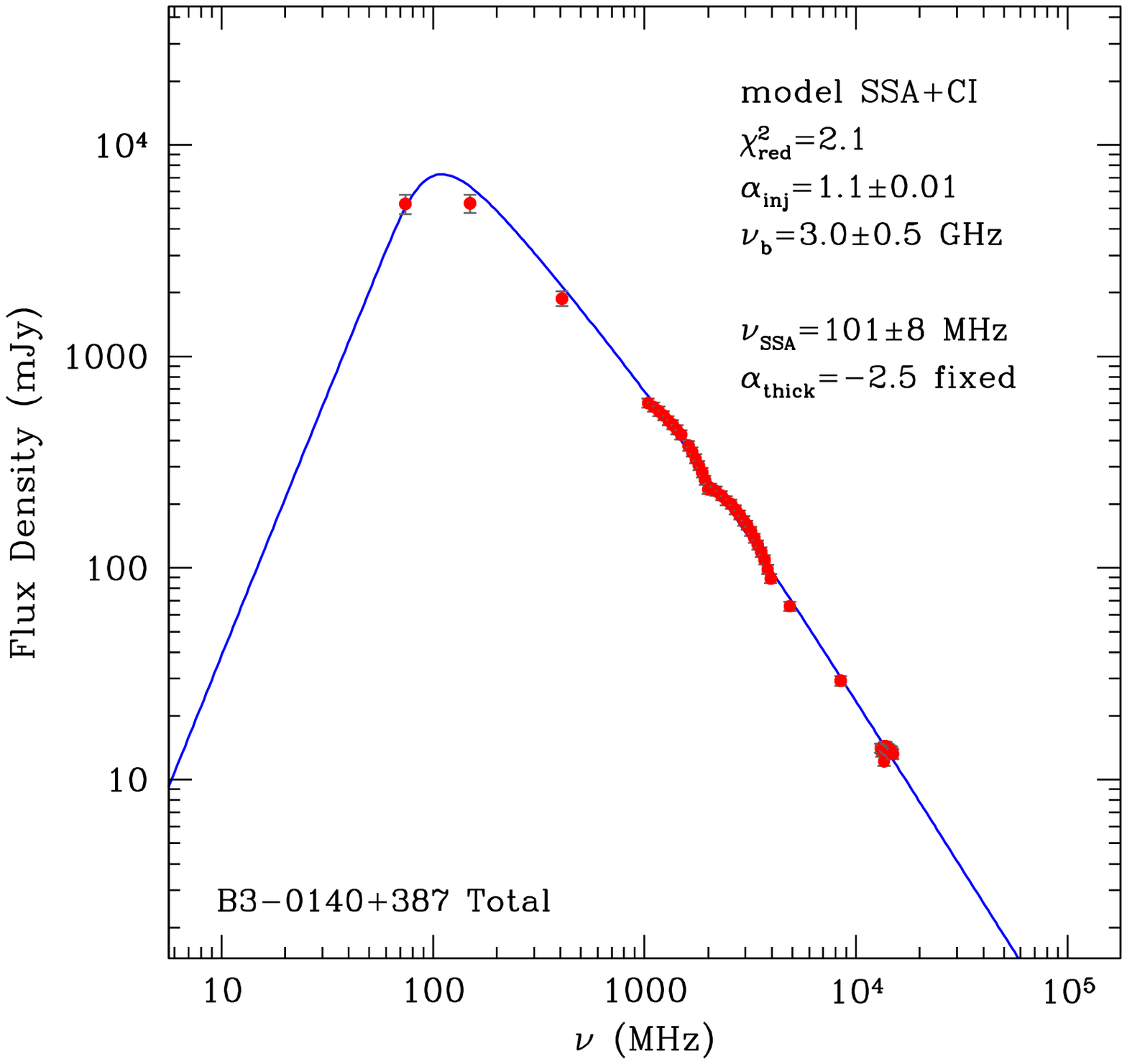}
\includegraphics{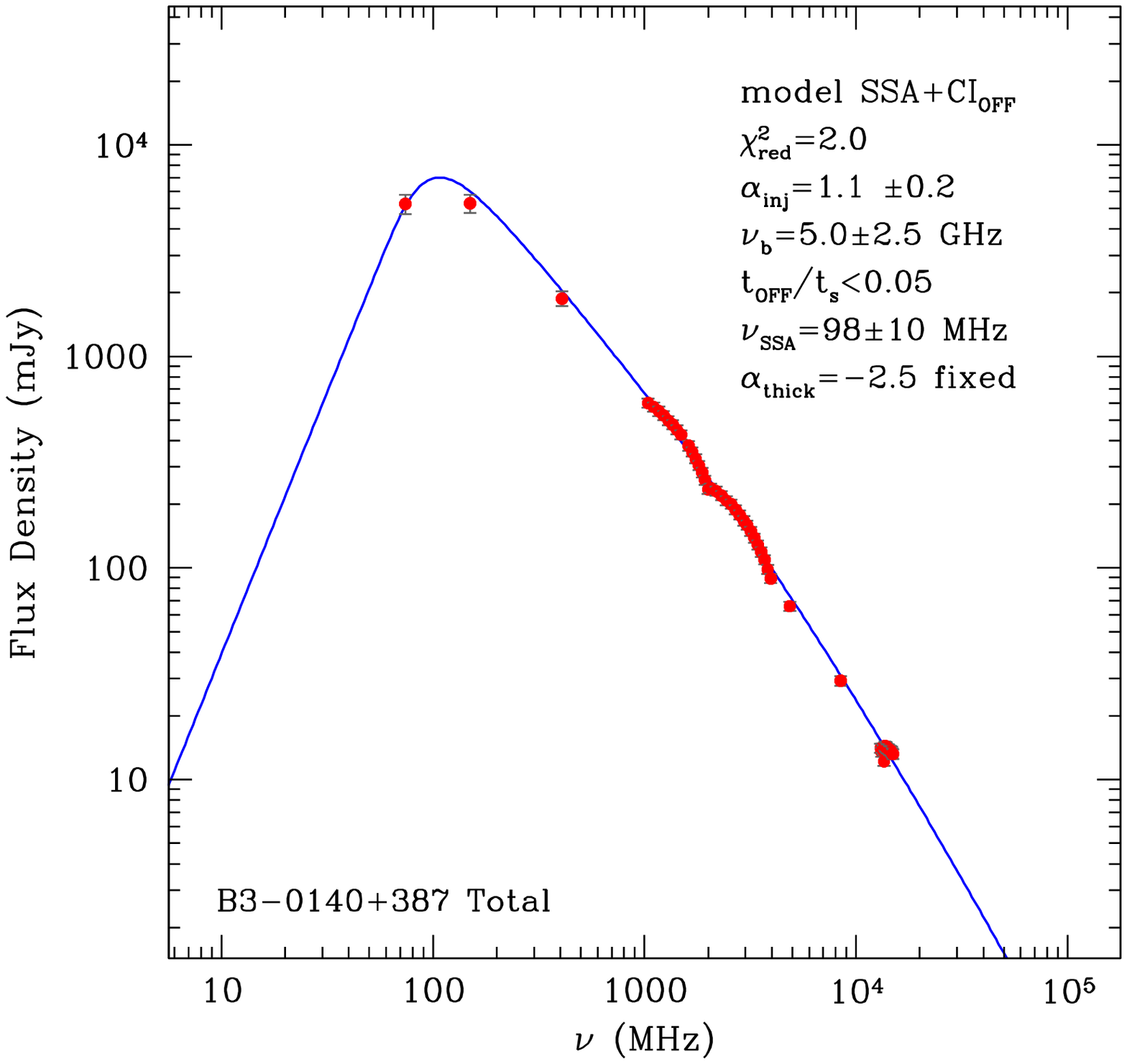}
\includegraphics{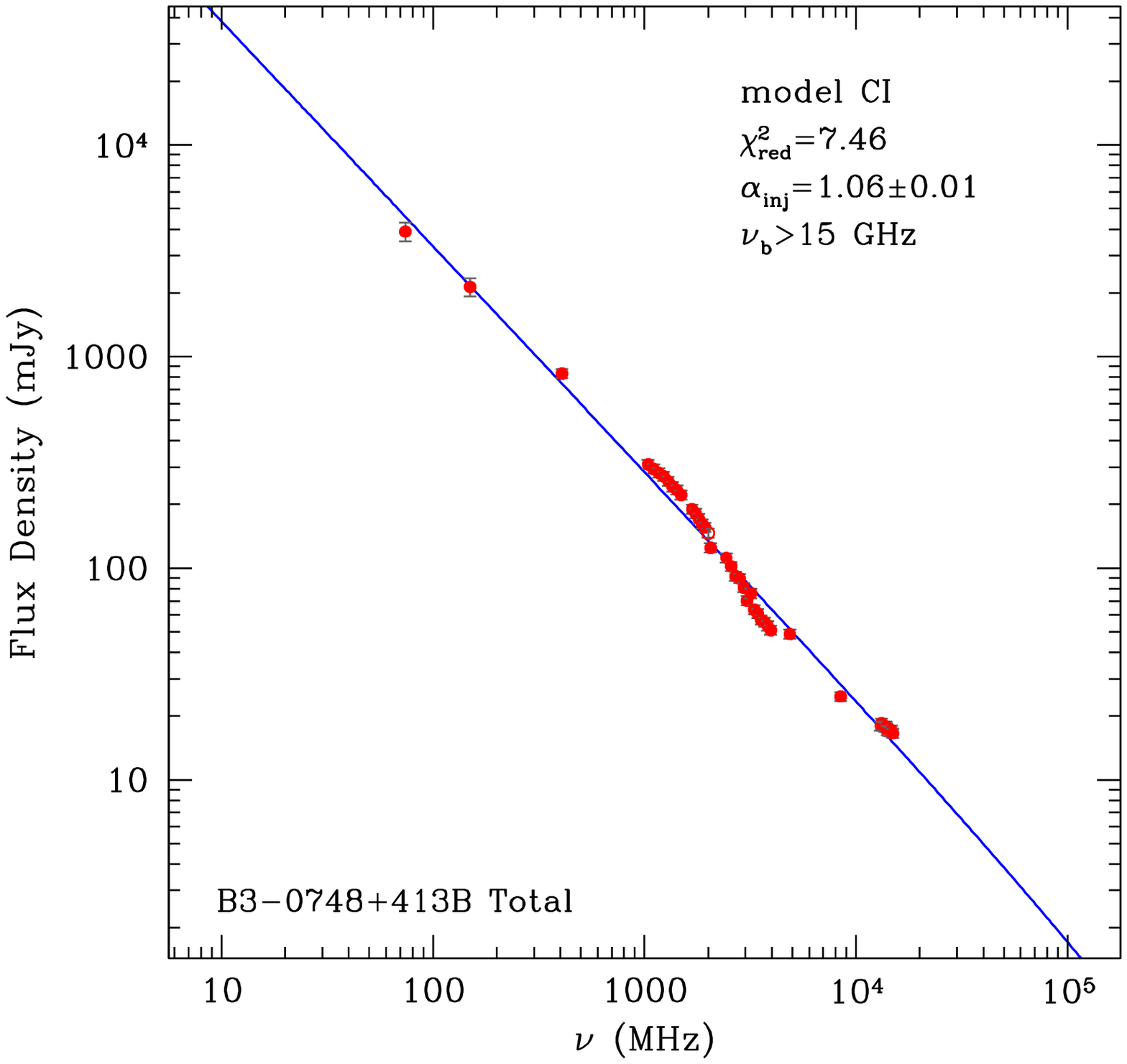}
\includegraphics{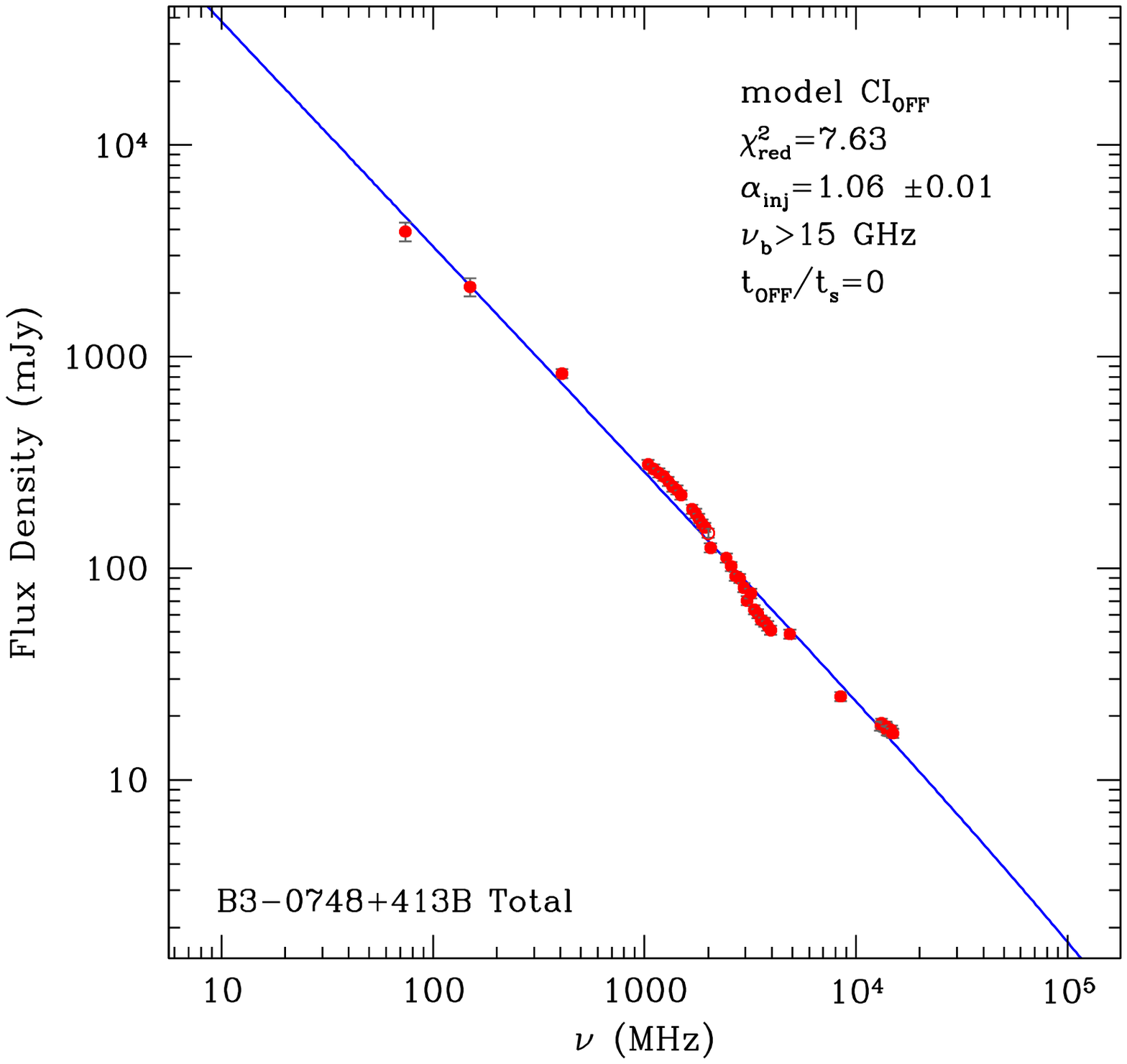}
\includegraphics{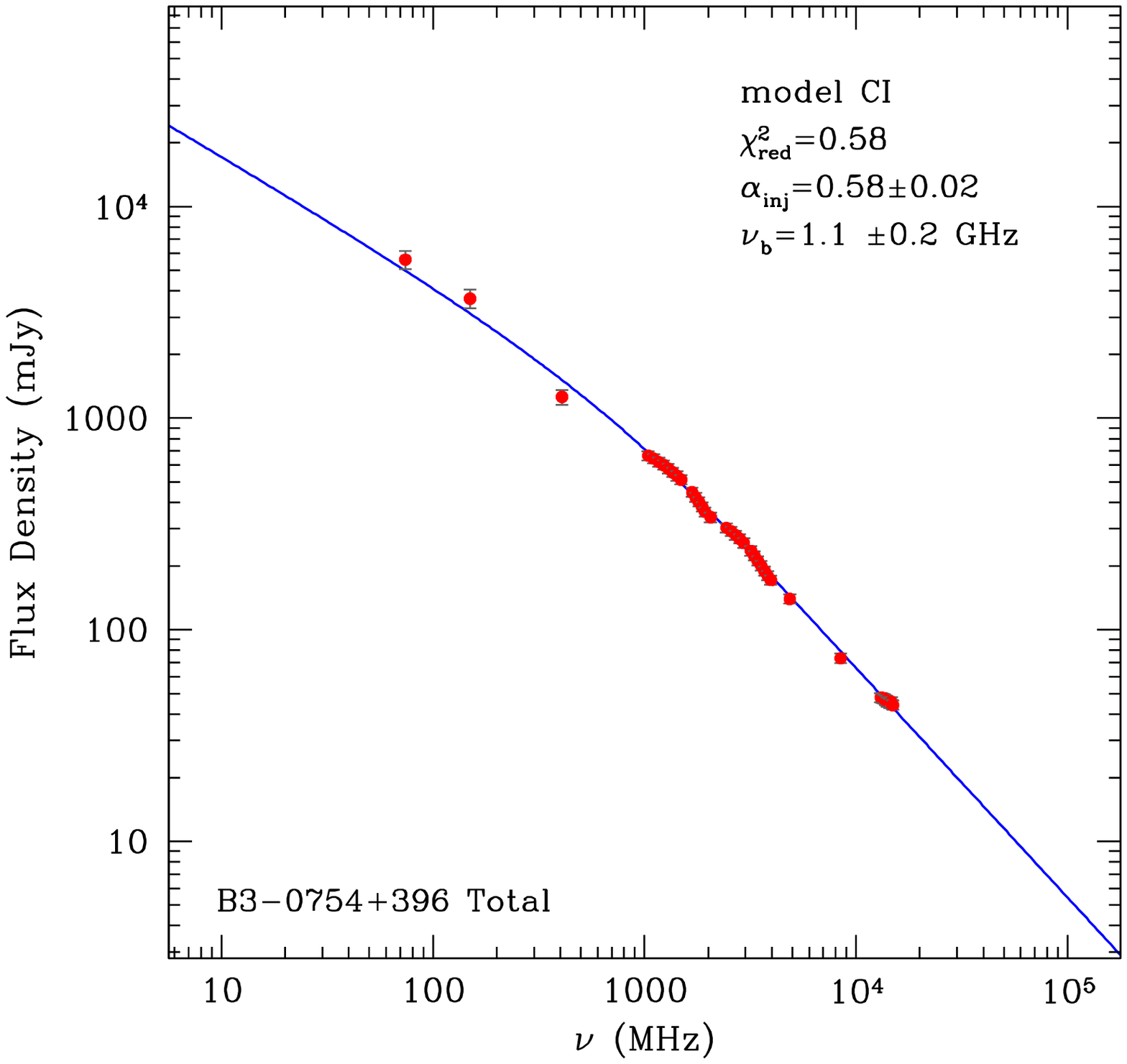}
\includegraphics{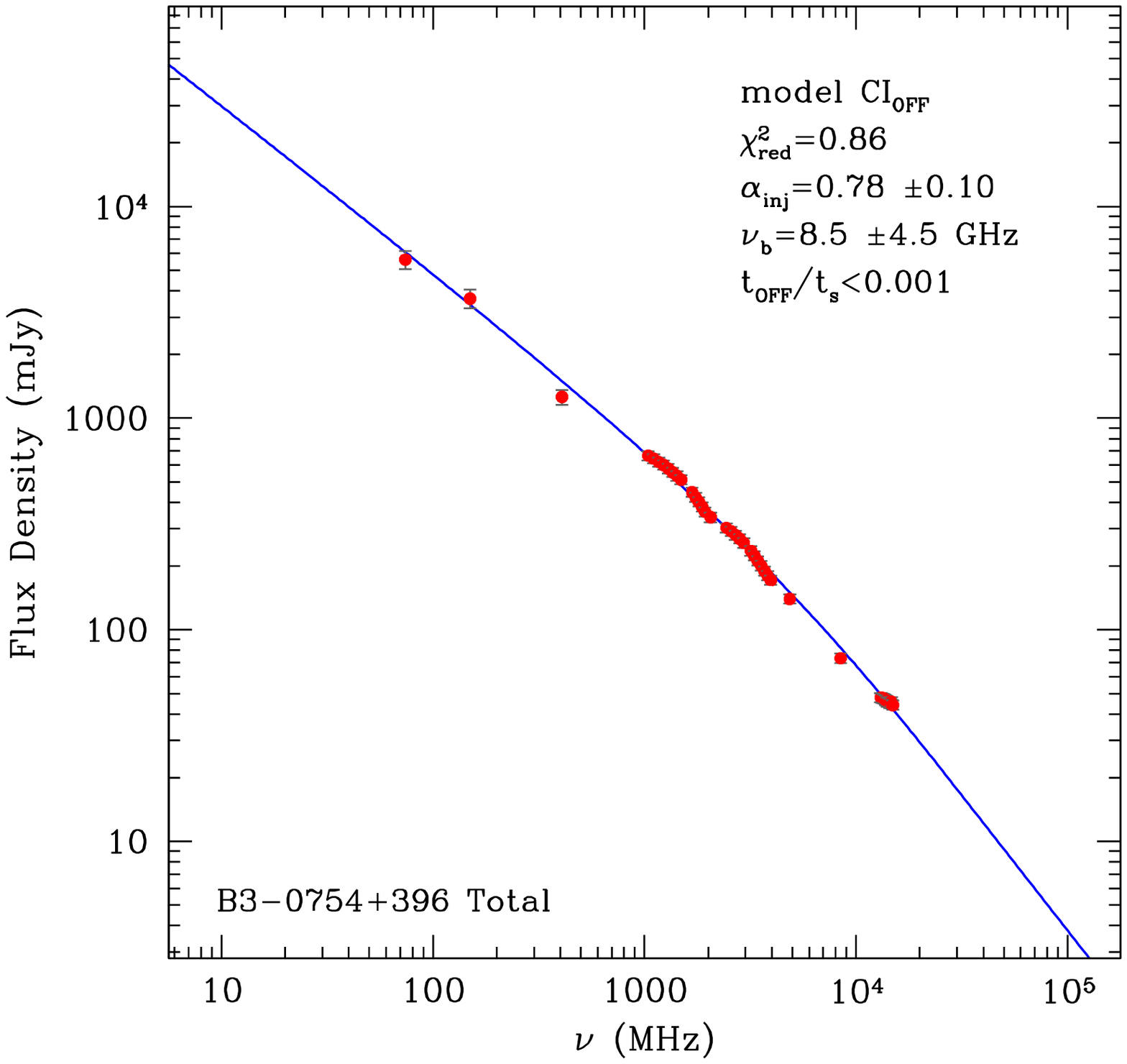}
\vspace{23cm}
\caption{Continued.}
\end{center}
\end{figure*}

\addtocounter{figure}{-1}
\begin{figure*}
\begin{center}
\includegraphics{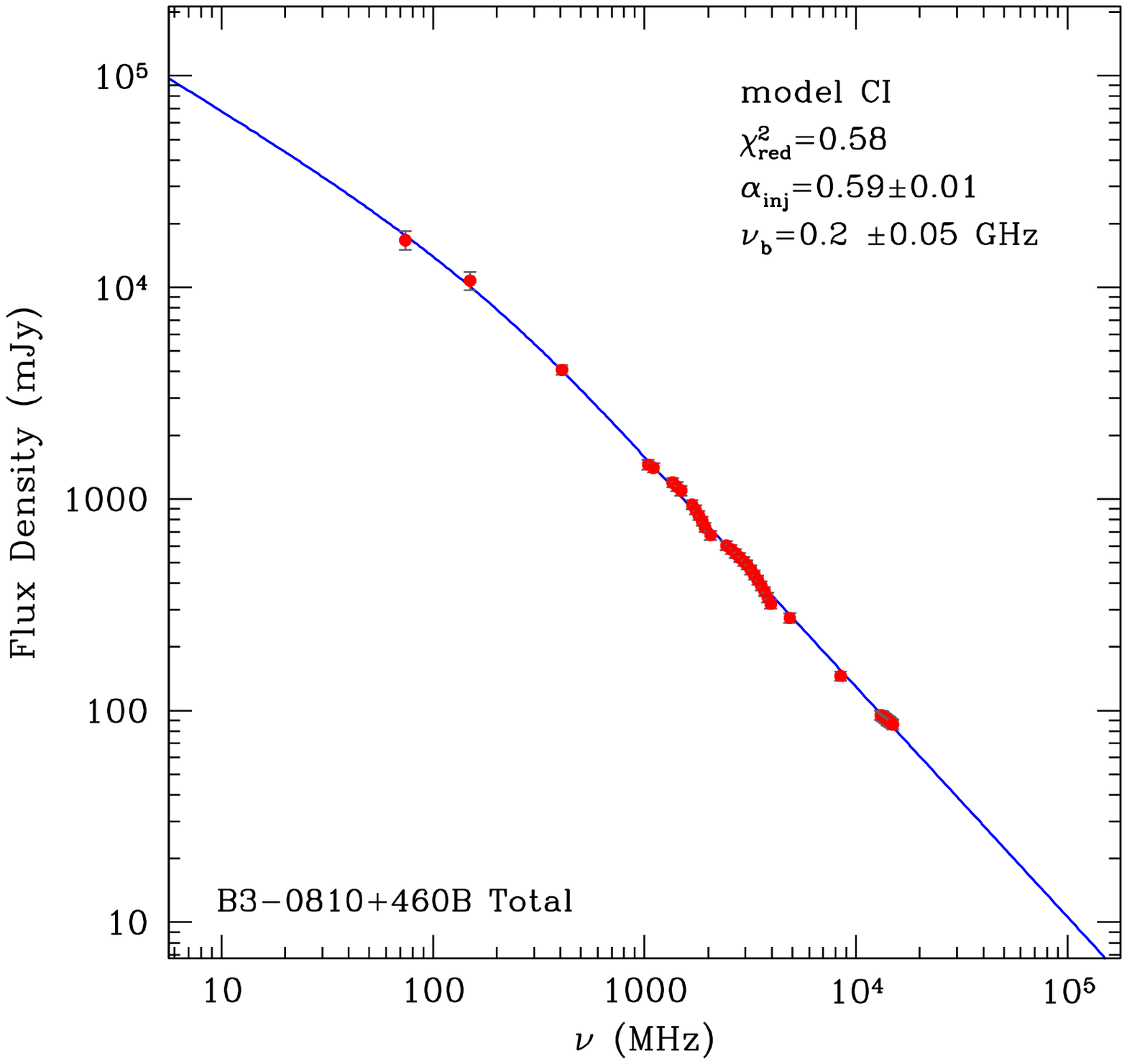}
\includegraphics{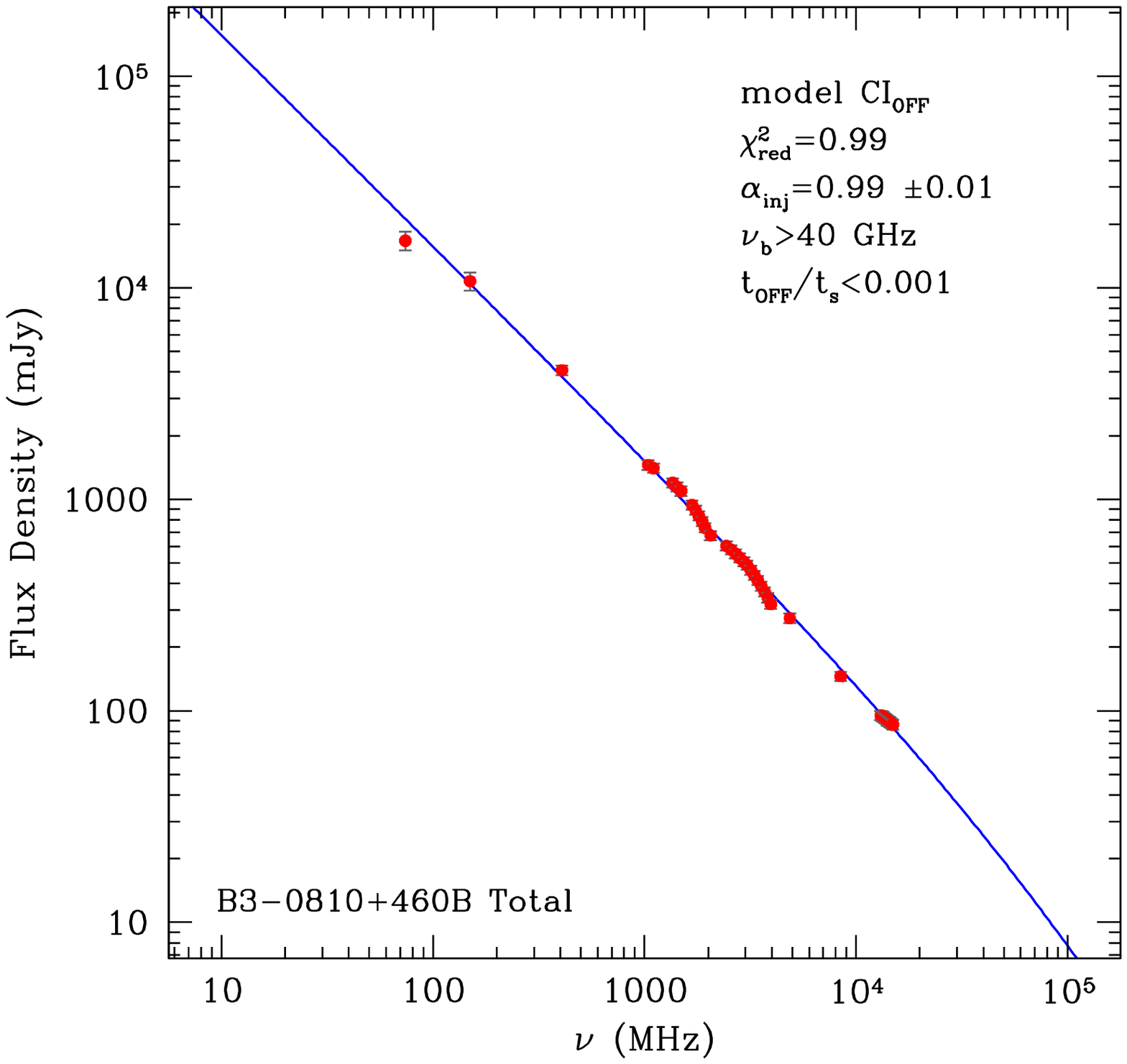}
\includegraphics{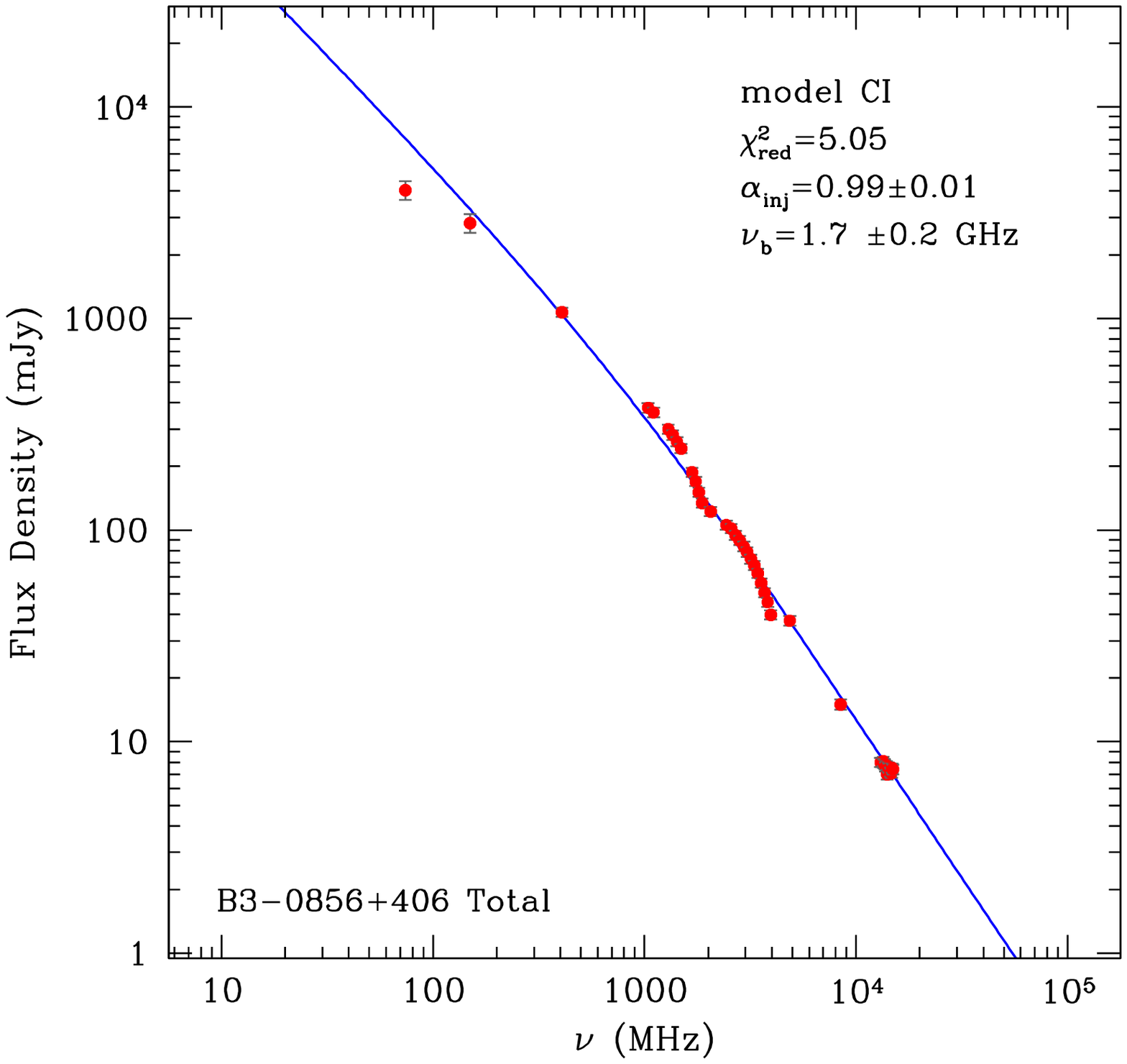}
\includegraphics{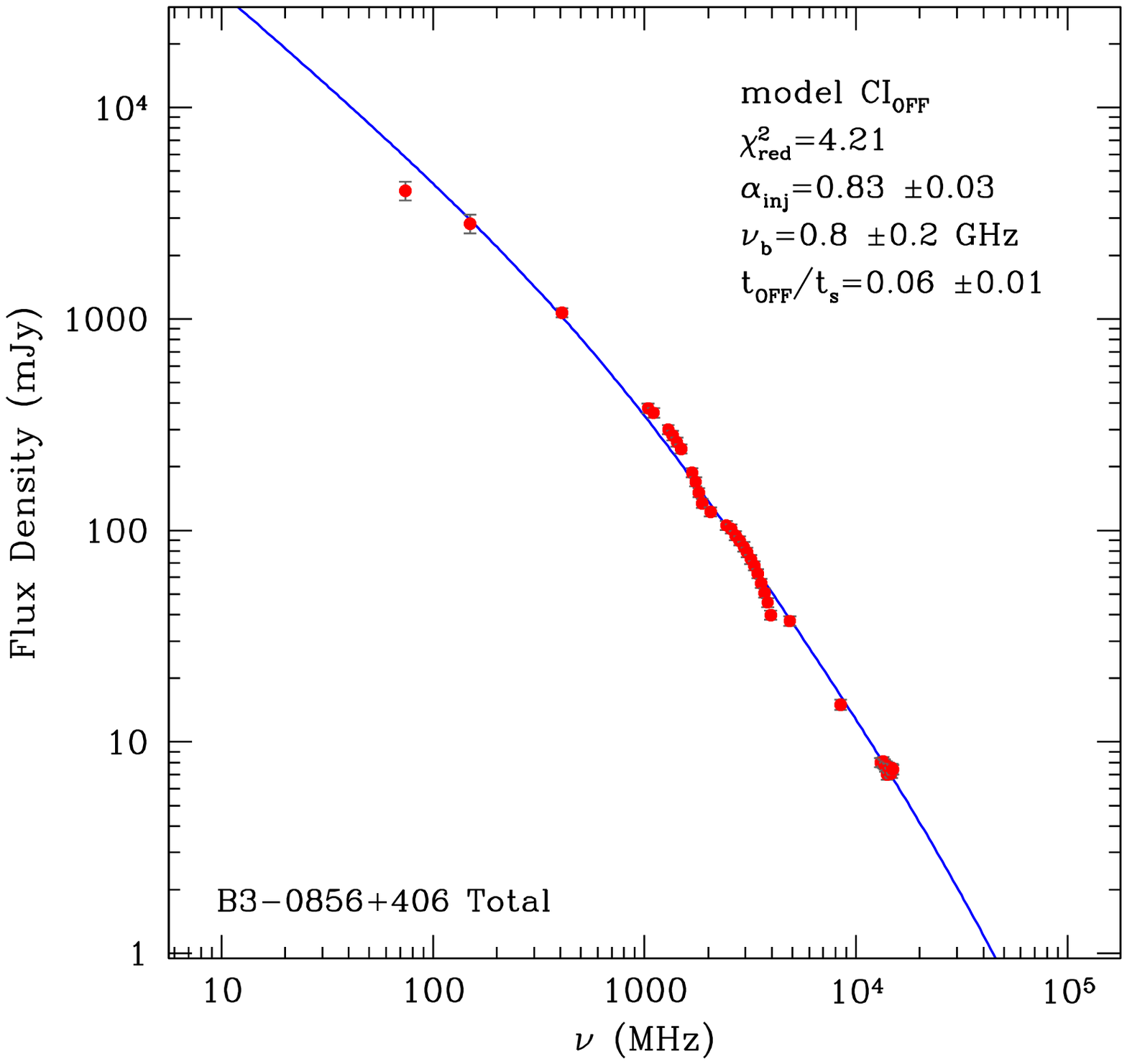}
\includegraphics{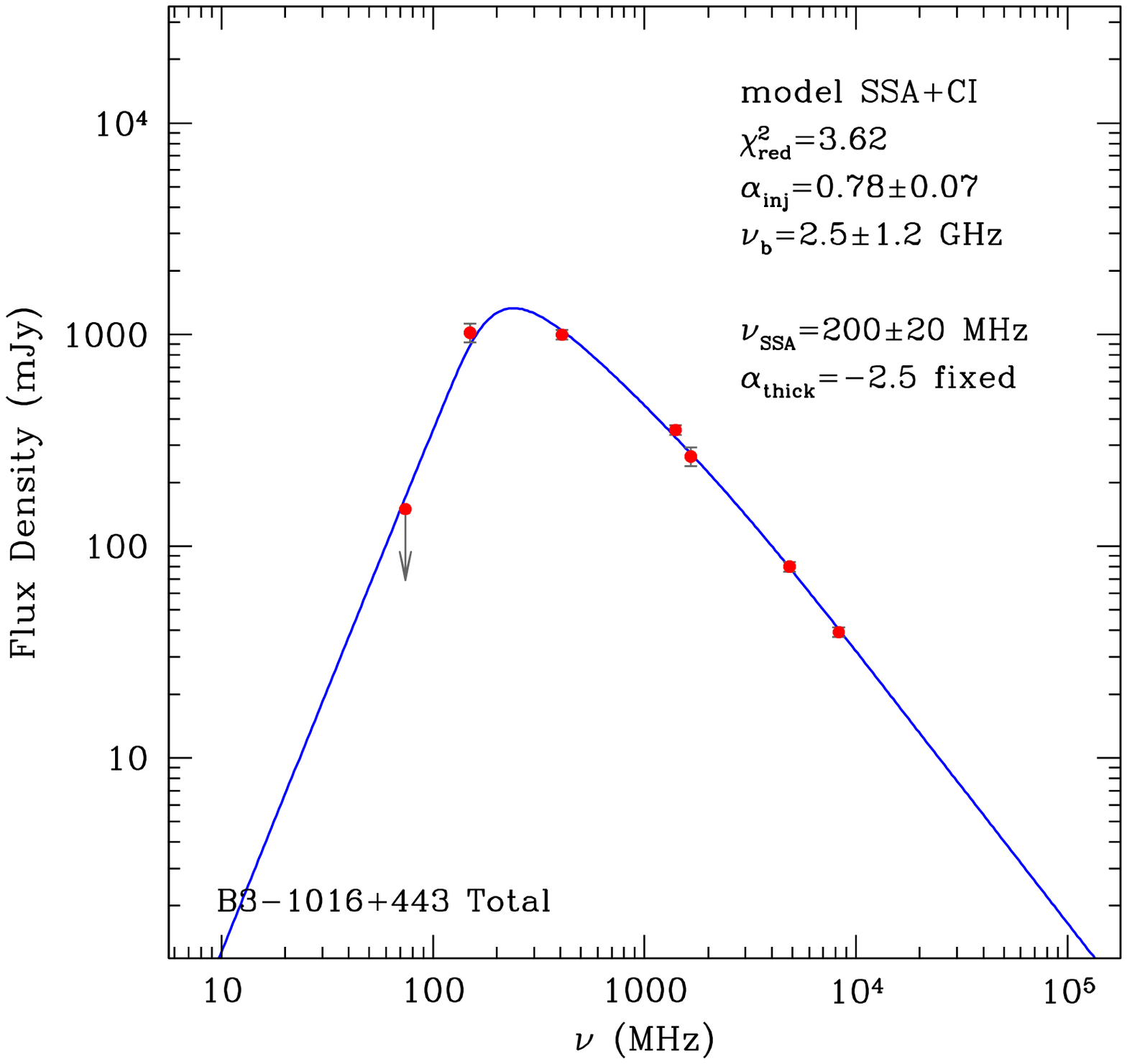}
\includegraphics{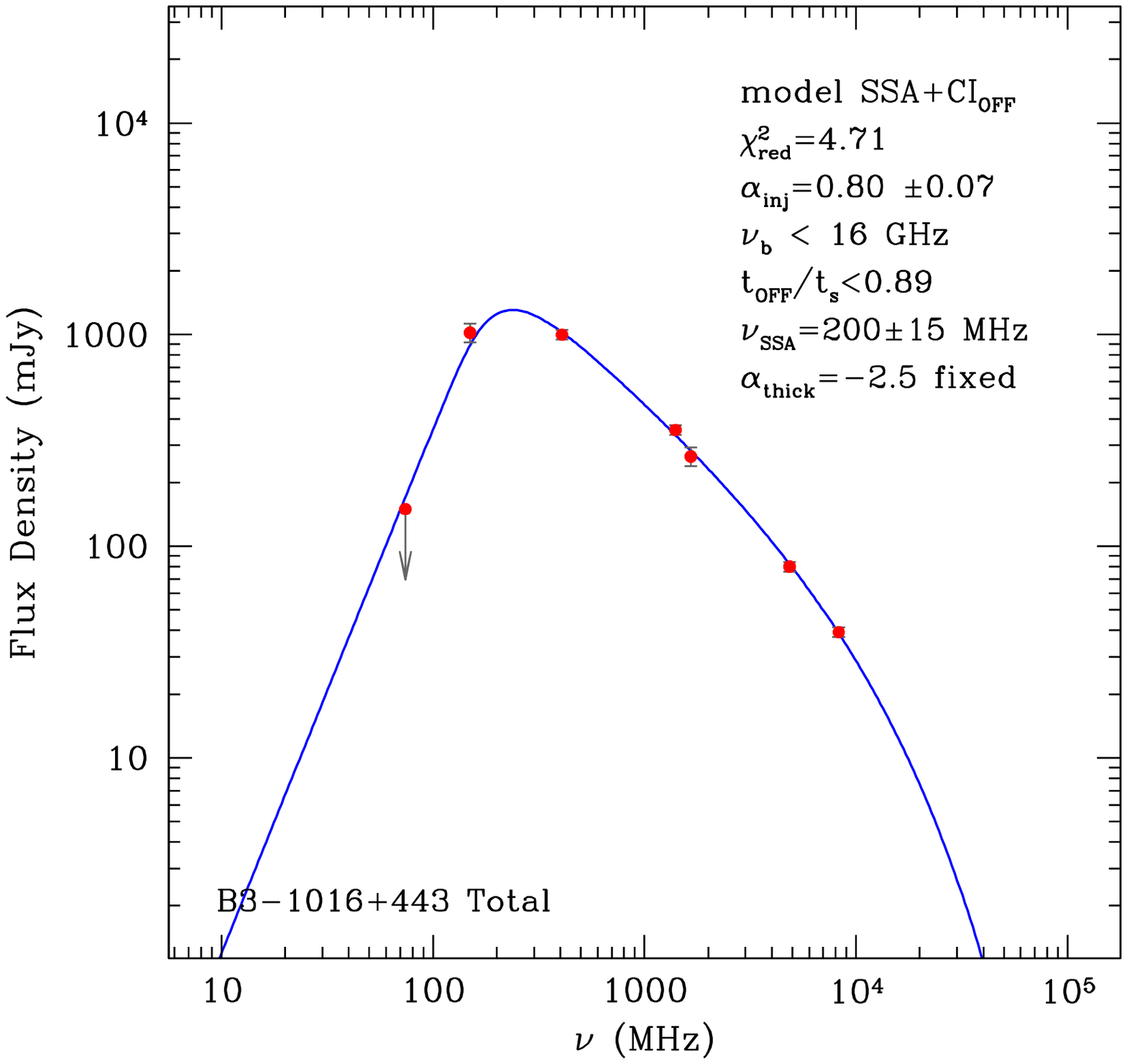}
\vspace{23cm}
\caption{Continued.}
\end{center}
\end{figure*}

\addtocounter{figure}{-1}
\begin{figure*}
\begin{center}
\includegraphics{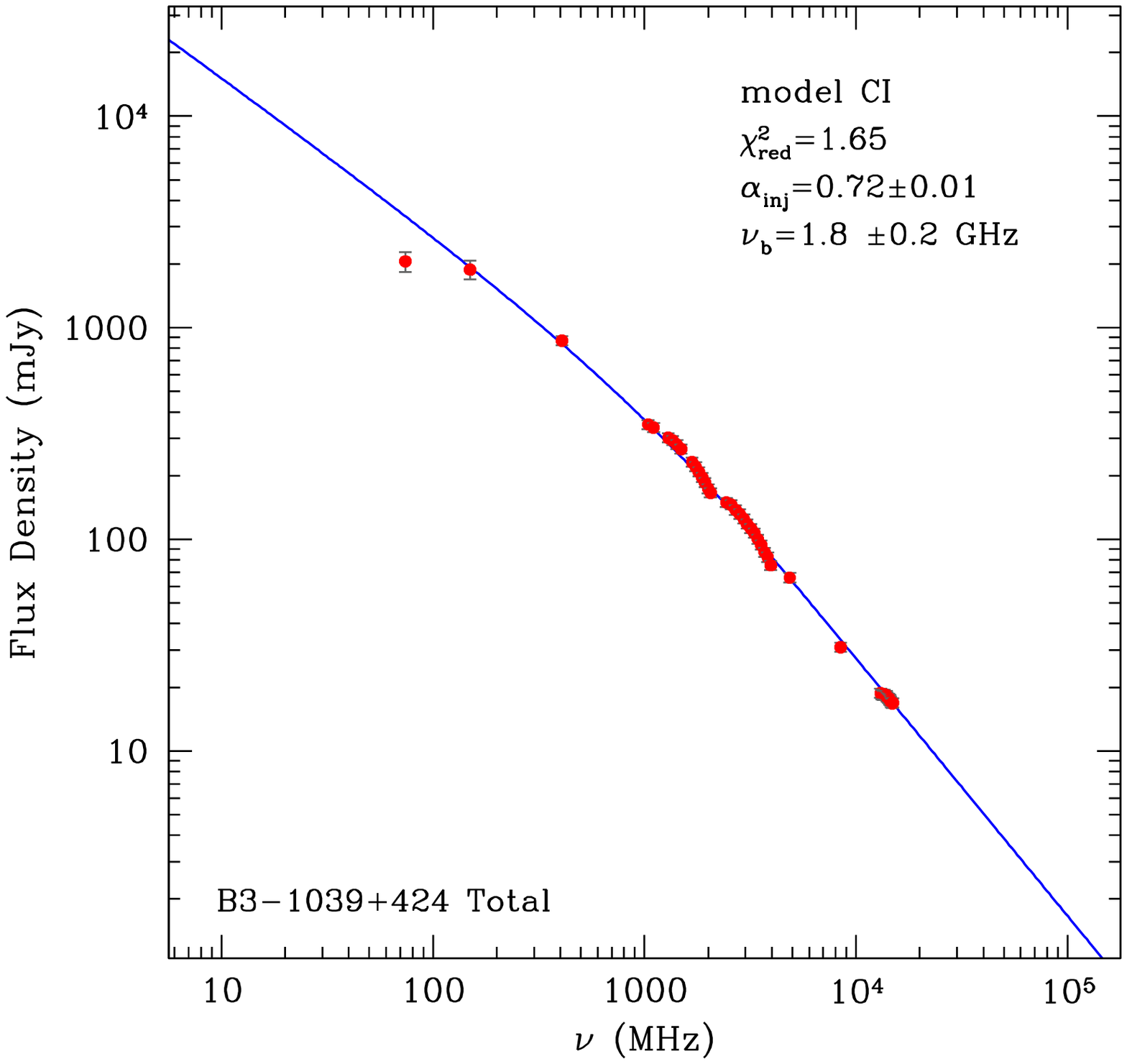}
\includegraphics{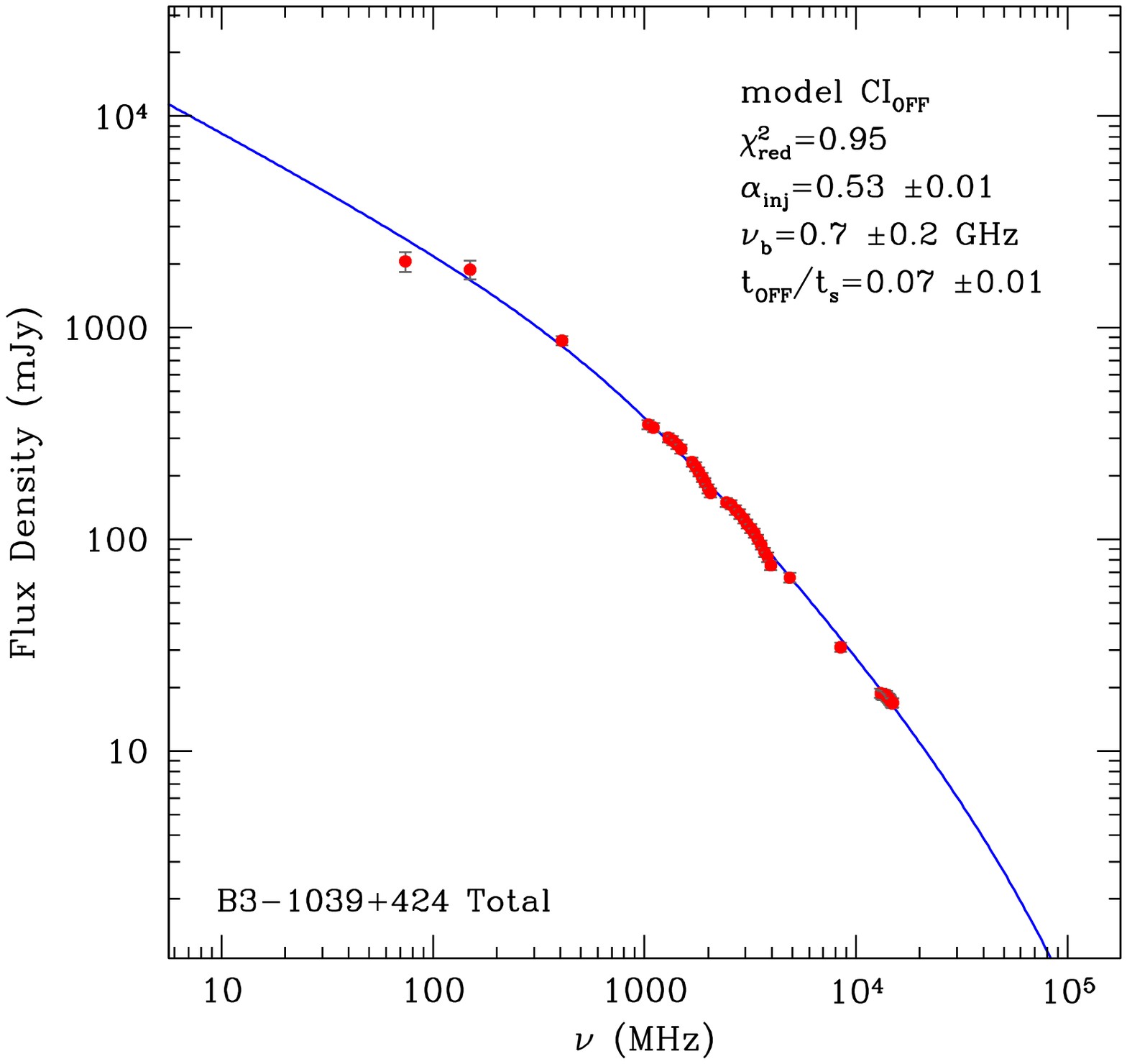}
\includegraphics{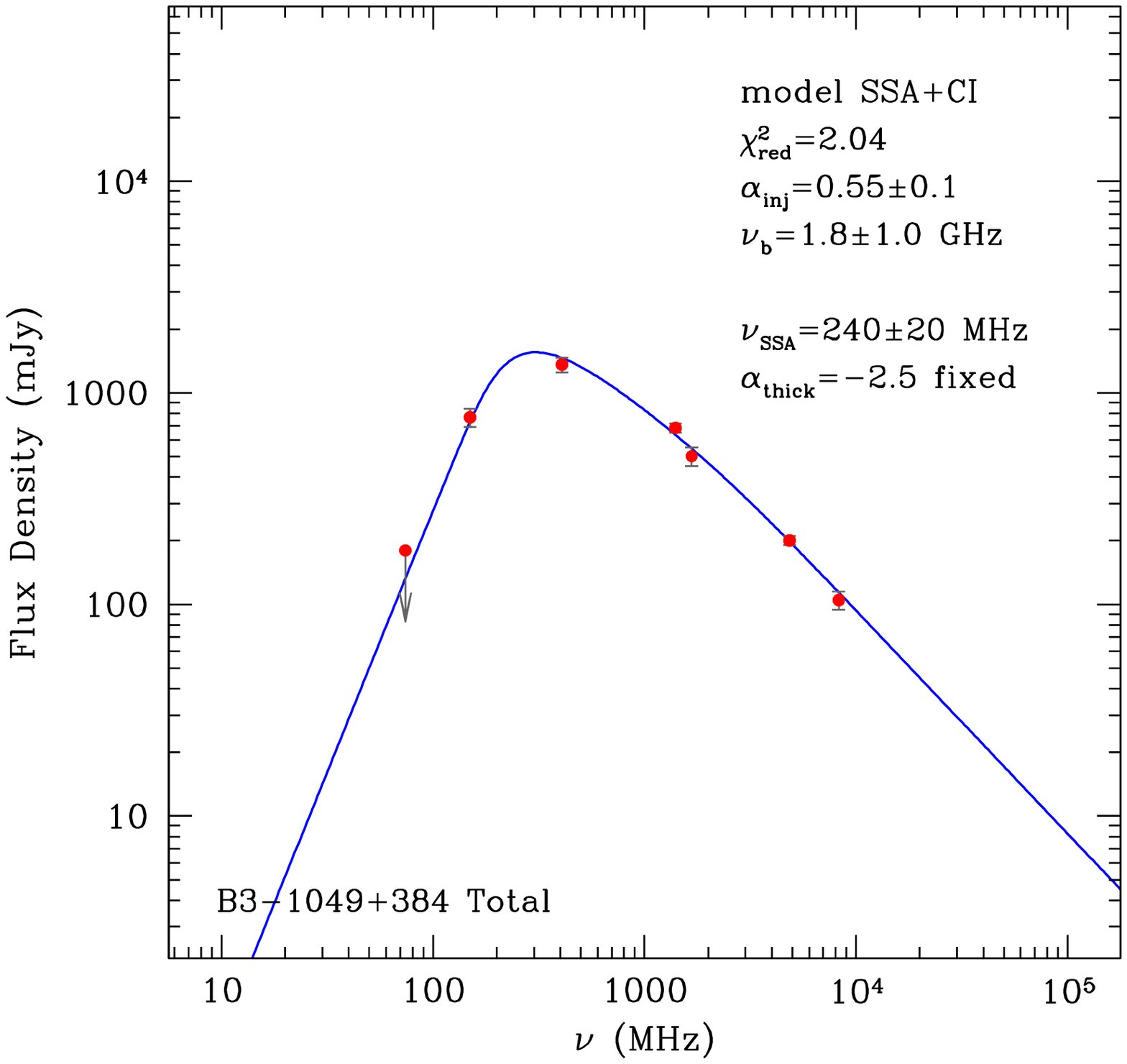}
\includegraphics{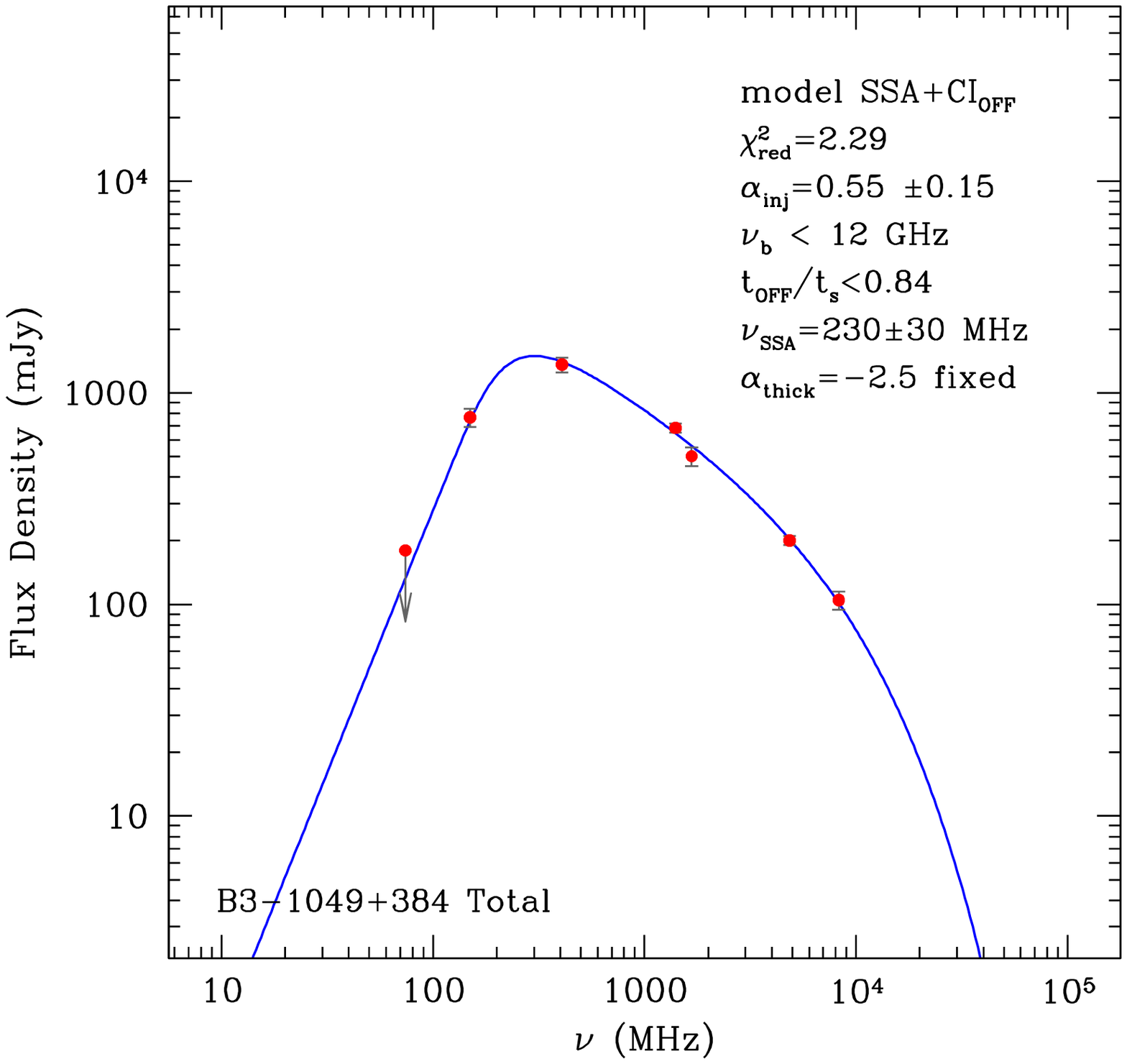}
\includegraphics{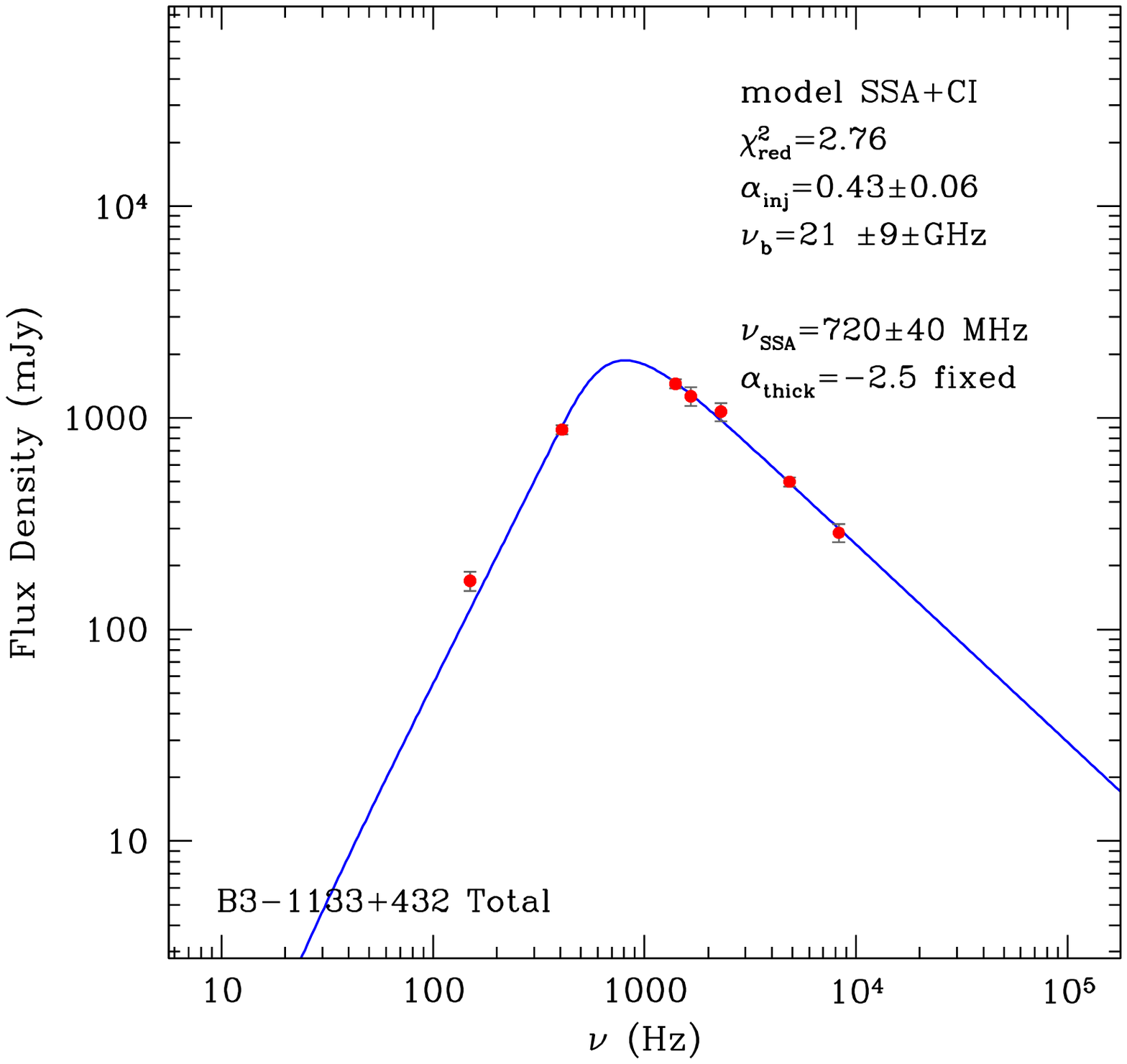}
\includegraphics{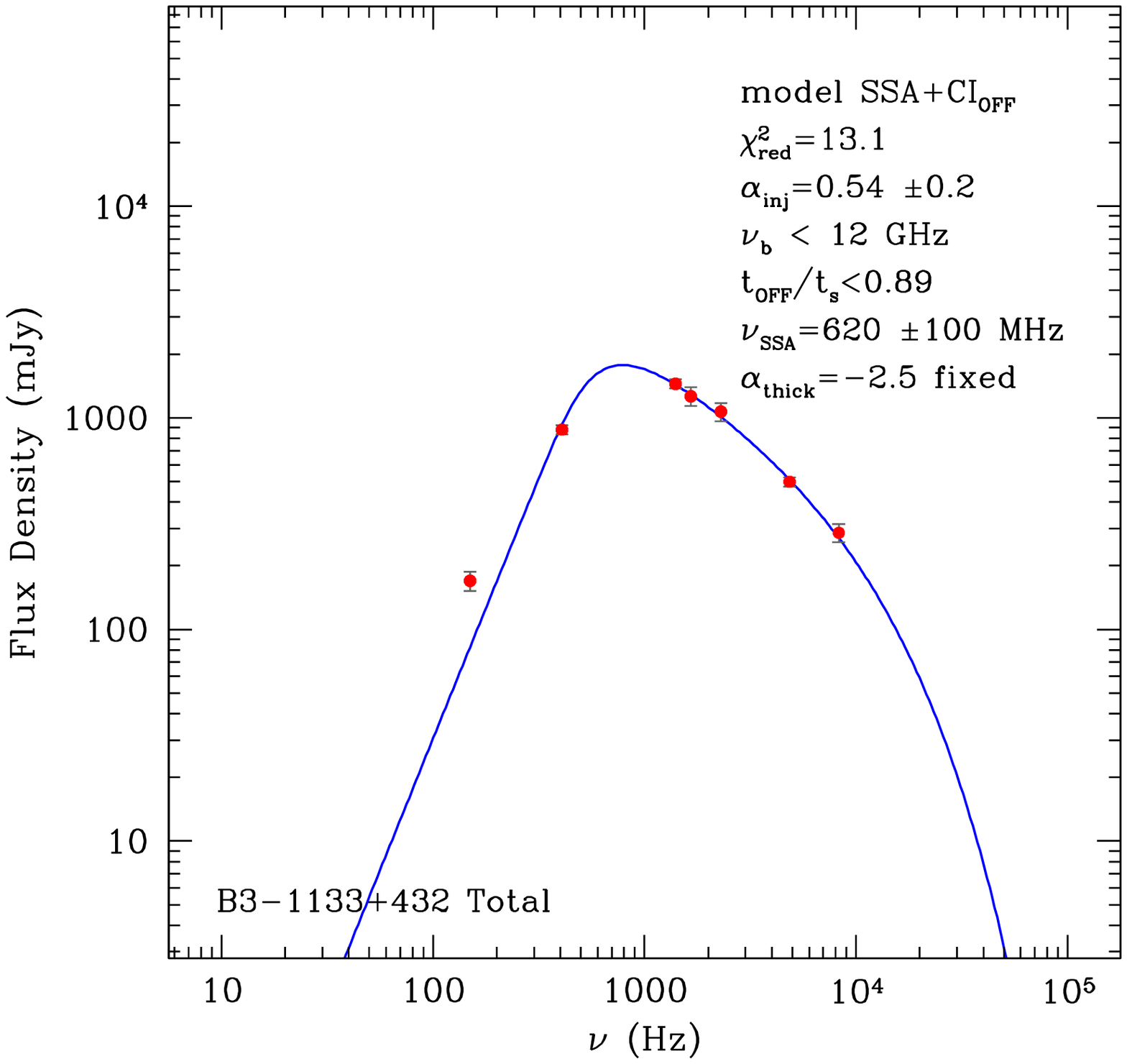}
\vspace{23cm}
\caption{Continued.}
\end{center}
\end{figure*}

\addtocounter{figure}{-1}
\begin{figure*}
\begin{center}
\includegraphics{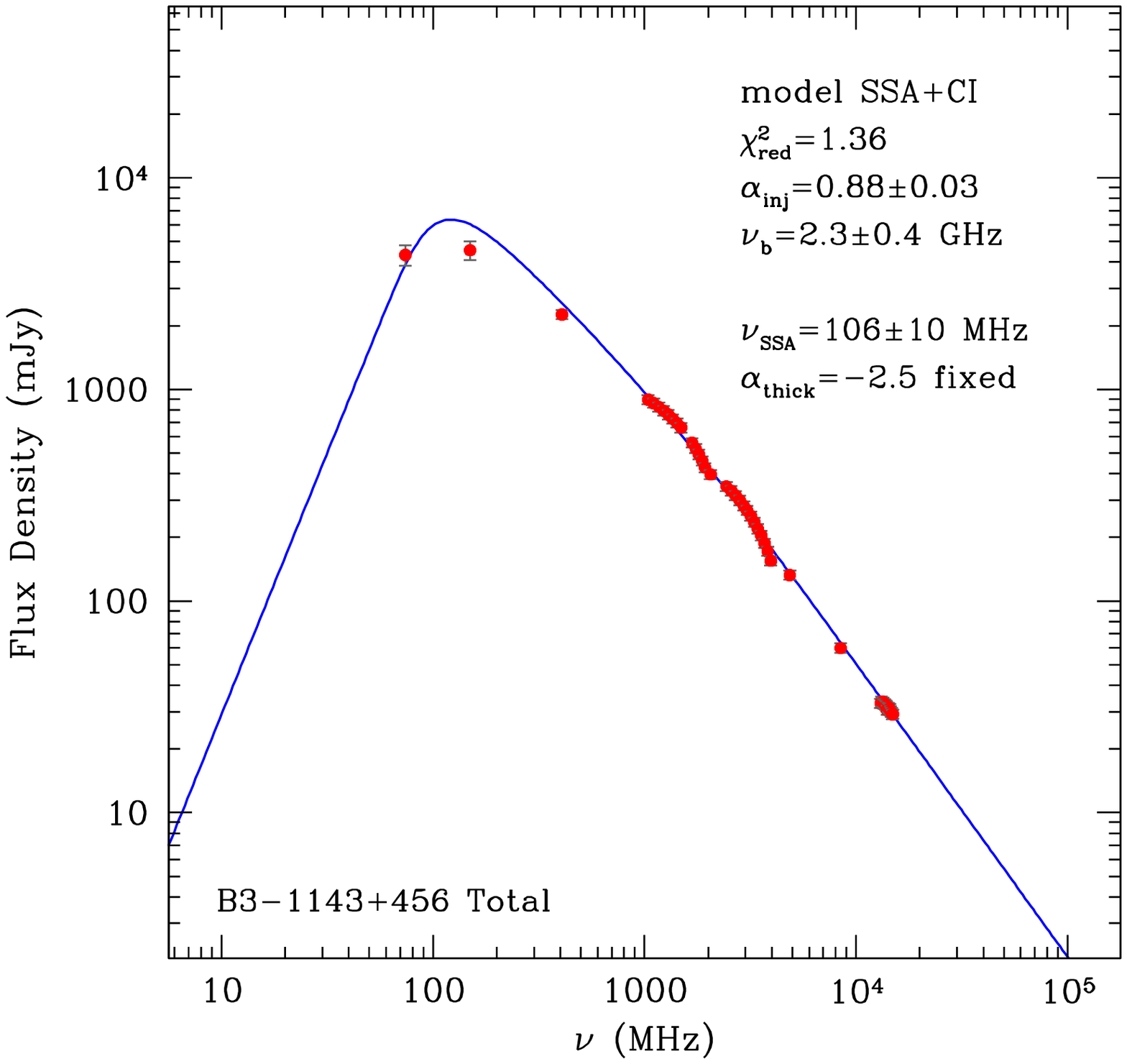}
\includegraphics{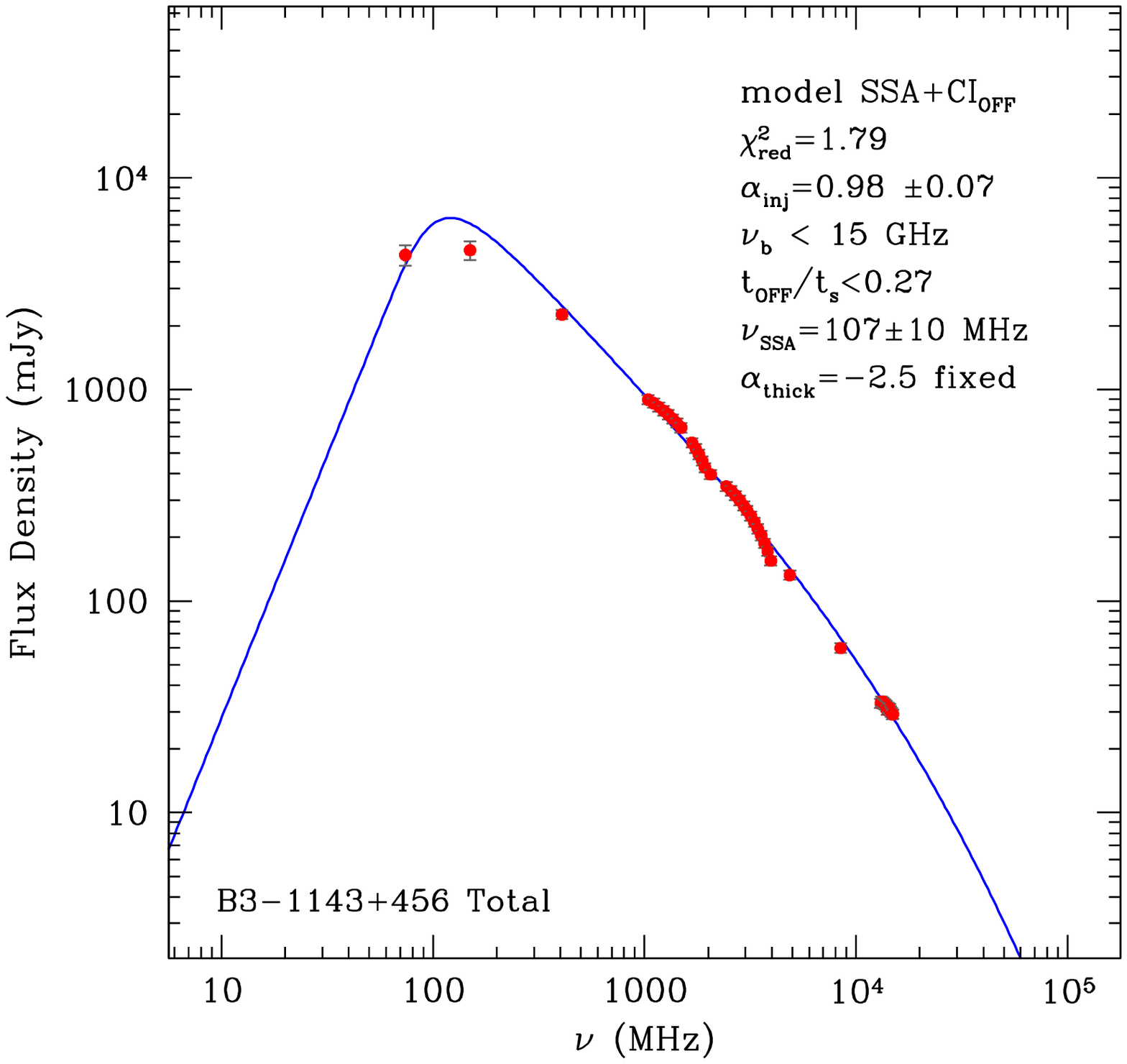}
\includegraphics{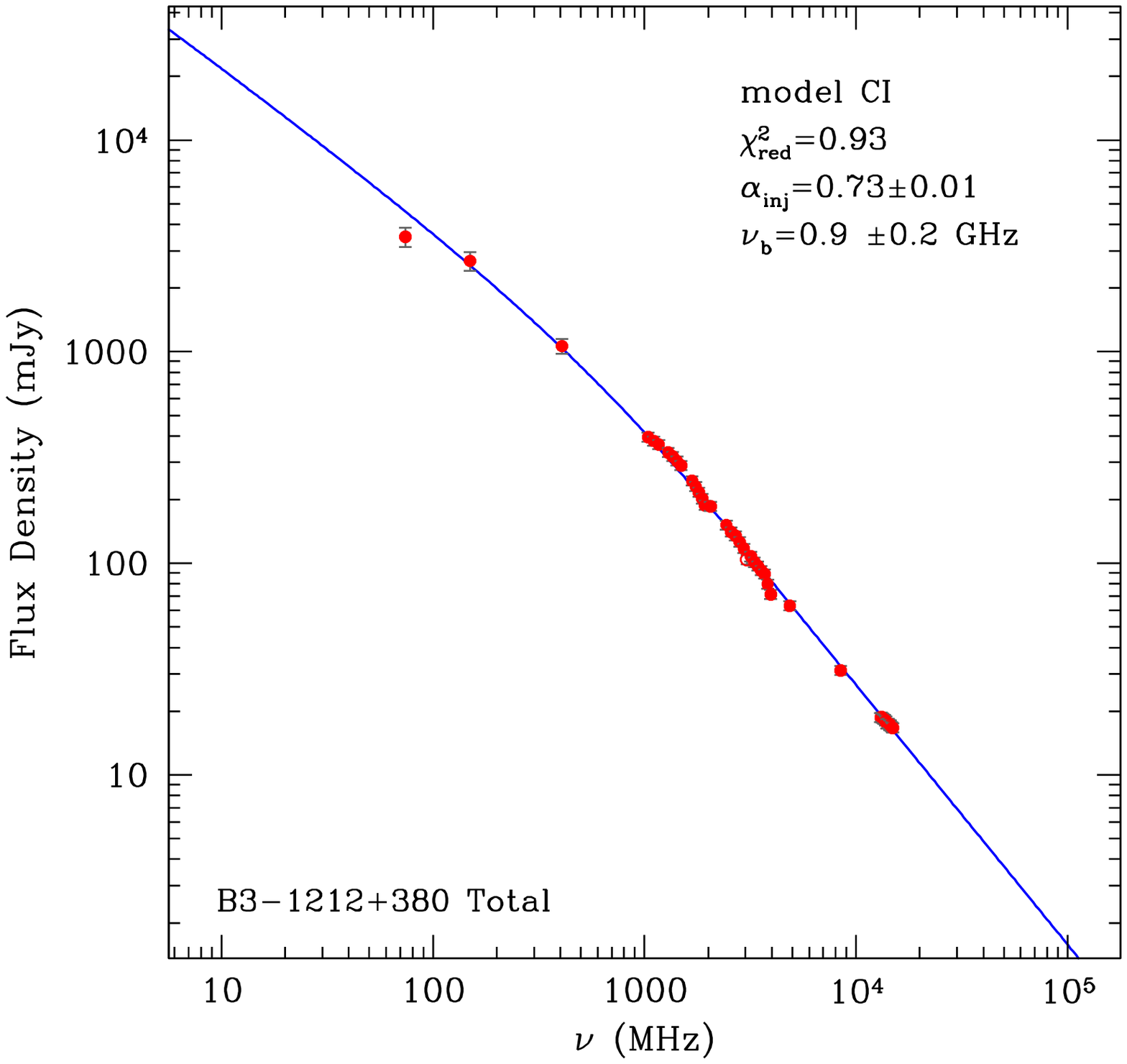}
\includegraphics{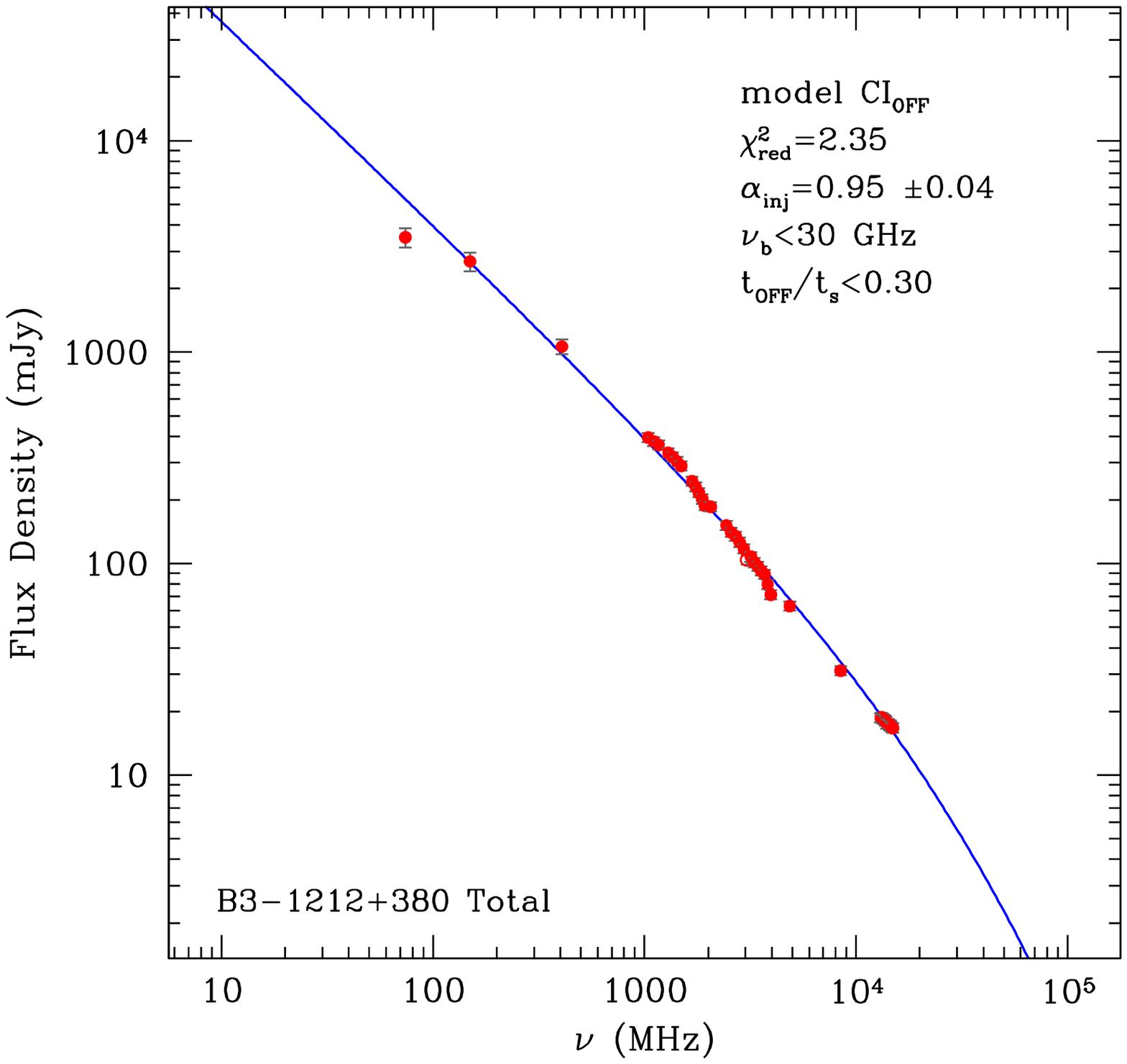}
\includegraphics{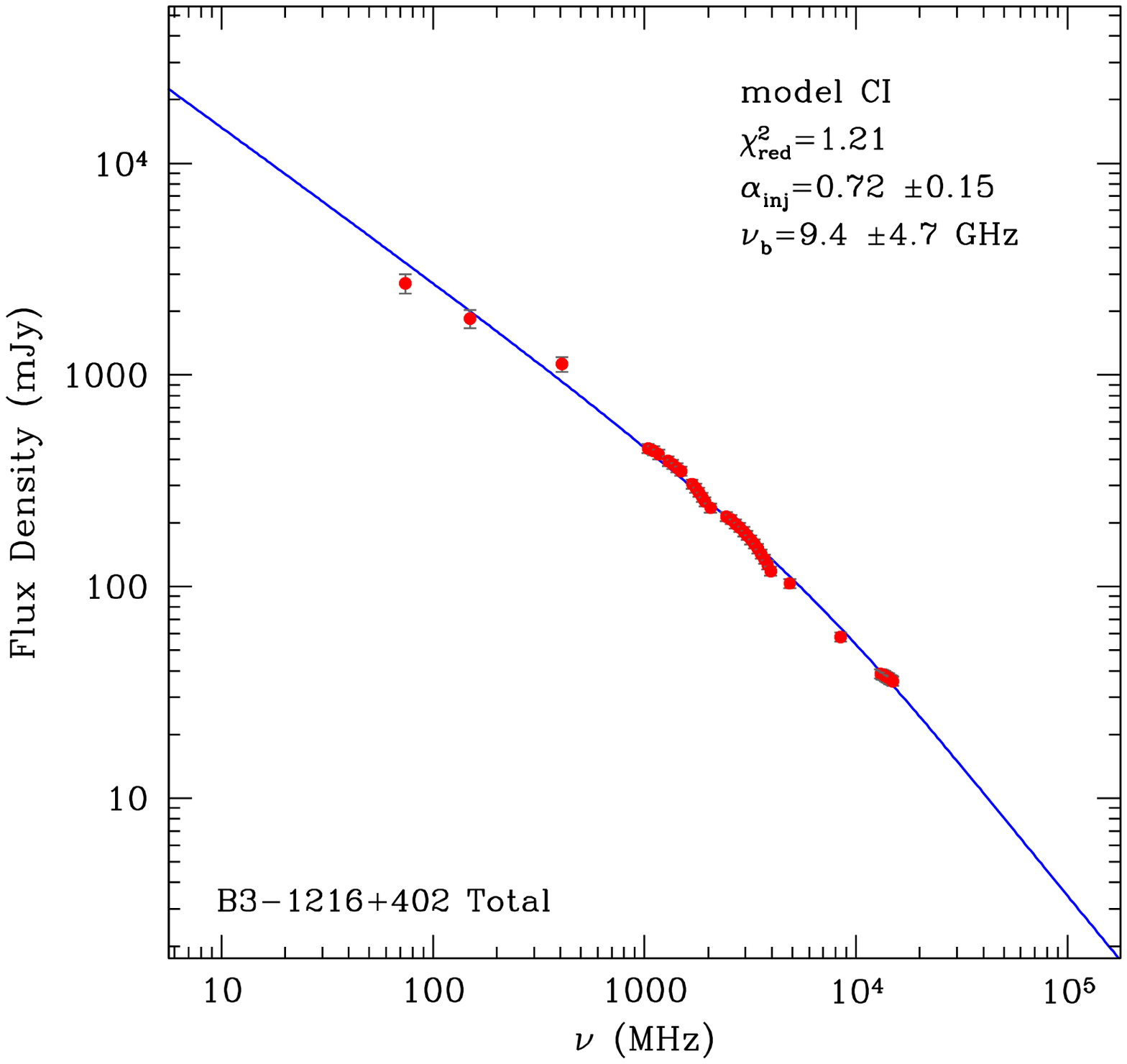}
\includegraphics{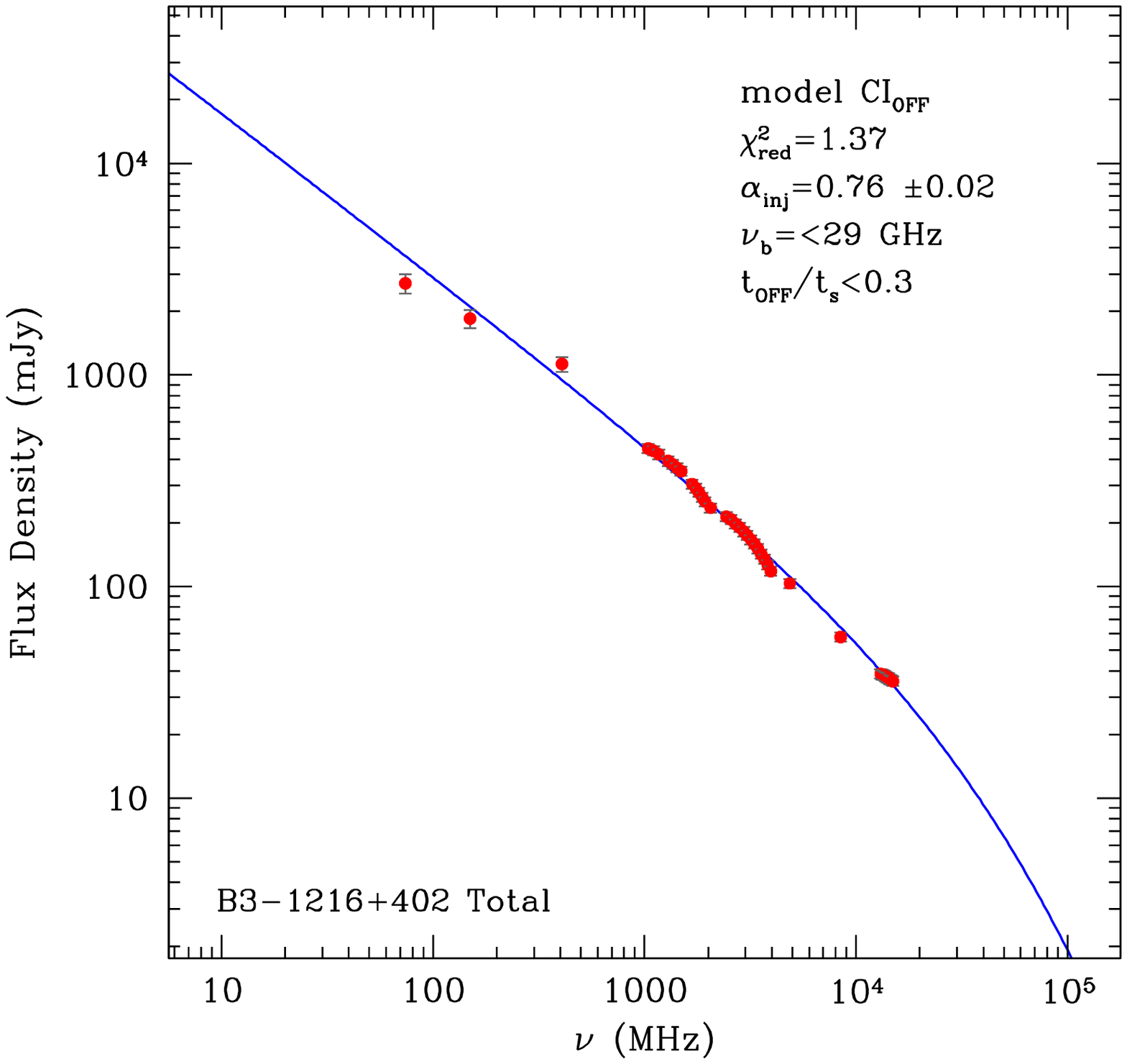}
\vspace{23cm}
\caption{Continued.}
\end{center}
\end{figure*}

\addtocounter{figure}{-1}
\begin{figure*}
\begin{center}
\includegraphics{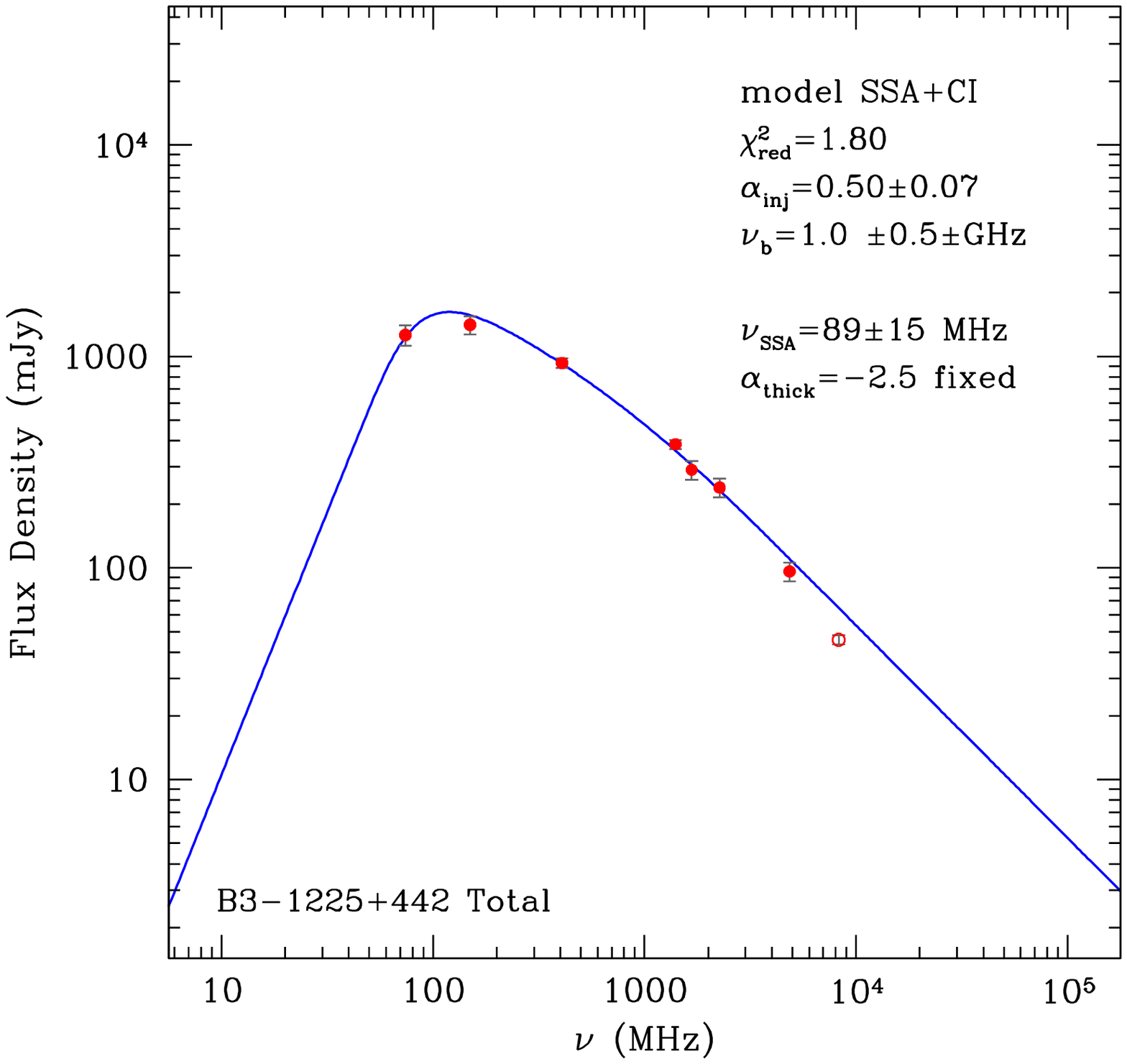}
\includegraphics{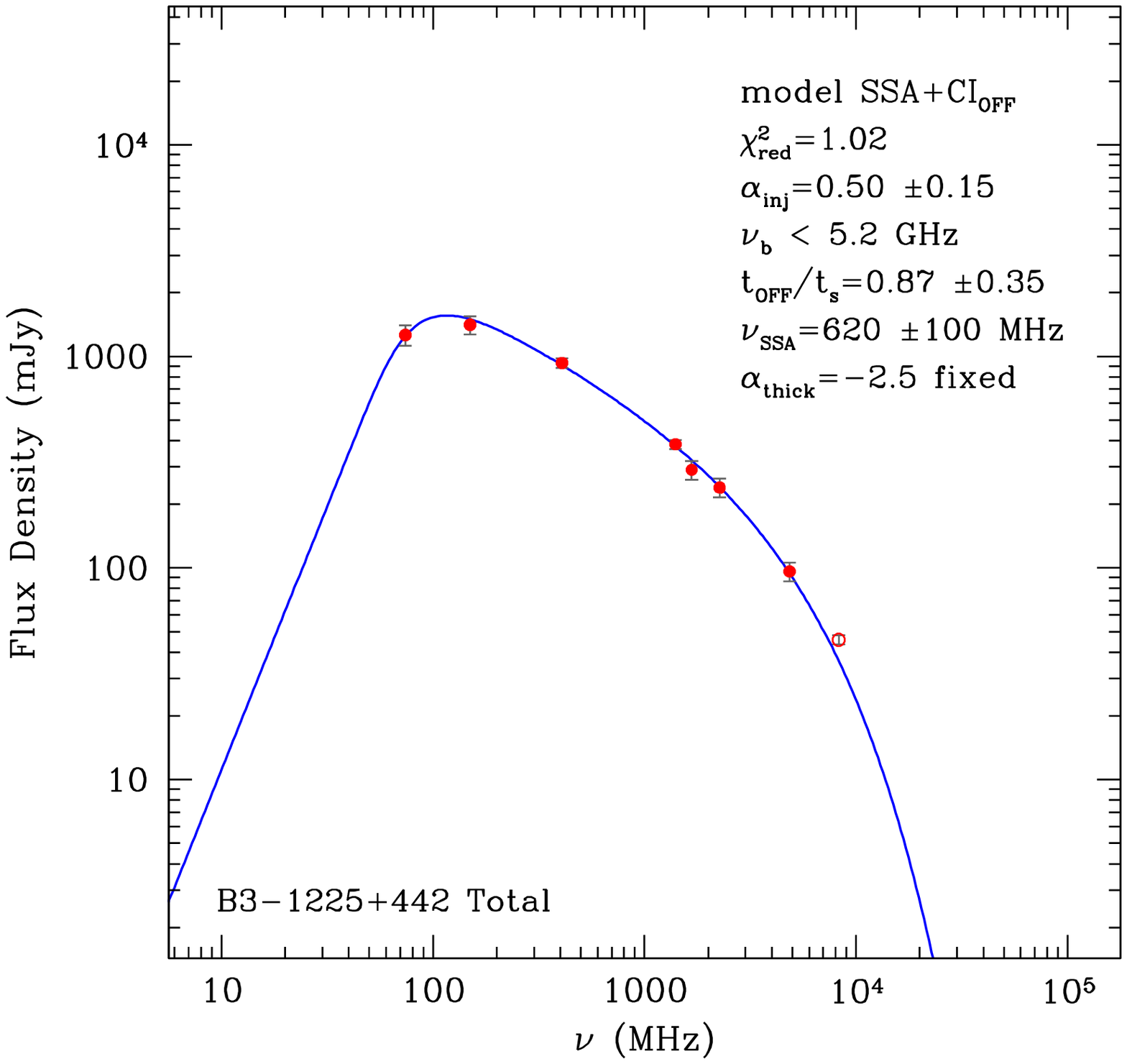}
\includegraphics{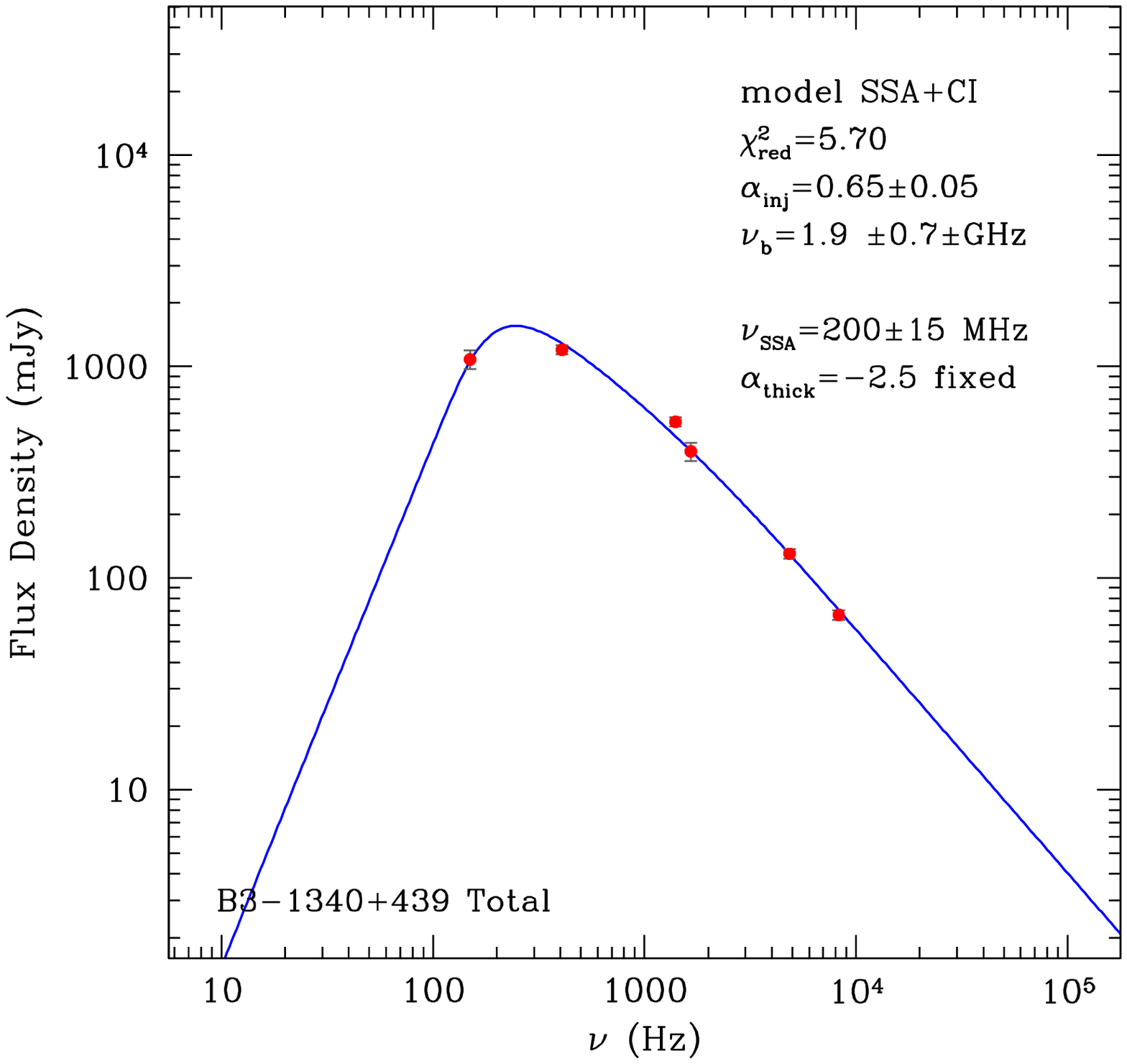}
\includegraphics{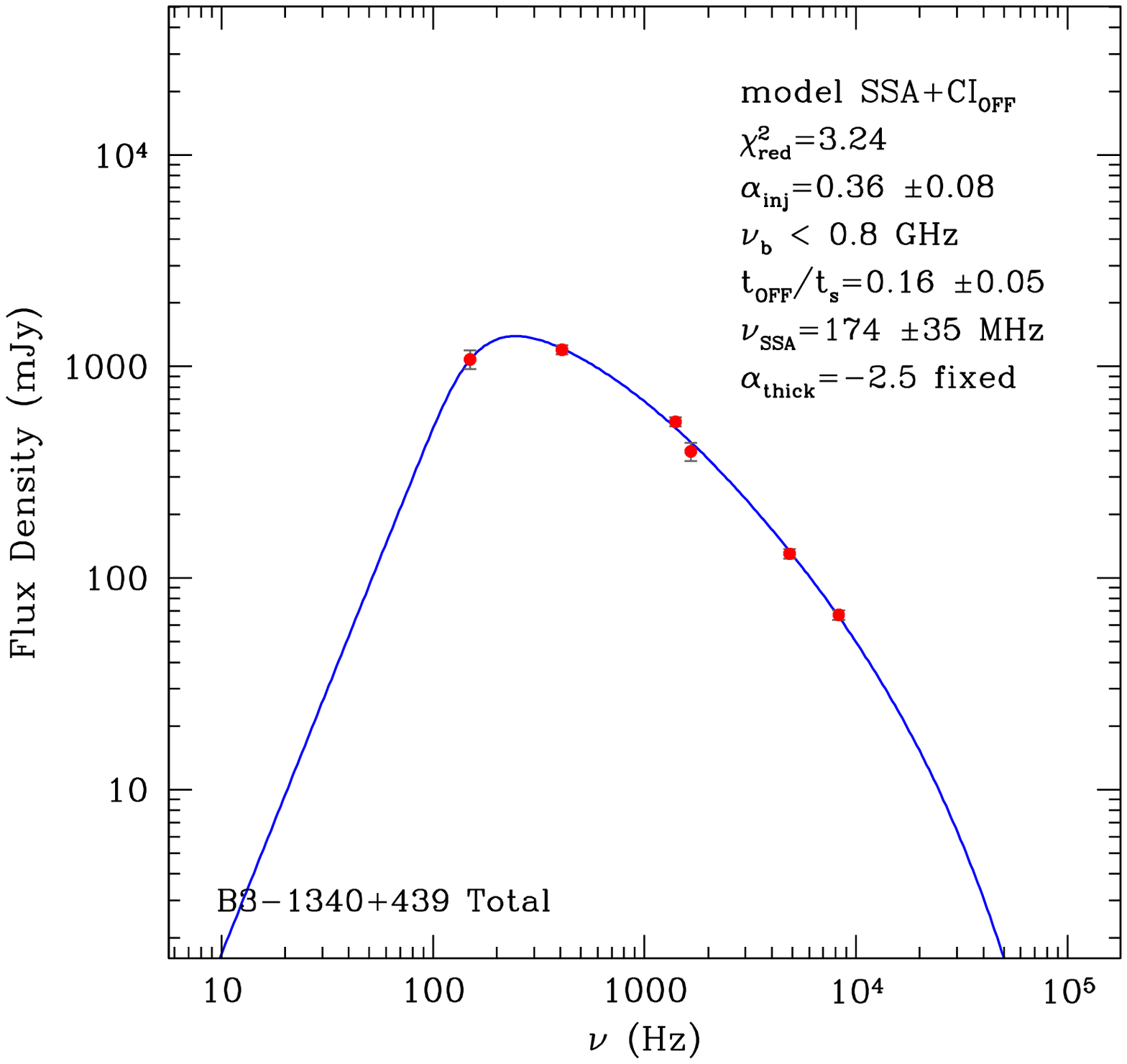}
\includegraphics{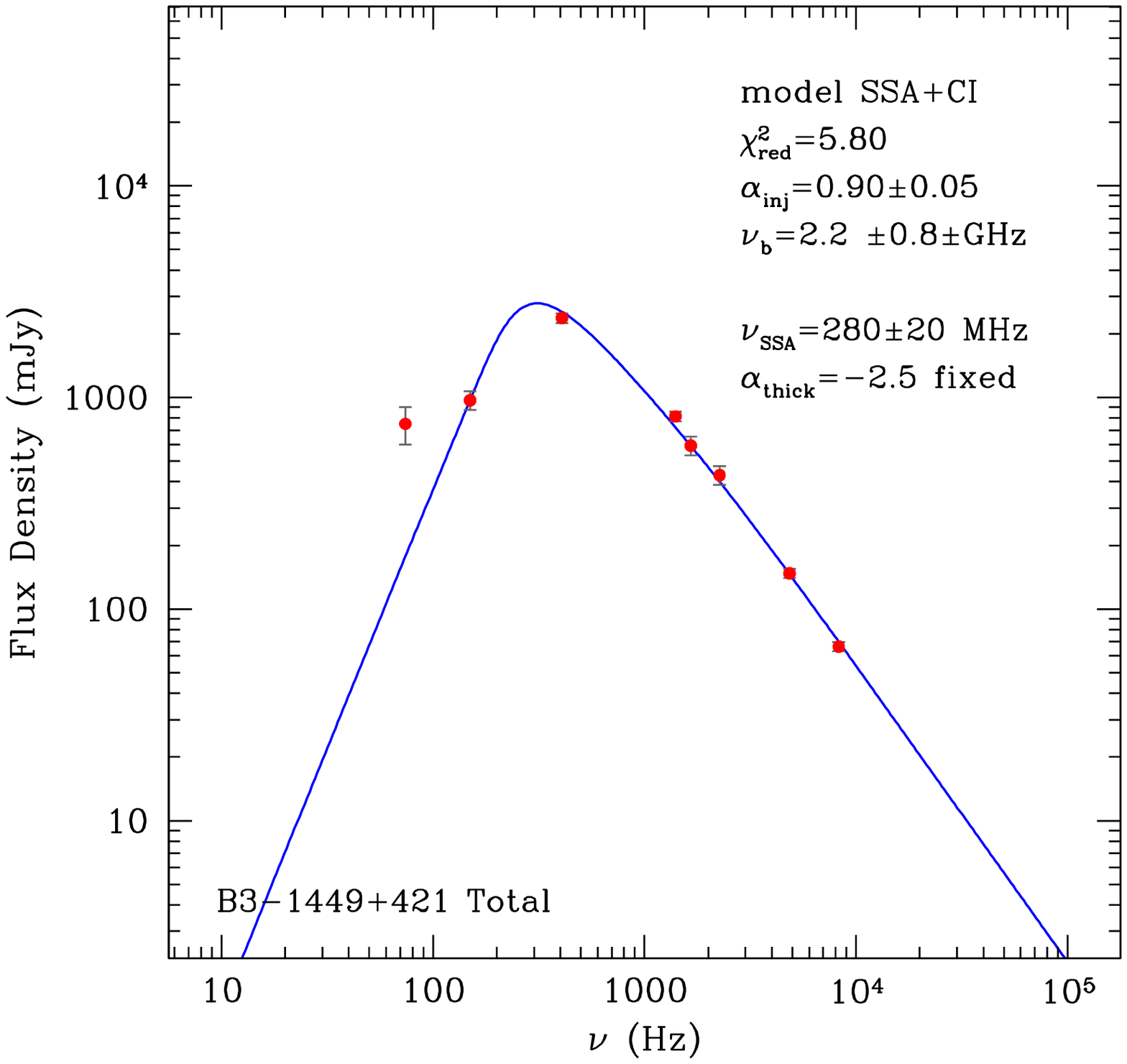}
\includegraphics{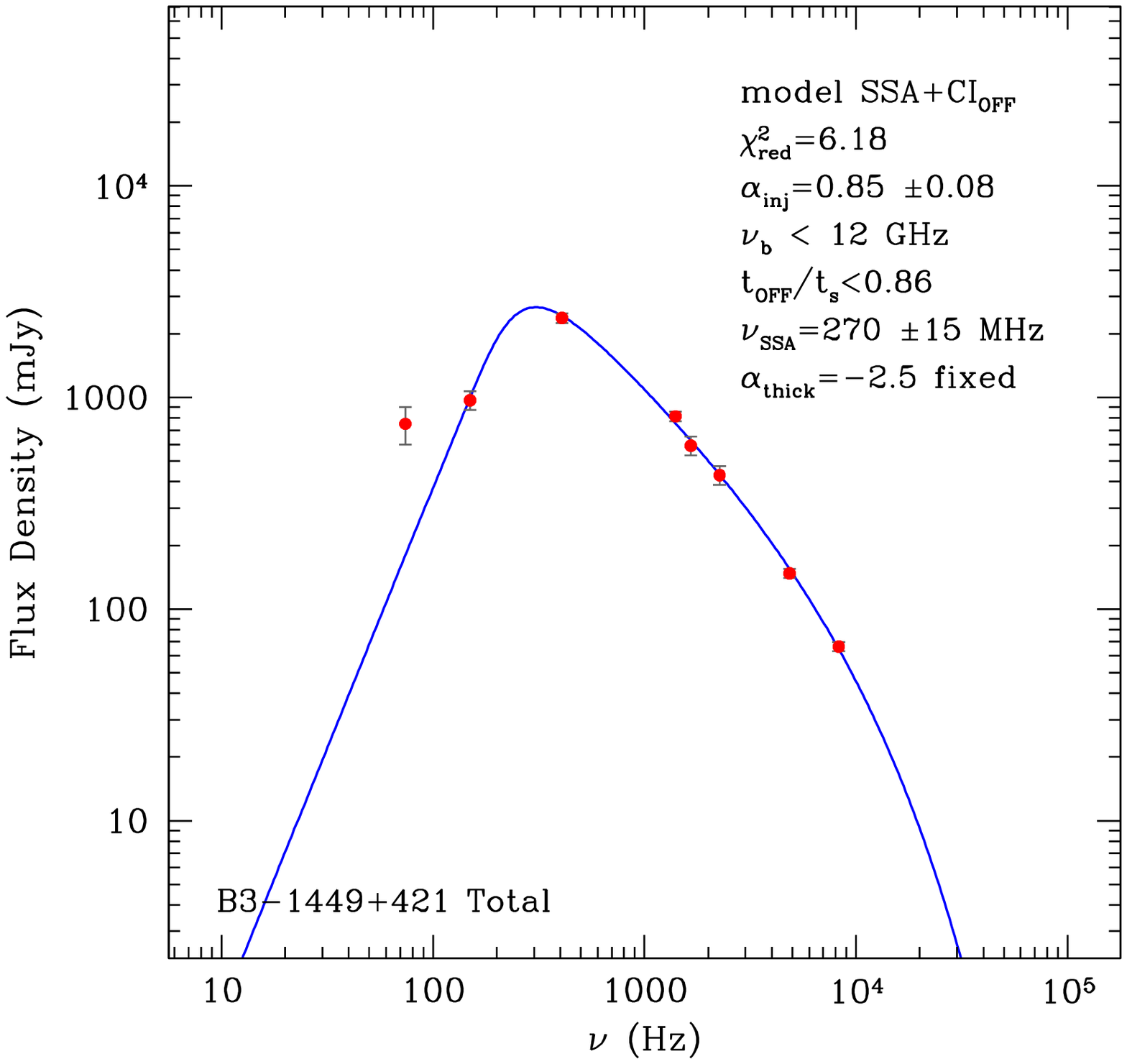}
\vspace{23cm}
\caption{Continued.}
\end{center}
\end{figure*}

\begin{table*}
  \caption{Best-fit parameters. Column 1: source name; columns 2, 3,
    4, and 5: injection spectral index, break frequency, turnover
    frequency, reduced chi-square, respectively, for the CI
    model. Columns 6, 7, 8, 9, and 10: injection spectral index, break
    frequency, the relic to total source age ratio, turnover
    frequency, reduced chi-square, respectively, for the CI OFF
    model.}
  \begin{center}
    \begin{tabular}{lccccccccc}
      \hline
      Source & \multicolumn{4}{c}{CI} & \multicolumn{5}{c}{CI OFF} \\
      & $\alpha_{\rm inj}$& $\nu_{\rm b}$ & $\nu_{\rm SSA}$ & $\chi^{2}_{\rm red}$ &
      $\alpha_{\rm inj}$ & $\nu_{\rm b}$ & $t_{\rm OFF}/t_{\rm s}$ &
      $\nu_{\rm SSA}$ & $\chi^{2}_{\rm red}$ \\
      & & GHz & MHz & & & GHz & & MHz & \\
\hline
B3-0003+387 & 0.81$\pm$0.02 & 2.6$\pm$0.3 & - & 8.1 & 0.5$\pm$0.1 &
1.0$\pm$0.4 & 0.16$\pm$0.02 & - & 0.76 \\
B3-0034+444 & 0.83$\pm$0.02 & 10.8$\pm$3.3 & - & 0.96 & 0.84$\pm$0.03 &
31$^{+2}_{-26}$ & $<$0.5 & - & 0.97 \\
B3-0128+394 & 0.68$\pm$0.03 & 2.2$\pm$0.4 & - & 1.2 & 0.57$\pm$0.10 &
1.4$\pm$0.6 & 0.06$\pm$0.03 & - & 0.87 \\
B3-0140+387 & 1.10$\pm$0.01 & 3.0$\pm$0.5 & 101$\pm$8 & 2.1 &
1.1$\pm$0.2 & 5.0$\pm$2.5 & $<$0.05 & 98$\pm$10 & 2.0 \\
B3-0748+413B& 1.06$\pm$0.01 & $>15$ & - & 7.46 & 1.06$\pm$0.01 & $>15$
& - & - & 7.63 \\
B3-0754+396 & 0.58$\pm$0.02 & 1.1$\pm$0.2 & - & 0.58 & 0.78$\pm$0.10 &
8.5$\pm$4.5 & $<$0.001 & - & 0.86 \\
B3-0810+460B& 0.59$\pm$0.01 & 0.2$\pm$0.05 & - & 0.58 & 0.99$\pm$0.01
& $>40$ & $<$0.001 & - & 0.99 \\
B3-0856+406 & 0.99$\pm$0.01 & 1.7$\pm$0.2 & - & 5.05 & 0.83$\pm$0.03 &
0.8$\pm$0.2 & 0.06$\pm$0.01 & - & 4.21 \\
B3-1016+443 & 0.78$\pm$0.07 & 2.5$\pm$1.2 & 200$\pm$20 & 3.62 &
0.80$\pm$0.07 & $<$16 & $<$0.89 & 200$\pm$15 & 4.71 \\
B3-1039+424 & 0.72$\pm$0.01 & 1.8$\pm$0.2 & - & 1.65 & 0.53$\pm$0.01 &
0.7$\pm$0.2 & 0.07$\pm$0.01 & - & 0.95 \\
B3-1049+384 & 0.55$\pm$0.10 & 1.8$\pm$1.0 & 240$\pm$20 & 2.04 &
0.55$\pm$0.15 & $<$12 & $<$0.84 & 230$\pm$30 & 2.29 \\
B3-1133+432 & 0.43$\pm$0.06 & 21$\pm$9 & 720$\pm$40 & 2.76 &
0.54$\pm$0.20 & $<$12 & $<$0.89 & 620$\pm$100 & 13.1 \\
B3-1143+456 & 0.88$\pm$0.03 & 2.3$\pm$0.4 & 106$\pm$10 & 1.36 &
0.98$\pm$0.07 & $<$15 & $<$0.27 & 107$\pm$10 & 1.79 \\
B3-1212+380 & 0.73$\pm$0.01 & 0.9$\pm$0.2 & - & 0.93 & 0.95$\pm$0.04 &
$<$30 & $<$0.30 & - & 2.35 \\
B3-1216+402 & 0.72$\pm$0.15 & 9.4$\pm$4.7 & - & 1.21 & 0.76$\pm$0.02 &
$<$29 & $<$0.3 & - & 1.37 \\
B3-1225+442 & 0.50$\pm$0.07 & 1.0$\pm$0.5 & 89$\pm$15 & 1.80 &
0.50$\pm$0.15 & $<$5.2 & 0.87$\pm$0.35 & 620$\pm$100 & 1.02 \\
B3-1340+439 & 0.65$\pm$0.05 & 1.9$\pm$0.7 & 200$\pm$15 & 5.70 &
0.36$\pm$0.08 & $<$0.8 & 0.16$\pm$0.05 & 174$\pm$35 & 3.24 \\
B3-1449+421 & 0.90$\pm$0.05 & 2.2$\pm$0.8 & 280$\pm$20 & 5.80 &
0.85$\pm$0.08 & $<$12 & $<$0.86 & 270$\pm$15 & 6.18 \\
\hline
    \end{tabular}
  \end{center}
  \label{synage_table}
  \end{table*}

\end{document}